\documentclass[12pt]{article}
\usepackage{amsmath,amssymb}
\usepackage{graphicx,psfrag,epsf}
\usepackage{enumerate}
\usepackage{natbib}
\usepackage{url} 
\usepackage{bm}
\usepackage{subfig}
\usepackage{algorithm}
\usepackage{hyperref,color}
\usepackage{multirow,multicol,array}

\graphicspath{{./}{./figs/}}

  \def\clap#1{\hbox to 0pt{\hss#1\hss}}

\providecommand{\mat}[1]{\bm{#1}}%
\renewcommand{\vec}[1]{\mathbf{#1}}

\providecommand{\mA}{\ensuremath{\mat{A}}}

\providecommand{\mC}{\ensuremath{\mat{C}}}
\providecommand{\mD}{\ensuremath{\mat{D}}}

\providecommand{\mH}{\ensuremath{\mat{H}}}
\providecommand{\mI}{\ensuremath{\mat{I}}}

\providecommand{\mW}{\ensuremath{\mat{W}}}

\providecommand{\va}{\ensuremath{\vec{a}}}
\providecommand{\vb}{\ensuremath{\vec{b}}}

\providecommand{\vg}{\ensuremath{\vec{g}}}

\providecommand{\vw}{\ensuremath{\vec{w}}}
\providecommand{\vx}{\ensuremath{\vec{x}}}
\providecommand{\vy}{\ensuremath{\vec{y}}}

\newcommand{\hLambda}{\hat{\Lambda}}

\newcommand{\hmW}{\hat{\mW}}

\newcommand{\bmat}[1]{\begin{bmatrix}#1\end{bmatrix}}

\newcommand{\brugg}{\text{brugg}}
\newcommand{\unit}[1]{\text{#1}}
\newcommand{\diag}{\operatorname{diag}}

\newcommand{\blind}{0}

\begin{document}

\def\spacingset#1{\renewcommand{\baselinestretch}%
{#1}\small\normalsize} \spacingset{1}


\if0\blind
{
\title{\bf Time-dependent global sensitivity analysis with active subspaces for a lithium ion battery model}
  \author{Paul G.~Constantine\thanks{
    We thank Dr. Mohammad Hadigol for providing us with the Li battery data set used in this work. This material is based upon (i) work of PC supported by the U.S. Department of Energy Office of Science, Office of Advanced Scientific Computing Research, Applied Mathematics program under Award Number DE-SC-0011077 and the Defense Advanced Research Projects Agency's Enabling Quantification of Uncertainty in Physical Systems and (ii) work of AD supported by the U.S. Department of Energy Office of Science, Office of Advanced Scientific Computing Research, under Award Number DE-SC0006402 and NSF grant CMMI-1454601. }\hspace{.2cm}\\
    Department of Applied Mathematics and Statistics\\ 
    Colorado School of Mines\\
    and\\
    Alireza Doostan\\
    Department of Aerospace Engineering Sciences\\ 
    University of Colorado Boulder}
  \maketitle
} \fi

\if1\blind
{
  \bigskip
  \bigskip
  \bigskip
  \begin{center}
    {\LARGE\bf Time-dependent global sensitivity analysis with active subspaces for a lithium ion battery model}
\end{center}
  \medskip
} \fi

\bigskip
\begin{abstract}
Renewable energy researchers use computer simulation to aid the design of lithium ion storage devices. The underlying models contain several physical input parameters that affect model predictions. Effective design and analysis must understand the sensitivity of model predictions to changes in model parameters, but global sensitivity analyses become increasingly challenging as the number of input parameters increases. Active subspaces are part of an emerging set of tools for discovering and exploiting low-dimensional structures in the map from high-dimensional inputs to model outputs. We extend linear and quadratic model-based heuristic for active subspace discovery to time-dependent processes and apply the resulting technique to a lithium ion battery model. The results reveal low-dimensional structure and sensitivity metrics that a designer may exploit to study the relationship between parameters and predictions. 
\end{abstract}

\noindent%
\vfill
\hfill {\tiny technometrics tex template (do not remove)}

\newpage
\spacingset{1.45} 

\section{Introduction}
\label{sec:intro}

With $\$27$ billion in yearly sales, lithium (Li) batteries are the most widely used rechargeable batteries in small portable electronics (e.g., laptops and cell-phones), satellite power systems, and the automotive industry~\citep{WiBr:04}. Assemblies of several cells in series and parallel configurations are commonly used to address issues of precise energy delivery. Although Li batteries are appealing for their high energy density and high operating voltage, their usage is restricted to low-to-medium power applications, because they often have short lifetimes and safety issues.

\begin{figure}[htb]
\centering
\includegraphics[trim =  0mm 0mm 5mm 0mm,clip,width=2.3in]{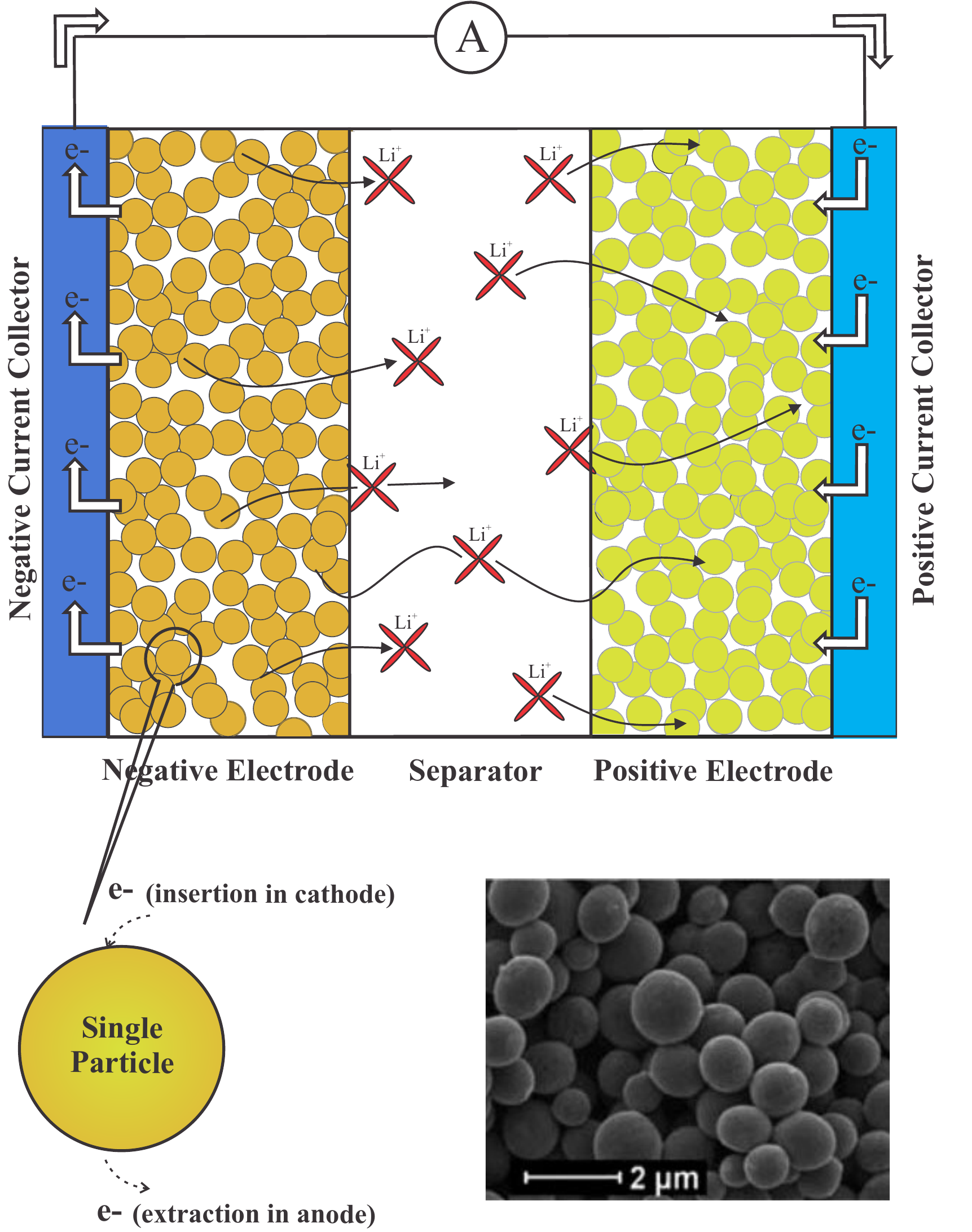}
\put(-95,0){\tiny Image from \cite{Doherty10}} 
\put(-115,35){\tiny\bf Micro }  
\put(-92,197){\tiny\bf Macro }   
\caption{Schematic of a Li battery cell identifying macro- and micro-scale models.}
\label{fig:battery}
\end{figure}

From a modeling perspective, an Li battery is a multi-scale and multi-physics system. The spatial length scales of interest range from the thickness of the overall cell (millimeters) to nanometer sized particles in electrode porous media (see Figure \ref{fig:battery}). Physical phenomena include transport processes, mechanical deformations and fracture, and electrochemical reactions. According to the physics, the dominant cause for the shortened lifetime of Li batteries is the chemical and mechanical degradation of the electrode particles as a consequence of electrical, chemical, and mechanical interactions during charge and discharge cycles. Accurate simulation of such phenomena---as well as their interactions---to predict battery performance is especially challenging for several reasons: (i) models contain many input parameters that must be set based on sparse experimental data,  (ii) manufactured cells vary due to inherent process variability, and (iii) several physical assumptions are needed to simplify the complex physics. One approach to address these difficulties is to develop data-driven battery models that account for various sources of uncertainty---e.g., parameter variability or simplified physics---and to quantify the impact of such uncertainties on the quantities of interest (QoIs), as recently advocated by \cite{Shriram07,Dua10,Hadigol2015}, among others. The emerging field of uncertainty quantification (UQ) studies systematic methodologies for the data-driven approach. 

In the probabilistic framework for UQ that we employ, system uncertainties are represented using a set of random variables $\vx\in\mathbb{R}^{m}$ with a joint probability density function $\rho(\vx)$, where $\rho$ is chosen and/or derived from data and expert opinion. The system's output quantity of interest $f(\vx)$ is a random variable; for simplicity, we consider scalar-valued quantities of interest, $f\in\mathbb{R}$. The objective of UQ is to statistically characterize the random model output---e.g., by estimating moments or a density function of $f$. For multi-physics and/or multi-scale models, such as Li batteries, the number of random variables $m$ needed to parameterize the system's uncertainty may be large. In such cases, characterizing $f(\vx)$ remains challenging due to the curse of dimensionality, where, to achieve a desired accuracy, the number of required realizations of $f(\vx)$ grows exponentially in $m$. To tackle this issue, several recent methods have relied on exploiting known structures in the mapping $\vx\rightarrow f(\vx)$, including low-rank representations, \citep{Doostan07b,Doostan09,Nouy10,Matthies12,Hadigol14}, sparse basis expansions, \citep{Doostan11a,Blatman11,Mathelin12a,Hampton15a,Hampton16}, and low-dimensional active subspaces, \citep{Constantine2014,asm2015,constantine2015computing}, to name some. 

In this paper, we demonstrate the existence of low-dimensional active subspaces in a multi-scale, multi-physics, electrochemical model of Li batteries---namely Newman's model~\citep{Newman1975,Doyle93}---incorporating several sources of uncertainty. The active subspace is the span of important directions in the input parameter space; the directions are not necessarily aligned with the input space's coordinate directions. Perturbing the inputs $\vx$ along the active subspace changes $f$ more, on average, than perturbing $\vx$ orthogonally to the active subspace. When the number of these important directions is small, $f(\vx)$ is effectively low-dimensional; methods for statistical characterization can exploit the low-dimensional structure to use fewer realizations of $f(\vx)$. Moreover, the components of the basis vectors for the active subspace can be used as sensitivity metrics to identify the input parameters that system outputs are most sensitive to. For the Li battery model, the sensitivity analysis we derive from active subspaces associated with two QoIs---cell capacity and voltage---is particularly useful in quality control and design of battery systems for improved performance.

The remainder of this manuscript is organized as follows. Section \ref{sec:sensitivity} introduces active subspaces for both stationary and time-dependent models, including how the subspaces generate sensitivity information for ranking input parameters. Section \ref{sec:asm} describes the Li battery model and the associated data set based on the numerical simulations of \cite{Hadigol2015} that are used in this work. The results of the active subspace-based sensitivity analysis applied to the battery data set are in Section \ref{sec:results}. We summarize our findings and conclusions in Section \ref{sec:conclusion}.

\section{Active subspaces}
\label{sec:sensitivity}

\subsection{Stationary models}

We present the concepts and methodologies for an abstract deterministic function of several input parameters $f(\vx)$ (e.g., the physical model's quantity of interest), and we follow the presentation in \cite{asm2015}. Let the probability density function of $\vx$, $\rho(\vx)$, be strictly positive on parameter regimes of interest, and assume that $\rho$ is such that 
\begin{equation}
\label{eq:rhoassump}
\int \vx\,\rho(\vx)\,d\vx=0 \quad\mbox{ and }\quad
\int \vx\,\vx^T\,\rho(\vx)\,d\vx=\mI,
\end{equation}
where $\mI$ is the $m\times m$ identity matrix. Note these assumptions are not restrictive; any given $\rho$ such that $\int \vx\vx^T\rho(\vx)\,d\vx$ is full admits a change of variables that satisfies \eqref{eq:rhoassump}. Assume that $f$ is differentiable with gradient vector $\nabla f(\vx)\in\mathbb{R}^m$, and assume that $f$ and its partial derivatives are square-integrable with respect to $\rho$. 

Define the $m\times m$ symmetric positive semidefinite matrix $\mC$ as
\begin{equation}
\label{eq:C}
\mC \;=\; \int \nabla f(\vx)\,\nabla f(\vx)^T\,\rho(\vx)\,d\vx.
\end{equation}
This matrix admits a real eigenvalue decomposition, 
\begin{equation}
\label{eq:Ceigs}
\mC = \mW\Lambda\mW^T,\qquad
\mW = \bmat{\vw_1 & \cdots & \vw_m},\qquad
\Lambda = \bmat{\lambda_1 & & \\ & \ddots & \\ & & \lambda_m},
\end{equation}
where the eigenvalues are in descending order. The eigenpairs are functionals of $f(\vx)$, and they reveal important properties. Lemma 2.1 from \cite{Constantine2014} shows
\begin{equation}
\lambda_i \;=\; \vw_i^T\mC\vw_i \;=\; \int \big(\nabla f(\vx)^T\vw_i\big)^2\,\rho(\vx)\,d\vx,\quad
i=1,\dots,m.
\end{equation}
In words, the $i$th eigenvalue is the average squared directional derivative of $f$ along the eigenvector $\vw_i$. Thus, $\lambda_i=0$ if and only if $f$ is constant along $\vw_i$. Moreover, if $\lambda_i$ is relatively small, then perturbations to $\vx$ along $\vw_i$ change $f$ relatively little, on average, compared to perturbations to $\vx$ along $\vw_j$ when $\lambda_j>\lambda_i$. Suppose that $\lambda_n>\lambda_{n+1}$ for some $n<m$. Then we can partition the eigenpairs as
\begin{equation}
\Lambda = \bmat{\Lambda_1 & \\ & \Lambda_2},\qquad
\mW = \bmat{\mW_1 & \mW_2},
\end{equation}
where $\Lambda_1=\diag(\lambda_1,\dots,\lambda_n)$, and $\mW_1$ contains the corresponding eigenvectors. The active subspace is the span of the columns of $\mW_1$. If the eigenvalues $\lambda_{n+1},\dots,\lambda_m$ are sufficiently small, then $f$ likely varies relatively little along the column span of $\mW_2$. In this case, we can justifiably approximate
\begin{equation}
\label{eq:ridgeapprox}
f(\vx) \;\approx\; g(\mW_1^T \vx),
\end{equation}
where $g:\mathbb{R}^n\rightarrow\mathbb{R}$. The right hand side $g(\mW_1^T\vx)$ is called a \emph{ridge function}~\citep{pinkus2015}, and it is constant along the column span of $\mW_2$. A special case of a ridge function model is used as the link function in projection pursuit regression~\citep{Friedman81,Diaconis84}. For details on the active subspace-based ridge approximation \eqref{eq:ridgeapprox}, see~\citet[Chapter 4]{asm2015} and~\citet{Constantine2014}. 

The simplest construction for $g$ is as follows. Suppose we have computed an estimate $\hmW_1$ of $\mW_1$, and suppose we have function evaluations $f_j=f(\vx_j)$ for $\vx_j$'s in $f$'s domain with $j=1,\dots,N$. In other words, we have run our computer model $N$ times with inputs $\vx_j$. Then the function $g=g(\vy)$ can be constructed from the pairs $(f_j,\vy_j)$, where $\vy_j=\hmW_1^T\vx_j$, using multivariate splines, polynomials, etc. This construction is in $n<m$ variables, so a budget of $N$ evaluations permits a higher order approximation along the relatively important directions in the input parameter space than a construction in all $m$ variables. This approach is comparable to the strategies discussed in Section 3.5 of \cite{ESL2009} using derived input directions; in this case the eigenvectors $\hmW_1$ provide the directions. 

\subsection{Model-based estimation of active subspaces}
\label{sec:modelbased}

The low-dimensional model \eqref{eq:ridgeapprox} requires an estimate of $\mW_1$. If the gradient $\nabla f(\vx)$ is available as a subroutine in the simulation code (e.g., via an adjoint solution or algorithmic differentiation~\citep{Griewank00}), then one strategy is to estimate the entries of $\mC$ from \eqref{eq:C} with a numerical integration rule and compute the numerical estimate's eigenpairs. If gradients are not available and evaluations of $f(\vx)$ are cheap enough, then a similar strategy may be employed with finite difference approximations. Finite differences are model-based approximations of the partial derivatives. For example, a first-order finite difference build a local linear model along each component of $\vx$ with two function evaluations; the slope of the local linear model provides the approximate gradient. \cite{constantine2015computing} and \citet[Chapter 3]{asm2015} analyze these strategies using simple Monte Carlo as the numerical integration method. However, many simulation codes---including the present battery simulation---do not have gradient capabilities, and the evaluations are too expensive to compute with finite differences. Therefore, we employ a model-based heuristic using a global linear model of $f(\vx)$, which we motivate as follows.

Assume that (i) $f(\vx)$ can be approximated by a linear function of $\vx$ and (ii) $f$'s gradient can be approximated by a constant vector,
\begin{equation}
\label{eq:linear}
f(\vx) \approx c + \vg^T\vx, \qquad \nabla f(\vx) \approx \vg.
\end{equation}
In this case, $\mC$'s eigendecomposition can be estimated as
\begin{equation}
\mC \;\approx\; \int \vg\,\vg^T\,\rho(\vx)\,d\vx
\;=\; \vg\,\vg^T
\;=\; \vw\,\lambda\,\vw^T,
\end{equation}
where $\lambda=\|\vg\|^2$ and $\vw=\vg/\|\vg\|$. The active subspace is approximated by the span of $\vw$---i.e., a one-dimensional subspace. We can compute $\vw$ given a set of model evaluations with Algorithm \ref{alg:lslin}.

\begin{algorithm}
Given $N>m$:
\begin{enumerate}
\itemsep0em
\item For $j$ from 1 to $N$, draw $\vx_j$ independently according to $\rho(\vx)$, and compute $f_j = f(\vx_j)$.
\item Compute $c_\ast\in\mathbb{R}$ and $\vg_\ast\in\mathbb{R}^m$ as minimizers of 
\begin{equation}
\label{eq:ls}
\underset{c,\vg}{\text{minimize}}\;\;\sum_{j=1}^N \left[
f_j - (c + \vg^T\vx_j)
\right]^2.
\end{equation}
\item Let 
\begin{equation}
\label{eq:w}
\vw \;=\; \vg_\ast\,/\,\|\vg_\ast\|,
\end{equation}
where $\|\cdot\|$ is the Euclidean norm.
\end{enumerate}
\caption{Estimate a one-dimensional active subspace with a global linear model.}
\label{alg:lslin}
\end{algorithm}

Algorithm \ref{alg:lslin} is the same as Algorithm 1.3 in~\citet[Chapter 1]{asm2015}. We emphasize that the goal of Algorithm \ref{alg:lslin} is to produce a single vector that represents an important direction in the space of $f$'s inputs. We do not use the least-squares-fit linear approximation of $f(\vx)$ as a predictive response surface. 

There are three ways that Algorithm \ref{alg:lslin} can fail to find an important direction in $f$'s input space, and all are related to the assumptions in \eqref{eq:linear}. The first is that $f(\vx)$ may vary significantly along more than one direction; the extreme example is a radially symmetric function, e.g., $f(\vx)=\vx^T\vx$. In this case, there are no directions in the input space that are more important---for any sense of importance---than any others with respect to $f$. The second case is best exemplified by the function $f(\vx)=(\va^T\vx)^2$ for some fixed $\va\in\mathbb{R}^m$, which is symmetric about the origin along $\va$. In this case, there is only one direction along which $f$ varies; perturbing $\vx$ orthogonally to $\va$ does not change $f$. However, since $f$ is symmetric about the origin along $\va$, the gradient of the least-squares-fit linear model converges to zero as the number $N$ of samples increases. Any nonzero gradient is due to finite samples, so the linear model gradient has no relation to the true $\va$. The third case is best exemplified by the function $f(\vx)=\exp(\gamma(\va^T\vx))$ for fixed $\va\in\mathbb{R}^m$ and $\gamma\gg 1$. In this case, there is again only one direction along which $f$ varies. However, for large $\gamma$ the vector $\vg_\ast$ from Algorithm \ref{alg:lslin} will be strongly influenced by samples of $\vx$ from Step 1 that produce large values of $\va^T\vx$. As the number $N$ of samples increases, $\vw$ from \eqref{eq:w} converges to a unit vector that points in the direction of $\va$. But for very large $\gamma$, $N$ might need to be very large for Algorithm \ref{alg:lslin} to produce a good estimate of the important direction. When assessing the suitability of $\vw$ from \eqref{eq:w} for describing the important direction in $f$'s input space, it is important to distinguish between the last case, where more samples gives a better estimate, and the first two cases, where any number of samples is insufficient---either because there is no one-dimensional structure in the map from $\vx\rightarrow f(\vx)$ or the method is not capable of recovering the one-dimensional structure. 

One heuristic approach to address the first two cases is to use a different model to estimate the gradient of $f$. Assume that (i) $f(\vx)$ can be approximated by a quadratic function of $\vx$ and (ii) $f$'s gradient can be approximated by as follows,
\begin{equation}
\label{eq:quad}
f(\vx) \approx c + \vg^T\vx + \frac{1}{2}\,\vx^T\mH\vx, \qquad \nabla f(\vx) \approx \vg + \mH\vx.
\end{equation}
In this case, $\mC$'s eigendecomposition can be estimated as
\begin{equation}
\label{eq:Cquad}
\mC \;\approx\; \int (\vg+\mH\vx)\,(\vg+\mH\vx)^T\,\rho(\vx)\,d\vx
\;=\; \vg\,\vg^T + \mH^2
\;=\; \hmW\,\hLambda\,\hmW^T,
\end{equation}
where the first equality follows from assumptions on $\rho(\vx)$. We treat the eigenpairs $\hmW$, $\hLambda$ like we treat the eigenpairs of $\mC$ in \eqref{eq:Ceigs}; a large gap between the $n$th and $(n+1)$th eigenvalues followed by small eigenvalues $\lambda_{n+1},\dots,\lambda_m$ provides evidence of an exploitable active subspace defined as the span of the first $n$ columns of $\hmW$. Algorithm \ref{alg:lsquad} outlines the associated algorithm.

\begin{algorithm}
Given $N>{m+2 \choose 2}$:
\begin{enumerate}
\itemsep0em
\item For $j$ from 1 to $N$, draw $\vx_j$ independently according to $\rho(\vx)$, and compute $f_j = f(\vx_j)$.
\item Compute $c_\ast\in\mathbb{R}$, $\vg_\ast\in\mathbb{R}^m$, and $\mH_\ast\in\mathbb{R}^{m\times m}$ as minimizers of 
\begin{equation}
\label{eq:lsq}
\underset{c,\vg,\mH}{\text{minimize}}\;\;\sum_{j=1}^N 
\left[
f_j - \left(c + \vg^T\vx_j + \frac{1}{2}\vx_j^T\mH\vx_j \right)
\right]^2.
\end{equation}
\item Compute the eigenpairs
\begin{equation}
\label{eq:Heigs}
\vg_\ast\,\vg_\ast^T + \mH_\ast^2
\;=\; \hmW\,\hLambda\,\hmW^T.
\end{equation}
\end{enumerate}
\caption{Estimate an active subspace with a global quadratic model.}
\label{alg:lsquad}
\end{algorithm}

Three notes about Algorithm \ref{alg:lsquad}. First, the least-squares problem in \eqref{eq:lsq} is written to emphasize the matrix $\mH$. In fact, this is a linear least-squares problem in all unknowns. Second, the number of samples needed to fit the global quadratic model in \eqref{eq:lsq} is ${m+2 \choose 2}$, which grows like $\mathcal{O}(m^2)$---asymptotically much larger than the $N>m$ samples needed for the linear model in Algorithm \ref{alg:lslin}. Third, the matrix $\vg_\ast\,\vg_\ast^T + \mH_\ast^2$ is symmetric and positive semidefinite. The relationship between the eigenspaces of this matrix and the eigenspaces of $\mH$ depends on (i) how $\vg$ relates to the eigenspaces of $\mH$ and (ii) how the eigenspaces of $\mH$ (the Hessian of the quadratic model), which is in general symmetric but not positive semidefinite, relate to the eigenspaces of $\mH^2$. There is no general formula that elucidates their relationship. Techniques derived from pure Hessians arise in the \emph{canonical analysis} of quadratic response surfaces from \cite[Section 6.3.1]{Myers1995} and the \emph{likelihood-informed subspaces} in Bayesian inverse problems from \cite{Cui2014} and \cite{Cui2016}.

We emphasize again that the least-squares-fit quadratic model is used to compute the eigenpairs $\hmW$, $\hLambda$. The fitted curve is not used as a predictive response surface for $f(\vx)$. The relationship between the computed $\hmW$ and $\hLambda$ and the eigenpairs of $\mC$ from \eqref{eq:Ceigs} depends on (i) the validity of the assumptions \eqref{eq:quad} and (ii) the number $N$ of samples in Algorithm \ref{alg:lsquad}.

\subsection{Validating important directions with summary plots}

Algorithms \ref{alg:lslin} and \ref{alg:lsquad} provide two model-based approaches for estimating active subspaces; the linear model-based approach can only estimate a one-dimensional active subspace. No matter how we estimate the vector $\vw$ that defines a one-dimensional active subspace---using \eqref{eq:w} in Algorithm \ref{alg:lslin} or the first eigenvector in \eqref{eq:Heigs} from Algorithm \ref{alg:lsquad} or any other approach---we can easily check its validity with a \emph{summary plot}. Summary plots were developed in the context of sufficient dimension reduction for statistical regression~\citep{cook2009regression}; we provide more details in the next subsection. In the current setting, a summary plot is a scatter plot of $\vw^T\vx_j$ versus $f_j$, where $(\vx_j,f_j)$ are the input/output pairs used to fit the multivariate curves in Algorithms \ref{alg:lslin} and \ref{alg:lsquad}. The summary plot may reveal a relationship between the particular linear combination of the inputs, $\vw^T\vx$, and the simulation model's output $f$. If the points in the scatter plot reveal a univariate or near univariate relationship, then the ridge approximation \eqref{eq:ridgeapprox} with $\mW_1=\vw$ is a good approximation to $f(\vx)$. If the points in the scatter plot do not reveal a functional relationship, then the model \eqref{eq:ridgeapprox} with $\mW_1=\vw$ may not be appropriate. 

\subsection{Comparison to sufficient dimension reduction in statistical regression}

We emphasize yet again that Algorithms \ref{alg:lslin} and \ref{alg:lsquad} are not meant to produce predictive response surfaces or multivariate approximations for $f(\vx)$. The fitted linear and quadratic models are used to estimate active subspaces for the map $\vx\rightarrow f(\vx)$---where the linear model-based Algorithm \ref{alg:lslin} is only capable of estimating a one-dimensional active subspace. How one uses the active subspaces---once estimated by whatever means---depends on the question one seeks to answer with $f(\vx)$. See \cite[Chapter 4]{asm2015} for a discussion of exploiting active subspaces for response surface construction, integration, optimization, and inverse calibration. 

The nearest problem set up in the statistical literature falls under the subfield of \emph{sufficient dimension reduction} for regression. \cite{cook2009regression} gives a complete exposition; we follow his notation and problem set up for comparison to active subspaces. The goal in sufficient dimension reduction is to find a subspace that is statistically sufficient to characterize the predictor/response relationship. More precisely, let $[y_j,\vx_j]^T\in\mathbb{R}^{m+1}$ with $j=1,\dots,N$ be independent samples from an unknown joint density $\pi(y,\vx)$. (Note the important difference between independent predictor/response samples and the data generation step in Step 1 of both Algorithms \ref{alg:lslin} and \ref{alg:lsquad}.) The goal is to identify a matrix $\mA\in\mathbb{R}^{m\times n}$ with $n\leq m$ such that
\begin{equation}
\label{eq:sdr}
F_{y|\vx}(t) \;=\; F_{y|\mA^T\vx}(t), \qquad t\in \mathbb{R},
\end{equation}
where $F_{y|\vx}(\cdot)$ is the conditional distribution of the response given the predictors and $F_{y|\mA^T\vx}(\cdot)$ is the conditional distribution of the response given a linear combination of predictors $\mA^T\vx$. The equality of the condition distributions implies that $\mA^T\vx$ is statistically sufficient to characterize $y$. 

Assuming that (i) such an $\mA$ exists and (ii) $\mA$ has one column (i.e., $n=1$), the summary plot $(\mA^T\vx_j,y_j)$ is a \emph{sufficient} summary plot, where the qualifier \emph{sufficient} is critical to the proper statistical interpretation of $\mA$ and the associated plot. In essence, \eqref{eq:sdr} implies that any perceived departure from a pure univariate relationship in a sufficient summary plot is due to the random variation in the response independent of the predictors---i.e., random noise. In contrast, there is no notion of sufficiency in the summary plots we produce with the vector $\vw$ from \eqref{eq:w} or the first eigenvector from \eqref{eq:Heigs}. Departure from a perceived univariate trend in the summary plot is not due to independent random noise because there is no such noise in $f(\vx)$. Instead, such variation is either because (i) $f(\vx)$ varies orthogonally to $\vw$ or (ii) the algorithm did not accurately estimate $\vw$. As such, we cannot rely on the deep and extensive theory developed in the statistical literature to test for sufficient dimension reduction. Such theory is not applicable to our problem set up. Therefore, conclusions drawn from Algorithms \ref{alg:lslin} and \ref{alg:lsquad} and their associated summary plots are qualitative and subjective. We discuss bootstrap-based heuristics for assessing the validity of such qualitative conclusions in Section \ref{sec:boot}. 

Though the problem set ups differ substantially, there are computational elements of Algorithms \ref{alg:lslin} and \ref{alg:lsquad} that are similar to methods proposed in the sufficient dimension reduction literature. The linear model-based approach is similar to the \emph{ordinary least squares} method proposed by \cite{Li1989}, where they discuss how recovering the coefficients of the linear model is robust to particular misspecifications of the link function. The quadratic model-based approach is related to the \emph{principal Hessian directions} method of \cite{Li1992}; see Corollary 3.2, in particular. 

More generally, a matrix similar to $\mC$ from \eqref{eq:C} that defines active subspaces has been studied in the context of regression by \cite{Hristache2001} and \cite{Samarov1993}, who called this matrix one of several \emph{average derivative functionals}. Both \cite{Xia2007} and \cite{Fukumizu2014} use gradients of kernel-based estimates of the regression function to estimate the dimension reduction subspace. If we place these computational methods in our context, they lead to methods similar to Algorithms \ref{alg:lslin} and \ref{alg:lsquad}, except the underlying model used to estimate the gradient is constructed with radial basis function approximation~\citep{wendland2004scattered} instead of a least-squares-fit polynomial. For a more in-depth comparison of sufficient dimension reduction to deterministic ridge approximation, see \cite{Glaws2017}.

\subsection{Sensitivity metrics from active subspaces}

If a near univariate relationship is present in the summary plot, then the components of $\vw$ provide sensitivity information. The normalization in \eqref{eq:w} and the eigenvectors from \eqref{eq:Heigs} implies that each component is between -1 and 1. A component with a relatively large magnitude indicates that the corresponding parameter is important in defining the important subspace. Often in practice, the eigenvector components with large magnitudes correspond to parameters with relatively large standard sensitivity metrics, e.g., Sobol' indices~\citep{constantine2015discovering,diaz2015global}. Moreover, if the functional relationship in the summary plot is monotonic---i.e., $f$ appears to be an increasing or decreasing function of $\vw^T\vx$ as assessed by the summary plot---then the sign of each eigenvector component reveals how $f$ changes in response to changes in the corresponding parameter, on average. For example, assume that $f$ is an increasing function of $\vw^T\vx$. And assume that $\vw$'s first component is negative with a relatively large magnitude. Then, on average, a positive perturbation to $x_1$ decreases $f$. Note that the signs should be treated relative to the perceived trend in the summary plot. 

In Section \ref{sec:results}, we apply Algorithm \ref{alg:lslin} to a Li battery model and create summary plots that give insight into the relationship between model's inputs and its outputs. However, the quantity of interest in this model also depends on a notion of time. We next extend the active subspace and ridge approximation to time-dependent quantities of interest. 

\subsection{Time-dependent models}
\label{sec:timedep}

Suppose that the quantity of interest depends on parameters $\vx\in\mathbb{R}^m$ and another independent coordinate $t\in\mathbb{R}$, which we may interpret as time. In other words, $f=f(\vx,t)$ is a parameter dependent temporal process. There are several ways to construct an active subspace for such a process. We could assume a finite time interval of interest $t\in[0,T]$ and treat $t$ as another parameter. The extended version of $\mC$ from \eqref{eq:C} becomes
\begin{equation}
\label{eq:Ct}
\mC \;=\; \frac{1}{T} \int \left(\int \bmat{f_t(\vx,t)^2 & f_t(\vx,t)\,\nabla f(\vx,t)^T \\ f_t(\vx,t)\,\nabla f(\vx) & \nabla f(\vx,t)\,\nabla f(\vx,t)^T} \,\rho(\vx)\,d\vx \right)\,dt,
\end{equation}
where $f_t(\vx,t)$ is the partial derivative of $f$ with respect to $t$. Note that $\mC$ from \eqref{eq:Ct} admits a block structure,
\begin{equation}
\label{eq:Cblock}
\mC \;=\; \bmat{a & \vb^T \\ \vb & \mD},\quad
a\in\mathbb{R}, \;\;\vb\in\mathbb{R}^m, \;\;\mD\in\mathbb{R}^{m\times m},
\end{equation}
where the blocks are apparent when comparing \eqref{eq:Ct} to \eqref{eq:Cblock}. We could construct a global subspace of $\mathbb{R}^m$ using the eigenvectors of the lower right block $\mD$, which is symmetric and positive semidefinite. This would be equivalent to averaging a time dependent analog of $\mC$ from \eqref{eq:C} over the time interval and computing its eigenvectors. If $f(t,\vx)$ is sufficiently smooth and bounded, this is equivalent to defining the scalar-valued quantity of interest to be the time-averaged $f(t,\vx)$. However, in many cases, the dynamics of the quantity of interest are important to the application.

Another approach is to treat $\mC$ from \eqref{eq:C} as a matrix whose elements depend on $t$, i.e.,
\begin{equation}
\mC(t) \;=\; \int \nabla f(\vx,t)\,\nabla f(\vx,t)^T \,\rho(\vx)\,d\vx \;=\;
\mW(t)\,\Lambda(t)\,\mW(t)^T,
\end{equation}
where the eigendecomposition is computed independently for each $t$. Spectral decompositions of parameter dependent linear operators are well studied; see~\cite{kato1966} for a complete treatment. We simplify the mathematics dramatically by considering a finite collection of points $t_k\in[0,T]$ with $k=1,\dots,P$. In effect, each of the $P$ sets of $m$ eigenpairs is indexed by $k$. \cite{loudon2016mathematical} use this approach to study the time dependent active subspace of a quantity of interest from a dynamical system model of HIV infection. The eigenvalues at time $t_k$ indicate low-dimensional structure in the map from inputs to output at $t_k$, so one can study how that structure changes over time. 

Similarly, we can use the model-based heuristics from Algorithms \ref{alg:lslin} and \ref{alg:lsquad} at each $t_k$, and we can create summary plots for each $t_k$. The result is an animation (one summary plot for each $t_k$) that reveals how well a one-dimensional subspace---defined by $\vw_k=\vw(t_k)$ from either \eqref{eq:w} or the first eigenvector in \eqref{eq:Heigs} at each $t_k$---captures the relationship between inputs and outputs over the time range defined by $t_1,\dots,t_P$. If the summary plots reveal univariate trends, then the components of $\vw_k$ can be plotted versus $t_k$ to study the sensitivity of $f$ with respect to $\vx$'s components over time. A dramatic change in sensitivities may reveal a transition between physical regimes of a system. 

\subsection{Assessing uncertainty in the active subspace}
\label{sec:boot}

Algorithms \ref{alg:lslin} and \ref{alg:lsquad} first collect pairs $(\vx_j,f_j)$. The $\vx_j$'s are typically chosen according to design-of-experiments criteria consistent with the given density $\rho(\vx)$, and some criteria lead to random designs. For example, simple Monte Carlo draws each $\vx_j$ independently according to $\rho(\vx)$. With random designs, it is natural to ask how robust the vector $\vw$ from \eqref{eq:w} is to the randomness. Practical uncertainty estimates are very difficult to derive for general nonlinear $f(\vx)$'s. We apply a nonparametric bootstrap~\citep{efron1994introduction} as a heuristic to assess the uncertainty in the vector $\vw$ used in the summary plot---computed from \eqref{eq:w} in Algorithm \ref{alg:lslin} or as the first eigenvector from \eqref{eq:Heigs} in Algorithm \ref{alg:lsquad}. In particular, we can compute the bootstrap standard error for each component of $\vw$ by following the standard sampling-with-replacement recipe. See \citet[Chapter 7]{efron1994introduction} for a related example applying the bootstrap to linear regression coefficients. We note that the interpretation of the standard error is not the same as in the statistical estimation, since the function values $f_j$ are not corrupted by random noise. Nevertheless, a large bootstrap standard error may indicate (i) the data are not sufficient to compute $\vw$ or (ii) the relationship between $\vx$ and $f(\vx)$ cannot be summarized with one linear combination of $\vx$. There is still work to be done to devise more precise characterizations of bootstrap standard errors in our setting with noiseless $f(\vx)$. 

In the time-dependent case, we can plot the bootstrap standard errors at each $t_k$ to see how they change over time. A large change from one point in time to the next may reveal that adequacy of the one-dimension subspace for capturing the relationship between $\vx$ and $f$ also changes. Such information may yield insight into the model's relationship between inputs and outputs. 

\cite{constantine2015computing} propose the bootstrap as a heuristic to assess uncertainty in estimates of the active subspace. \cite{constantine2014exploiting} use the bootstrap to assess uncertainty in the components $\vw$ from Algorithm \ref{alg:lslin} for analyzing parameter dependence in a numerical model of a scramjet-powered vehicle. 

\subsection{Advantages and limitations}

Executing Algorithm \ref{alg:lslin} and producing the summary plot is remarkably cheap. Only enough samples are needed to fit a linear model of $f(\vx)$; recall that $\vx$ has $m$ components. From a linear algebra perspective, this can be accomplished with $N=m+1$ runs if $f$ is a linear function of $\vx$---since the components of $\vx$ are assumed independent. However, there is no reason to think that the quantity of interest from a complex physical simulation is exactly linear---even if it can be well approximated by a linear model. Therefore, we advise oversampling---i.e., choosing $N>m+1$. Recent work by \cite{Hampton2015b} in deterministic least-squares approximation with random evaluations shows that choosing $N=\mathcal{O}(m)$ produces a linear model that behaves like the best linear approximation in the continuous, mean-squared sense. 

The downside of the linear model-based Algorithm \ref{alg:lslin} compared to the quadratic model-based \ref{alg:lsquad} is that the former can only estimate a one-dimensional active subspace. If the summary plot shows $f(\vx)$ departing from a univariate function of $\vw^T\vx$, then one is left wondering whether this departure is due to variation in $f$ orthogonal to $\vw$ or the algorithm's natural drawbacks; see Section \ref{sec:modelbased}. The benefit of the additional cost in terms of evaluations of $f(\vx)$ (roughly, quadratic in the input space dimension $m$) is that the quadratic model-based Algorithm \ref{alg:lsquad} provides an opportunity to assess the reasons for a summary plot's spread. 

The proper way to think of the heuristics from Algorithms \ref{alg:lslin} and \ref{alg:lsquad} and their associated summary plots is as a set of cheap tests for a particular type of low-dimensional structure, namely $f(\vx)\approx g(\vw^T\vx)$ for $g:\mathbb{R}\rightarrow\mathbb{R}$.  Similar to hypothesis tests, the test may return a false positive or a false negative. However, since the data is not corrupted by random noise, the formalism for regression hypothesis testing is not appropriate. In other words, the lack of randomness in the function evaluations implies that hypothesis tests do not have a statistically valid interpretation. Instead, we must consider the conditions that might lead to an incorrect conclusion or inconclusive results about the presence of the low-dimensional structure. 

Assume that the summary plot suggests a functional relationship between the active variable $\vw^T\vx$ and the output $f(\vx)$. The summary plot is equivalent to viewing the relationship between $\vx$ and $f(\vx)$ from one off-axis perspective defined by $\vw$. This view collapses the $m$-dimensional input space to a one-dimensional interval. Input values that appear to be close in the one-dimensional interval may be very far apart in the $m$-dimensional space. More precisely, for two inputs $\vx_1$ and $\vx_2$, $|\vw^T\vx_1-\vw^T\vx_2|$ may be small when $\|\vx_1-\vx_2\|$ is large. If $f$ varies dramatically in a small region of the input space, then it is possible that sparsely sampled input points\footnote{$\mathcal{O}(m)$ points in $m$ dimensions is very sparse for $m>2$.} used to construct the linear or quadratic model and produce the summary plot missed the small region of dramatic variability. Resolving the region of variability (assuming sufficient sampling were possible) might change the conclusions from summary plot. In other words, this is a false positive. One way to test for this error is to compute an independent testing set and see if it satisfies the perceived functional relationship. However, any testing set short of densely sampling the $m$ dimensional input space may produce the same false positive. In practice, we have never experienced such a false positive. We suspect this is because many physical models have outputs that vary smoothly with changes in the inputs; in other words, the dramatic variation that might lead to a false positive seems largely absent from physics-based simulation models. 

Assume that the summary plot does not suggest a functional relationship between $\vw^T\vx$ and $f(\vx)$. In other words, the plot of $\vw^T\vx_j$ versus $f_j$ looks like a collection of random points. There are two possible explanations for this. First, $f(\vx)$ may vary substantially along more than one direction; no matter how the direction $\vw$ is computed, the summary plot would always show variation in $f$ orthogonal to $\vw$. The eigenvalues from \eqref{eq:Heigs} in the quadratic model-based Algorithm \ref{alg:lsquad} may provide some indication of $f$'s variation orthogonal to $\vw$. Second is that the one-dimensional structure is present in $f(\vx)$ but the method for computing $\vw$ did not find the right direction. If the assumptions \eqref{eq:linear} or \eqref{eq:quad} that motivate Algorithms \ref{alg:lslin} and \ref{alg:lsquad}, respectively, are significantly violated, then the computed direction $\vw$ (\eqref{eq:w} in Algorithm \ref{alg:lslin} and the first eigenvector from \eqref{eq:Heigs} from Algorithm \ref{alg:lsquad}) may be nowhere near the true direction.

\section{Application to a Li battery model}
\label{sec:asm}

We apply the active subspace-based techniques to a data set derived from a numerical simulation of a Li battery model described by \cite{Hadigol2015}, who developed the simulation for an uncertainty quantification study. Our goal is to assess the existence of a time-dependent active subspace in the map from battery inputs to outputs of interest---namely, voltage and capacity as a function  parameters that control the physical and chemical processes. Given an active subspace, we quantify the time-dependent sensitivity of outputs with respect to each input parameter. We briefly summarize the most important aspects of the Li battery model for our purposes, and we refer the interested reader to \cite{Hadigol2015} for details and references that provide context in the battery modeling literature. 

Newman's model~\citep{Newman1975} is a coupled system of nonlinear differential equations that describes a Li battery cell as an \emph{anode} and \emph{cathode} separated by a \emph{separator}. We provide the governing equations in Appendix \ref{sec:goveq}. As the cell discharges, Li$^+$ ions diffuse from anode to cathode through the separator, and electrons flow through the external circuit from the negatively charged electrode (the anode) to the positively charged electrode (the cathode). The flow of electrons creates electrical current that powers an electronic device. Newman's model includes several parameters that characterize the physical processes and material properties affecting the cell's power generation. For our purposes, we identify the model's output quantities of interest and its inputs, and we apply the active subspace-based techniques to study the input/output relationships. In the notation of the previous section, the outputs are $f$ and the inputs are $\vx$. 

\subsection{Output quantities of interest}

A battery designer may use several model outputs as quantities of interest to characterize performance. For demonstration, we consider two quantities of interest: (i) capacity as a function of voltage and (ii) voltage as a function of time. Capacity is the available energy stored in a fully charged battery measured in milliampere-hours per square centimeter ($\unit{mAh}\cdot\unit{cm}^{-2}$), and is inversely proportional to the cell voltage. The battery's voltage---measured in volts ($\unit{V}$)---decreases as the battery discharges; we study the voltage over the discharge process. 

\subsection{Input parameters}

\cite{Hadigol2015} modeled 19 of the Newman model's input parameters as random variables to study the effects of input uncertainties on output quantities of interest (including capacity and voltage). They repeat this study for three different discharge rates: 0.25$\unit{C}$, 1$\unit{C}$, and 4$\unit{C}$, where $\unit{C}$ denotes the so-called $\unit{C}$-rate measuring the rate at which a battery discharges from its full capacity \cite[Chapter 7]{Pistoia2013}. Realistically, when a battery is connected to an electronic device, its discharge rate is not constant; the rate depends on the device's consumption. Studying three different constant rates allows \cite{Hadigol2015} to observe how the uncertainty quantification changes, globally, with the discharge rate. The data set for the present study is similar.

\begin{table}
\caption{Units, notation, and distributions for the parameters $\vx$ of the battery model. The distributions from \cite{Hadigol2015} represent a particular set of operating conditions for a LiC$_6$/LiCoO$_2$ cell.}
\label{tab:inputs}
\begin{center}
\begin{tabular}{llllll}
Name & Units & Notation & Nominal & Distribution \\ \hline
anode porosity & --- & $\epsilon_a$ & 0.485 & $U[0.46, 0.51]$ \\
anode Bruggeman coeff. & --- & $\brugg_a$ & 4 & $U[3.8,4.2]$ \\
anode solid diffusion coeff. & $\unit{m}^2\,\unit{s}^{-1}$ & $D_{s,a}$ & $3.9\times 10^{-14}$ & $U[3.51,4.29]\times 10^{-14}$ \\
anode conductivity & $\unit{S}\,\unit{m}^{-1}$ & $\sigma_a$ & 100 & $U[90,110]$ \\
anode reaction rate & $\unit{m}^4\,\unit{mol}\,\unit{s}$ & $k_a$ & $5.031\times 10^{-11}$ & $U[4.52,5.53]\times 10^{-11}$ \\
anode particle size & $\mu\unit{m}$ & $r_{s,a}$ & 2 & $N(2,0.1354)$ \\
anode length & $\mu\unit{m}$ & $L_a$ & 80 & $U[77,83]$ \\
cathode porosity & --- & $\epsilon_c$ & 0.385 & $U[0.36, 0.41]$ \\
cathode Bruggeman coeff. & --- & $\brugg_c$ & 4 & $U[3.8,4.2]$ \\
cathode solid diffusion coeff. & $\unit{m}^2\,\unit{s}^{-1}$ & $D_{s,c}$ & $1.0\times 10^{-14}$ & $U[0.90,1.10]\times 10^{-14}$ \\
cathode conductivity & $\unit{S}\,\unit{m}^{-1}$ & $\sigma_c$ & 100 & $U[90,110]$ \\
cathode reaction rate & $\unit{m}^4\,\unit{mol}\,\unit{s}$ & $k_c$ & $2.334\times 10^{-11}$ & $U[2.10,2.56]\times 10^{-11}$ \\
cathode particle size & $\mu\unit{m}$ & $r_{s,c}$ & 2 & $N(2,0.3896)$ \\
cathode length & $\mu\unit{m}$ & $L_c$ & 88 & $U[85,91]$ \\
separator porosity & --- & $\epsilon_s$ & 0.724 & $U[0.63, 0.81]$ \\
separator Bruggeman coeff. & --- & $\brugg_s$ & 4 & $U[3.2,4.8]$ \\
separator length & $\mu\unit{m}$ & $L_s$ & 25 & $U[22,28]$ \\
Li$^+$ transference number & --- & $t_+^0$ & 0.363 & $U[0.345,0.381]$ \\
salt diffusion coeff.~in liquid & $\unit{m}^2\,\unit{s}^{-1}$ & $D$ & $7.5\times 10^{-10}$ & $U[6.75,8.25]\times 10^{-10}$ 
\end{tabular}
\end{center}
\end{table}

Table \ref{tab:inputs} summarizes the battery model parameters and their associated distributions from \cite{Hadigol2015}. These parameter distributions were taken from the available modeling literature for LiC$_6$/LiCoO$_2$ cells. In some cases, characterizations of the parameters' variability is not available in the literature, so \cite{Hadigol2015} made modeling choices consistent with engineering expertise. Short descriptions of each parameter follow. Additionally, when available, we provide engineering intuition on how changes in the parameter affect the quantity of interest. 

\subsubsection{Porosity, $\epsilon$}

\emph{Porosity} is the ratio of pore volume to bulk volume. Each component of the cell model---anode, cathode, and separator---has its own porosity parameter: $\epsilon_a$, $\epsilon_c$, and $\epsilon_s$. Increasing porosity in any component increases power and lowers capacity. 

\subsubsection{Solid particle size, $r$}

The flux of Li$^+$ ions is affected by the surface area of the solid particles in the anode and cathode; a larger surface area permits faster reaction. Therefore, a larger particle size---where all particles are modeled as spheres---leads to higher power. One parameter controls the average particle size for each electrode: $r_a$ for the anode and $r_c$ for the cathode.

\subsubsection{Bruggeman coefficient, $\brugg$}

Bulk transport properties in the cell (e.g., bulk ion transport and diffusion) are affected by a geometric property of the particles called \emph{tortuosity}, $\tau$. The Bruggeman relation expresses tortuosity as porosity raised to a power, $\tau = \epsilon^{(1-\brugg)}$, where $\brugg$ is the Bruggeman coefficient. Roughly, the smaller the Bruggeman coefficient, the faster the transport and higher the power. Each component of the cell has its own Bruggeman coefficient. 

\subsubsection{Salt diffusion coefficient, $D$}

The salt diffusion coefficient $D$ is a bulk measure of friction between ions and solvents. As salt diffuses, ions travel more freely. Thus, larger $D$ produces higher power. There is one salt diffusion coefficient for the model.

\subsubsection{Solid diffusion coefficient, $D_a$ and $D_c$}

The solid diffusion coefficient characterizes how quickly ions diffuse from the particles in the electrodes, so larger diffusion coefficient leads to higher voltage. The diffusion coefficient for the anode is $D_a$, and the diffusion coefficient for the cathode is $D_c$. 

\subsubsection{Solid conductivity, $\sigma$}

Increasing conductivity in the electrodes increases power. The anode's conductivity is $\sigma_a$, and the cathode's conductivity is $\sigma_c$. 

\subsubsection{Reaction rate, $k$}

Faster chemical reactions---corresponding to higher rate constants---are preferred for Li batteries. The anode's reaction rate is $k_a$, and the cathode's reaction rate is $k_c$. 

\subsubsection{Component length, $L$}

We treat the length of each component as an independent parameter that can vary in the prescribed range. The component lengths are denoted $L_a$, $L_s$, and $L_c$ for anode, separator, and cathode, respectively. 

\section{Data set and results}
\label{sec:results}

The data set we use includes input/output pairs for 3600 runs of the simulation model for each of the three discharge rates for a total of 10800 runs. These runs were executed by \cite{Hadigol2015} for the uncertainty quantification study. The total set of runs creates approximately 67MB of text data. Each run uses a realization of the input parameters drawn independently according the distributions in Table \ref{tab:inputs}. Given a realization of the inputs, a finite difference method approximates the solution to the system of transport equations that comprises Newman's model; see Appendix \ref{sec:goveq} for details on the governing equations. The spatial and temporal discretizations are chosen such that PDE approximation errors are negligible for all parameter values. Each simulation produces (i) 50 voltage/capacity pairs and (ii) 50 time/voltage pairs; the collected pairs from all simulations produce the quantities of interest.

Given initial conditions, each simulation was run until the voltage reached a cutoff of 2.8V. However, the time to reach the threshold depends on the input parameters. Instead of comparing voltage at the same physical time, following \cite{Hadigol2015}, we introduce a scaled time coordinate that depends on the time to reach the 2.8V threshold. Define $t^\ast=100$ at physical time $t=0$, and let $t^\ast=0$ at the physical time when the voltage reaches 2.8V. In effect, $t^\ast$ represents a charge meter from 100 to 0 for each run. We compare voltages at different parameter values for the same scaled time coordinate $t^\ast$. 

The input parameters for the simulation were sampled according to the distributions in Table \ref{tab:inputs}. However, for the active subspace-based analysis, we shift and scale the inputs to the hypercube $[-1,1]^{19}$. The analysis proceeds on the normalized inputs, which, as in Section \ref{sec:sensitivity}, we denote $\vx$.

The simulation models were run in parallel on the University of Colorado Boulder's JANUS supercomputer. Each run took approximately 20 minutes on 1 core. The total computing time to generate the data was 3600 CPU hours. We developed and executed Python scripts to compute the active subspace weights and produce summary plots. The time to generate these analyses was negligible relative to the computing time used to run the simulations. We executed these scripts on a dual core MacBook Pro with 16GB of memory. The data set and scripts to generate the plots are available at  \href{https://bitbucket.org/paulcon/time-dependent-gsa-for-batteries}{\color{blue}https://bitbucket.org/paulcon/time-dependent-gsa-for-batteries}. 


\subsection{Active subspace results}

As described in Section \ref{sec:timedep}, we apply Algorithm \ref{alg:lslin} at each discharge rate (0.25C, 1C, 4C) (i) at each of the 50 voltage values for capacity and (ii) at each of the 50 $t^\ast$ values for the voltage history. Additionally, we apply Algorithm \ref{alg:lsquad} to a few select voltage/capacity and $t^\ast$/voltage pairs; the computed eigenvalues validate and support the structures we observe in the summary plots generated with $\vw$ from \eqref{eq:w} (i.e., Algorithm \ref{alg:lslin}). Each application uses all 3600 evaluations. In other words, for the least squares problems \eqref{eq:ls} \eqref{eq:lsq}, $N=3600$. This is more than sufficient to estimate (i) the 20 coefficients of the linear model from Algorithm \ref{alg:lslin} and (ii) the 210 coefficients of the quadratic model from Algorithm \ref{alg:lsquad}. Again, we emphasize that the polynomial models are not used as a predictive surrogate response surfaces for the map from physical inputs to output quantities of interest. For Algorithm \ref{alg:lslin}, the fitted linear model's gradient---normalized to have unit 2-norm---provides the 19-component vector $\vw$ from \eqref{eq:w} that is the candidate basis for a one-dimensional active subspace. For Algorithm \ref{alg:lsquad}, the coefficients of the quadratic model form the matrix whose eigenpairs estimate the active subspace eigenpairs; see \eqref{eq:Heigs} and its motivation in \eqref{eq:Cquad}. The summary plots using $\vw$ from \eqref{eq:w} also use all 3600 runs. We stress that these results are for (i) the particular model used to the generate the data set and (ii) the assumptions on parameter variability from Table \ref{tab:inputs}. Strictly speaking, changing any of these assumptions would require a fresh analysis. However, an experienced battery designer may derive insights into other cases and conditions from these results; we do not attempt such extrapolation. 

\begin{figure}[!h]
\centering
\subfloat[Components of $\vw$ from Algorithm \ref{alg:lslin} as function of voltage]{
\includegraphics[width=0.99\textwidth]{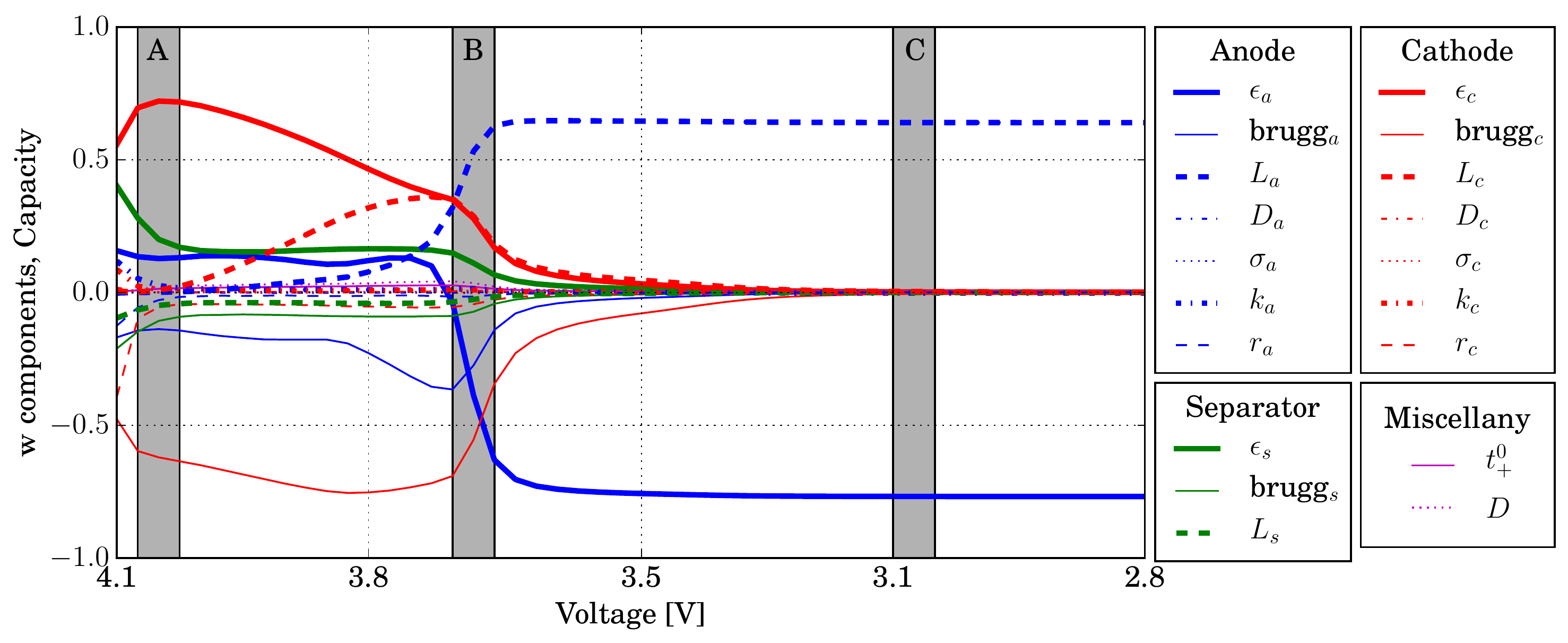}%
}
\\
\subfloat[A (4.1V)]{
\includegraphics[width=0.33\textwidth]{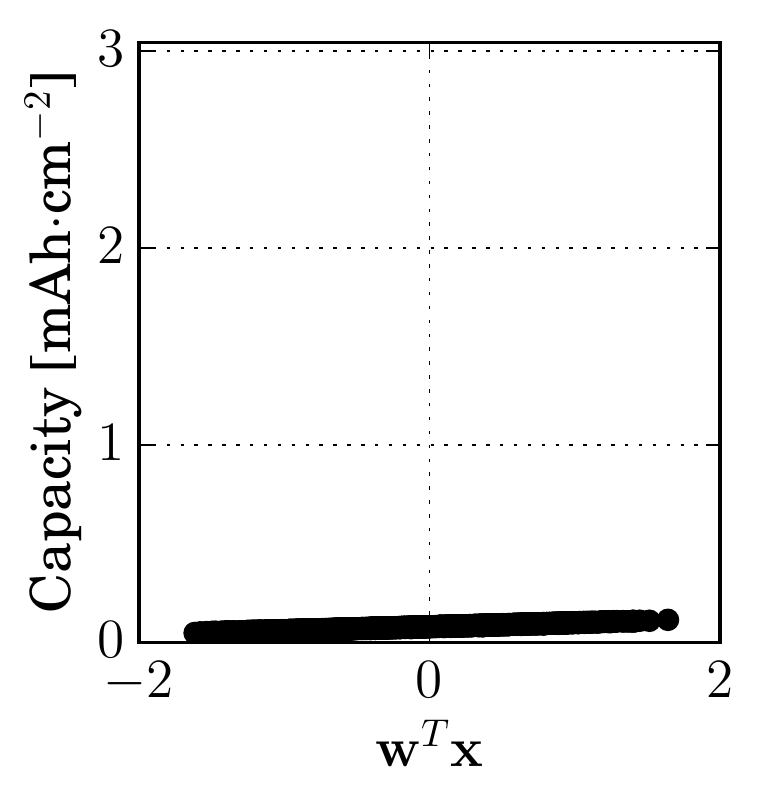}%
}
\subfloat[B (3.7V)]{
\includegraphics[width=0.33\textwidth]{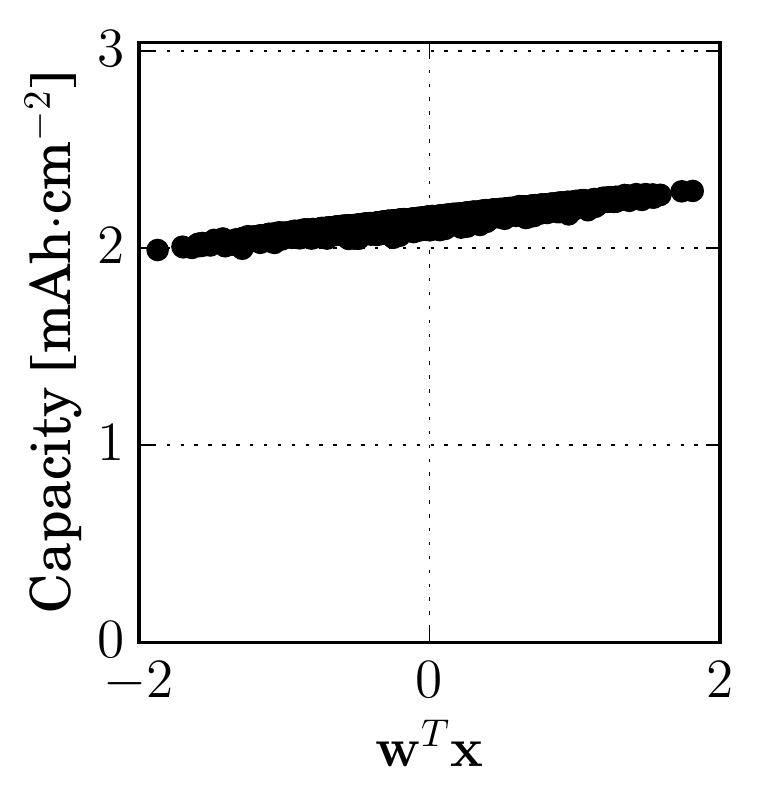}%
}
\subfloat[C (3.1V)]{
\includegraphics[width=0.33\textwidth]{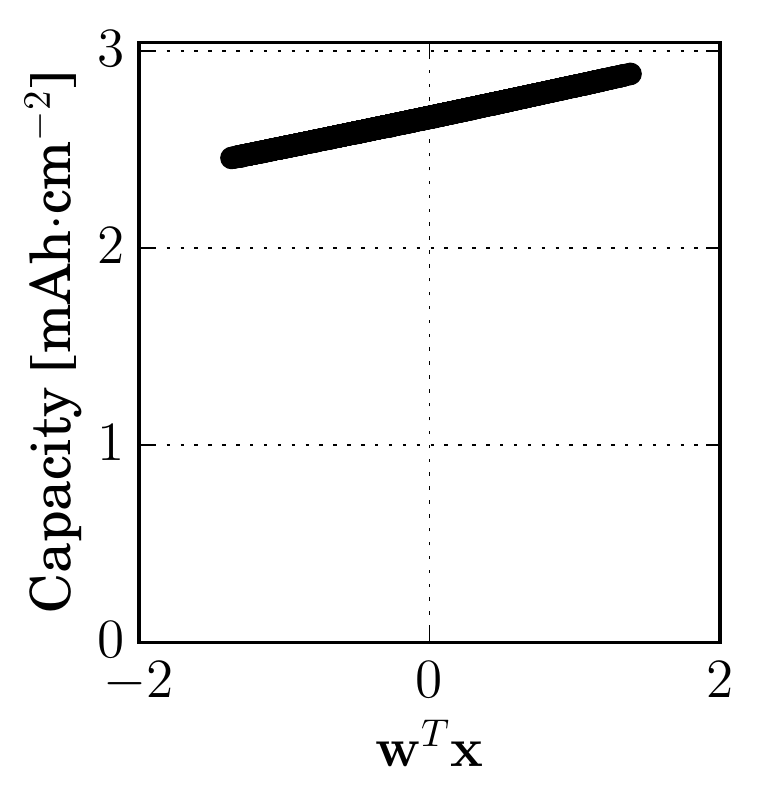}%
}
\\
\subfloat[A (4.1V)]{
\label{fig:cap025Azoom}
\includegraphics[width=0.33\textwidth]{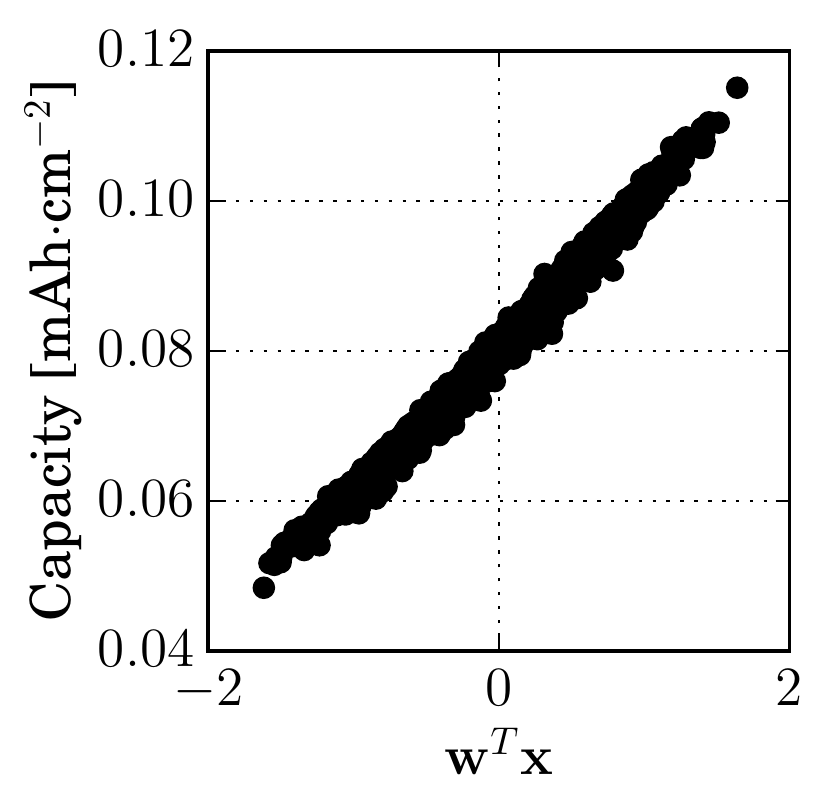}%
}
\subfloat[B (3.7V)]{
\label{fig:cap025Bzoom}
\includegraphics[width=0.33\textwidth]{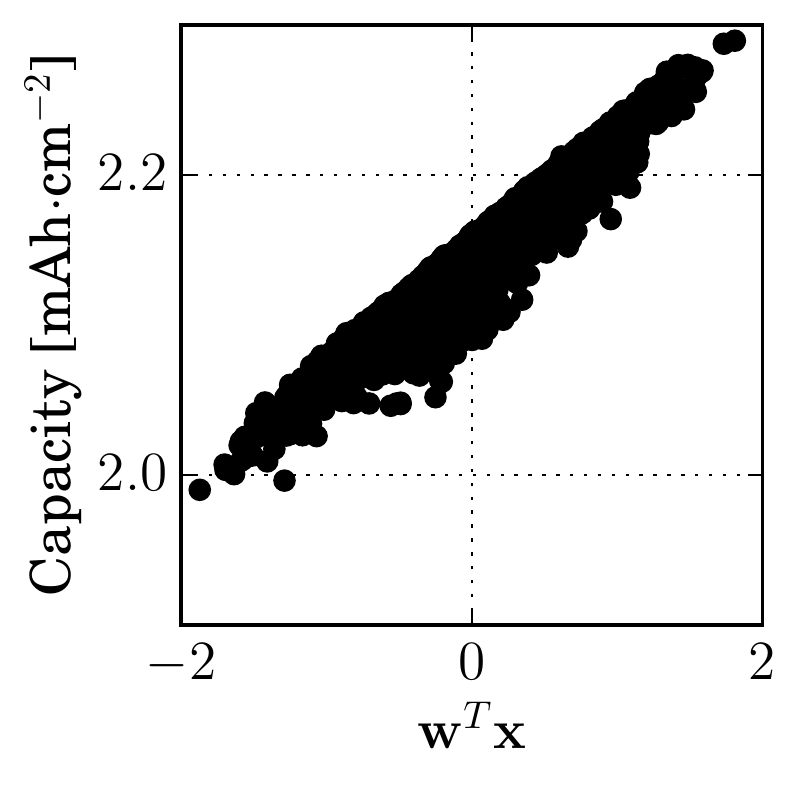}%
}
\subfloat[C (3.1V)]{
\label{fig:cap025Czoom}
\includegraphics[width=0.33\textwidth]{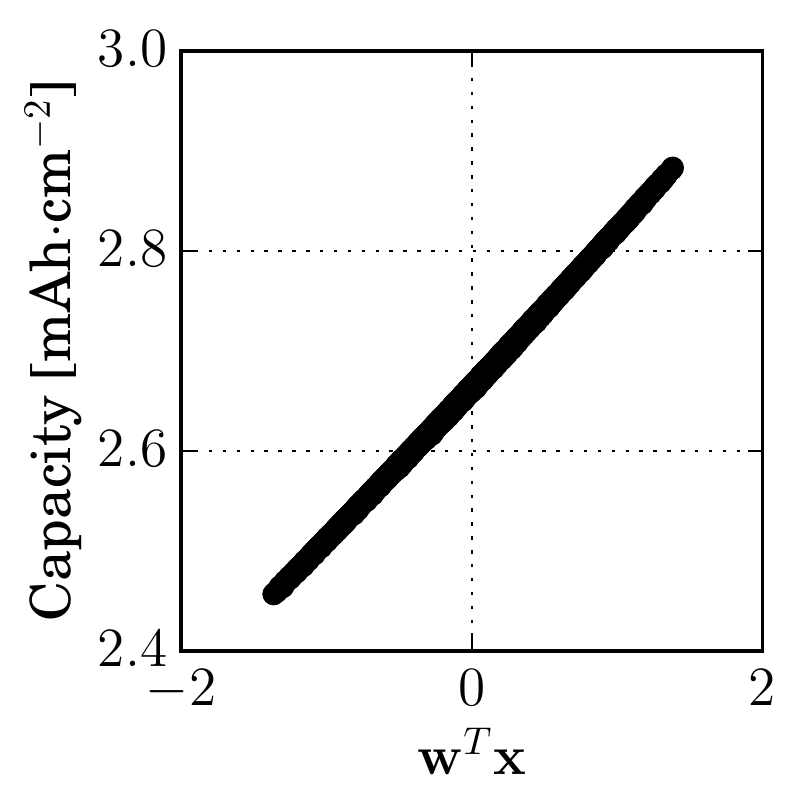}%
}
\caption{Results for capacity at discharge rate 0.25C. The top figure shows the components of $\vw$ from Algorithm \ref{alg:lslin} as function of voltage. The middle row shows summary plots corresponding to the voltages labeled A, B, and C in the top figure. The bottom row is identical to the top row with the vertical axis zoomed to elucidate the relationship.}
\label{fig:capacity025C}
\end{figure}

\begin{figure}[!h]
\centering
\subfloat[Active subspace weights as function of voltage]{
\includegraphics[width=0.99\textwidth]{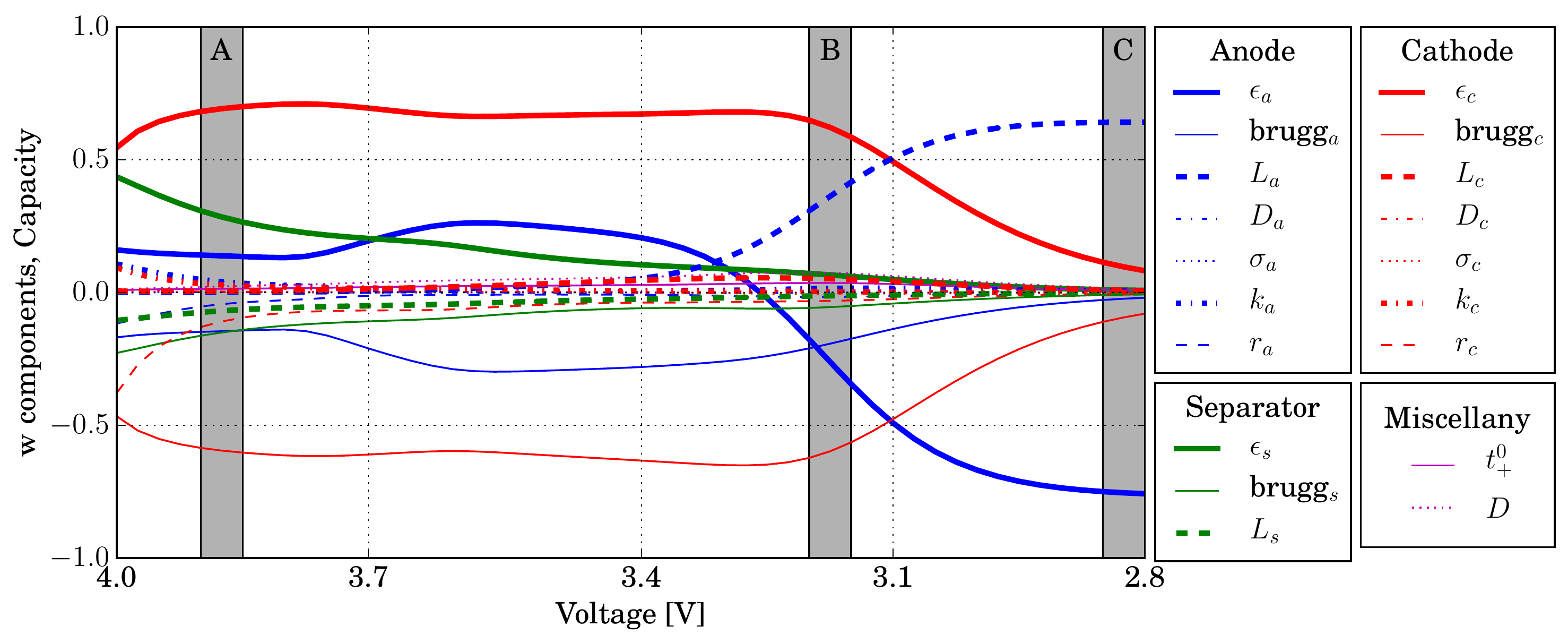}%
}
\\
\subfloat[A (3.9V)]{
\includegraphics[width=0.33\textwidth]{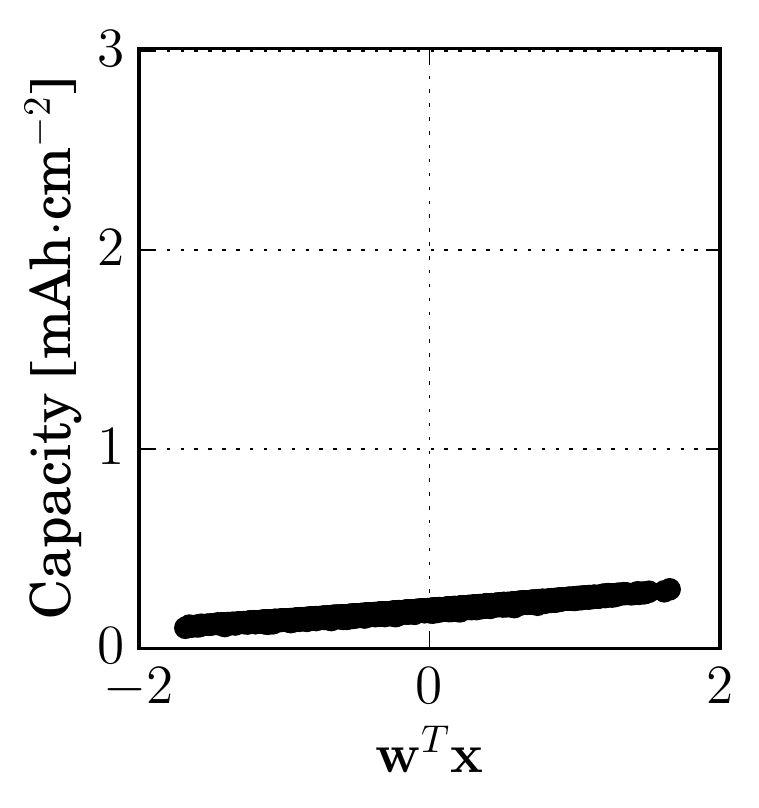}%
}
\subfloat[B (3.2V)]{
\includegraphics[width=0.33\textwidth]{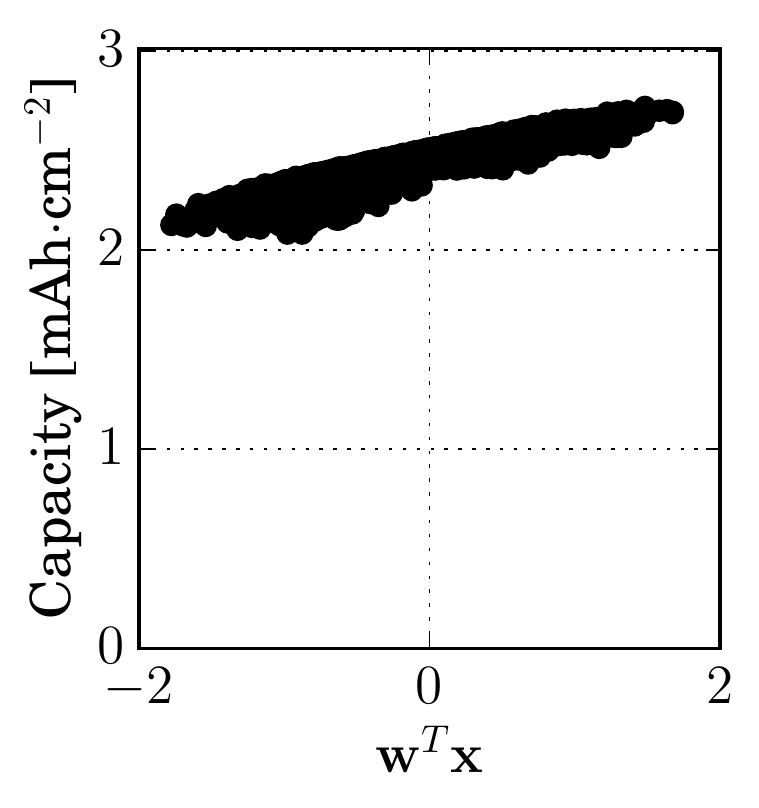}%
}
\subfloat[C (2.8V)]{
\includegraphics[width=0.33\textwidth]{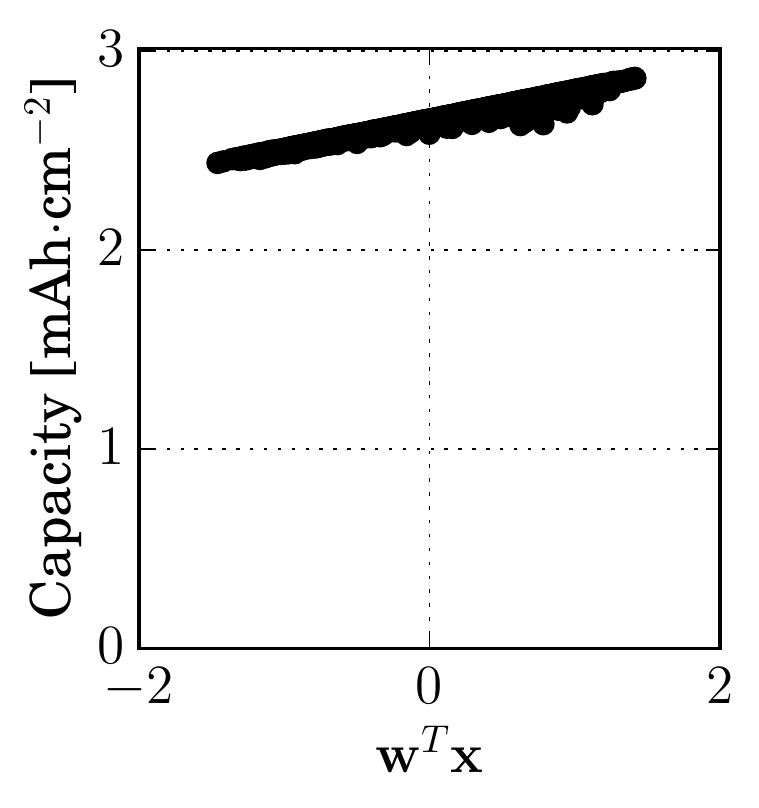}%
}
\\
\subfloat[A (3.9V)]{
\includegraphics[width=0.33\textwidth]{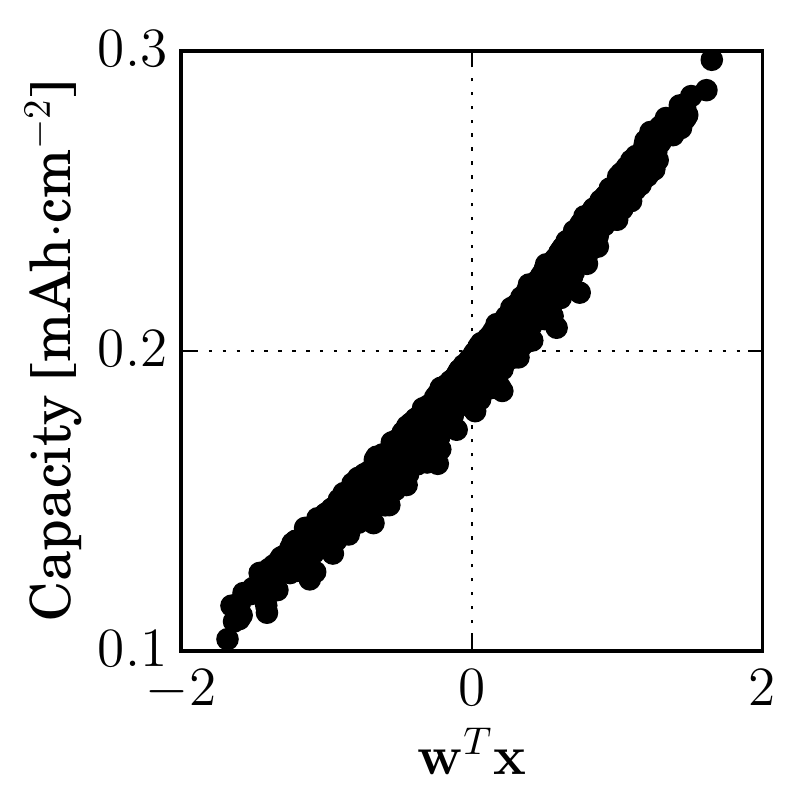}%
}
\subfloat[B (3.2V)]{
\label{fig:cap1Bzoom}
\includegraphics[width=0.33\textwidth]{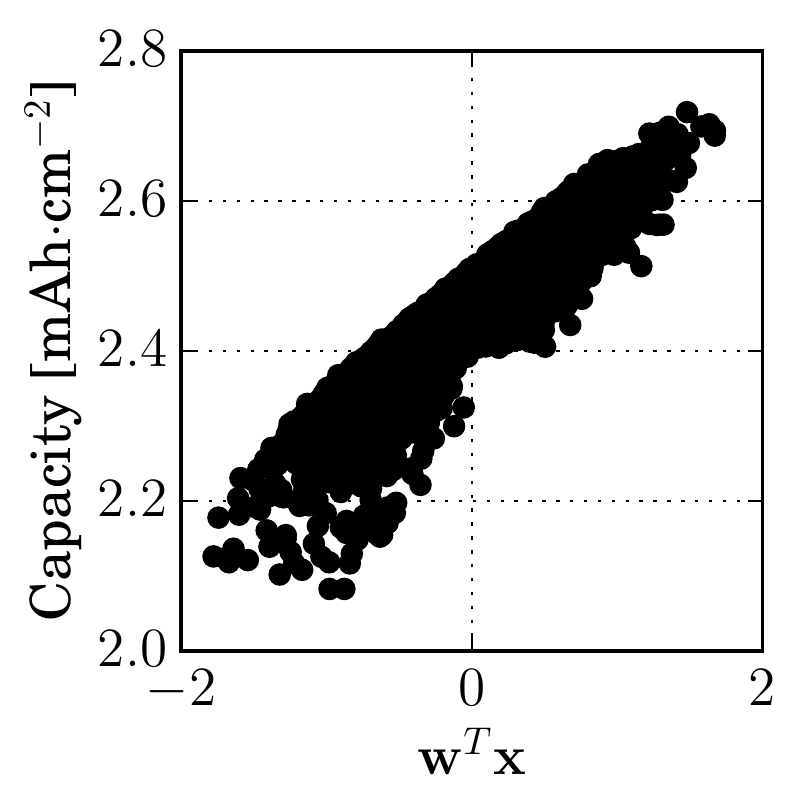}%
}
\subfloat[C (2.8V)]{
\includegraphics[width=0.33\textwidth]{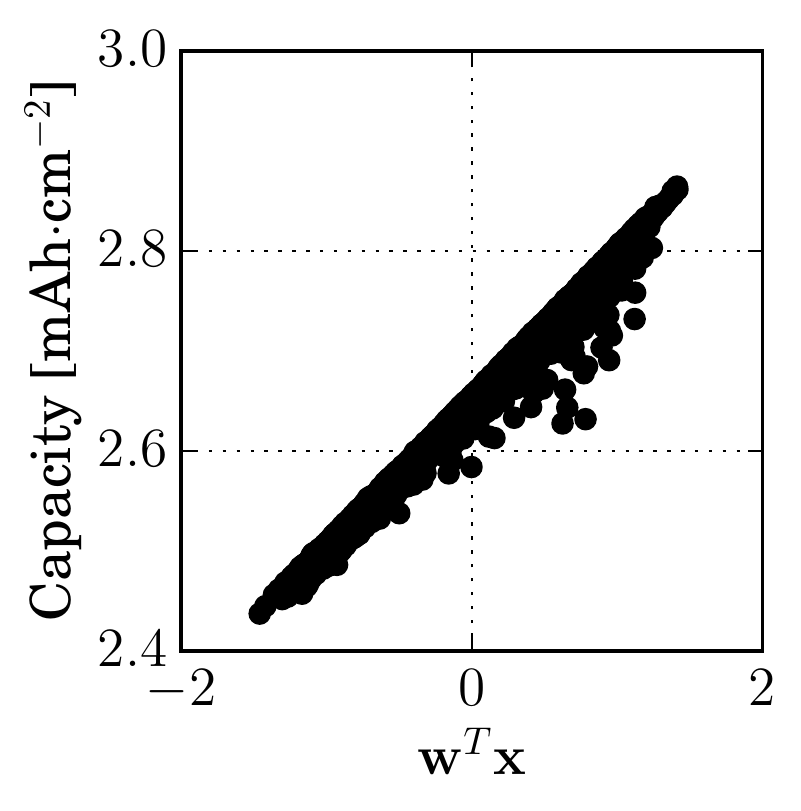}%
}
\caption{Results for capacity at discharge rate 1C. The top figure shows the components of $\vw$ from Algorithm \ref{alg:lslin} as function of voltage. The middle row shows summary plots corresponding to the voltages labeled A, B, and C in the top figure. The bottom row is identical to the top row with the vertical axis zoomed to elucidate the relationship.}
\label{fig:capacity1C}
\end{figure}

\begin{figure}[!h]
\centering
\subfloat[Active subspace weights as function of voltage]{
\includegraphics[width=0.99\textwidth]{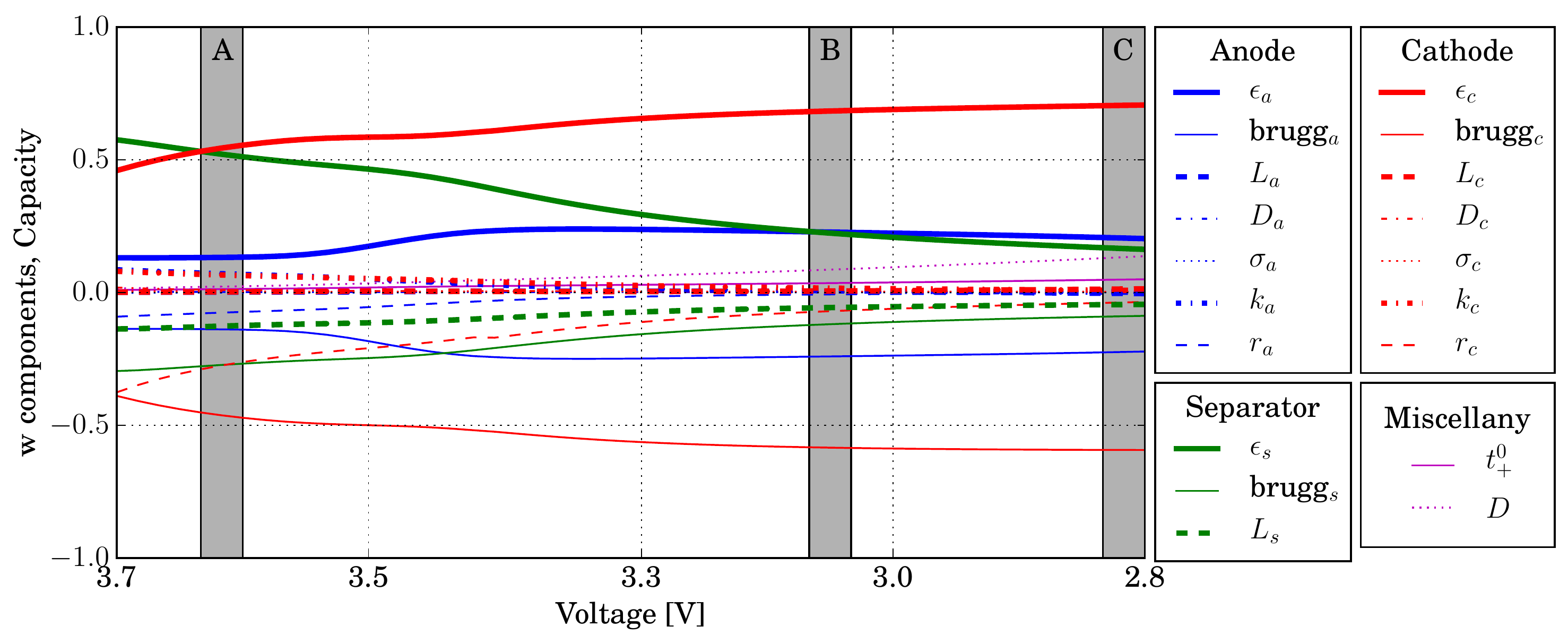}%
}
\\
\subfloat[A (3.6V)]{
\includegraphics[width=0.33\textwidth]{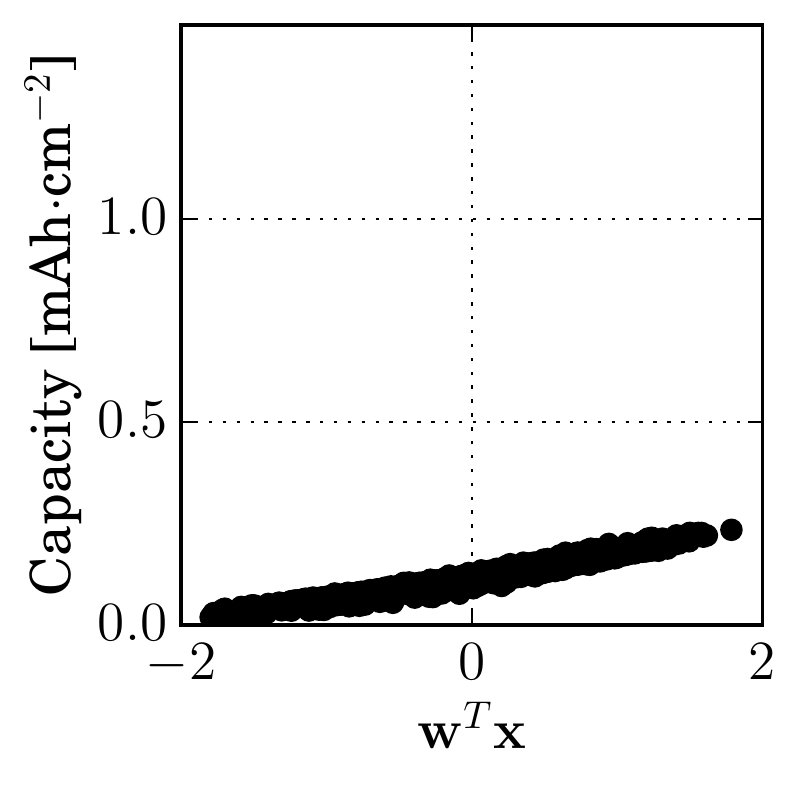}%
}
\subfloat[B (3.1V)]{
\includegraphics[width=0.33\textwidth]{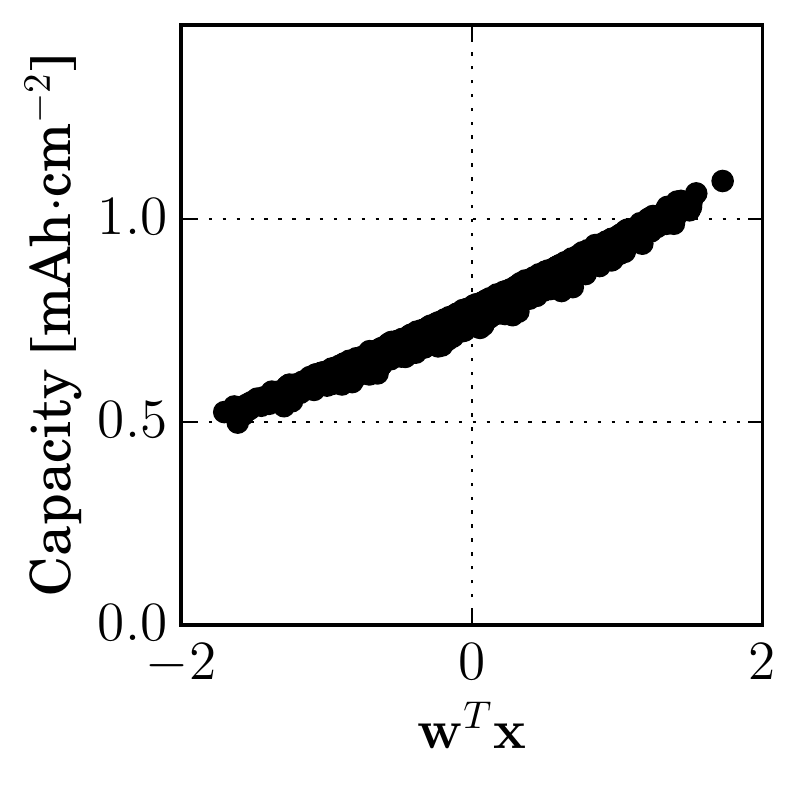}%
}
\subfloat[C (2.8V)]{
\includegraphics[width=0.33\textwidth]{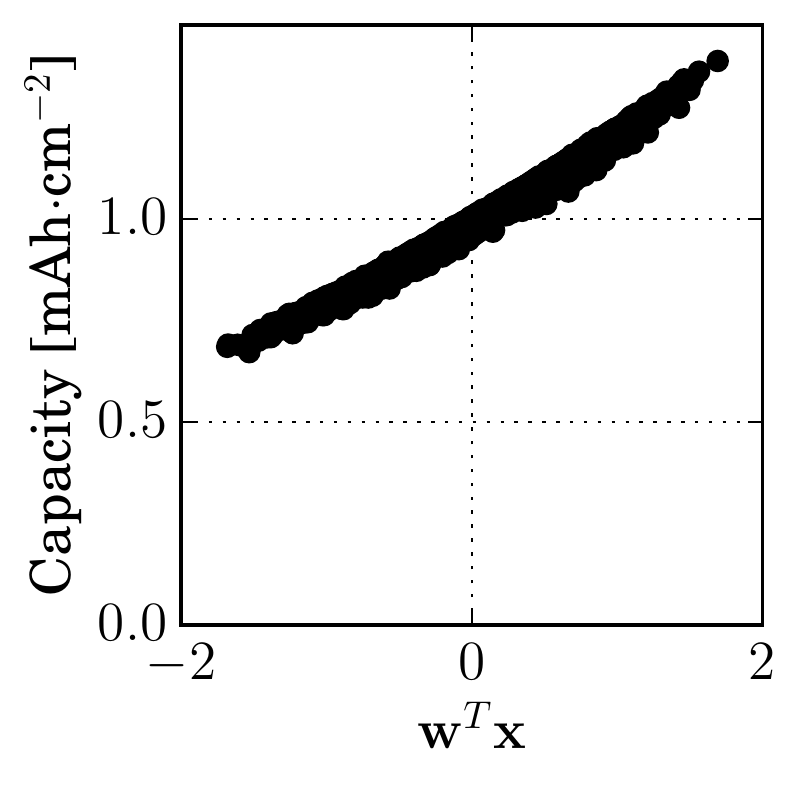}%
}
\\
\subfloat[A (3.6V)]{
\includegraphics[width=0.33\textwidth]{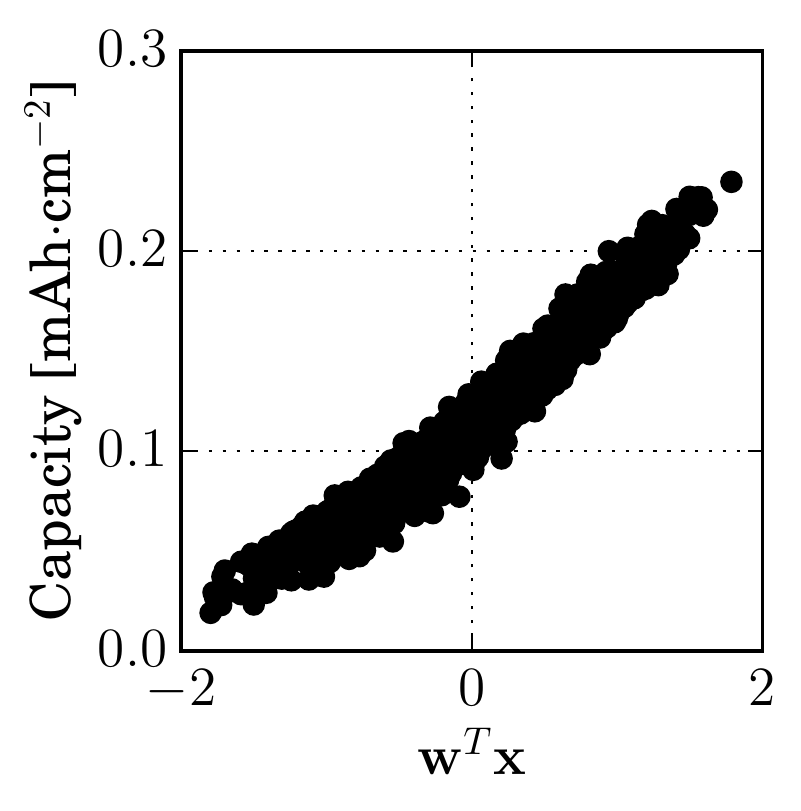}%
}
\subfloat[B (3.1V)]{
\includegraphics[width=0.33\textwidth]{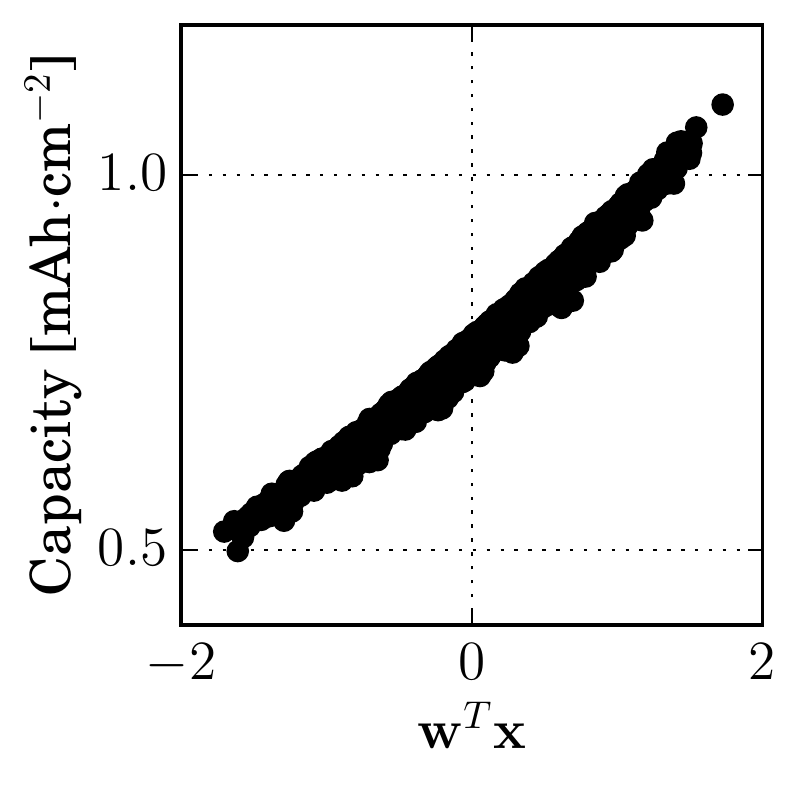}%
}
\subfloat[C (2.8V)]{
\includegraphics[width=0.33\textwidth]{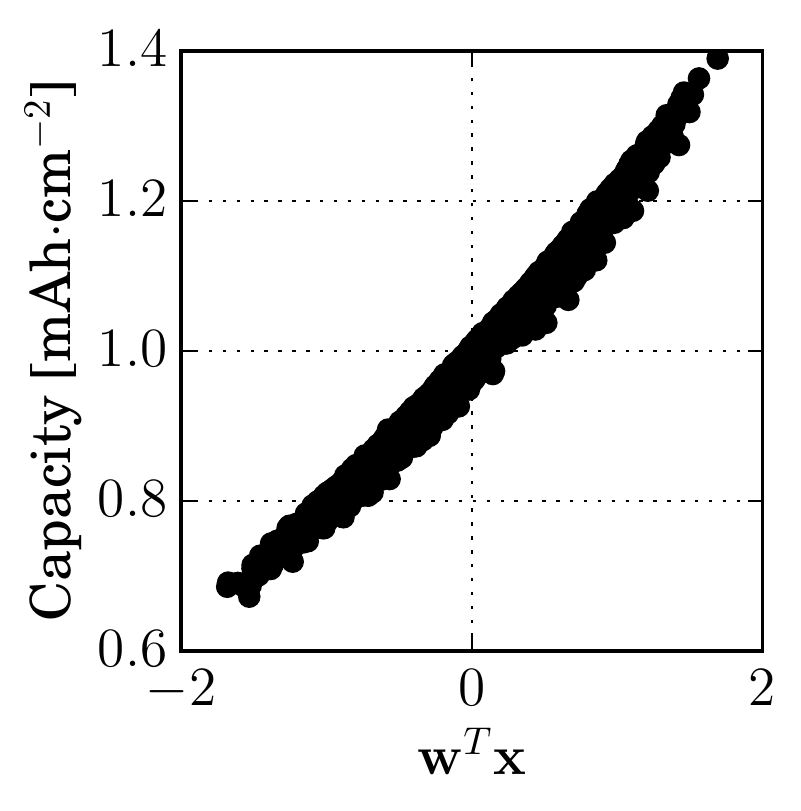}%
}
\caption{Results for capacity at discharge rate 4C. The top figure shows the components of $\vw$ from Algorithm \ref{alg:lslin} as function of voltage. The middle row shows summary plots corresponding to the voltages labeled A, B, and C in the top figure. The bottom row is identical to the top row with the vertical axis zoomed to elucidate the relationship.}
\label{fig:capacity4C}
\end{figure}

Figures \ref{fig:capacity025C}, \ref{fig:capacity1C}, and \ref{fig:capacity4C} show the results for capacity as a function of voltage. In each figure, the top subfigure shows $\vw$'s components at each of the 50 voltage values. Each component is associated with one of the 19 model input parameters; the legends to the right match the input parameter with its line style in the plot. For parameter names and units, refer to Table \ref{tab:inputs}. Three gray shaded regions labeled A, B, and C identify three voltages of interest, chosen according to interesting features in the voltage/capacity weights relationships. The middle row of subfigures shows summary plots corresponding to the voltages A, B, and C, from left to right. The vertical axis scale is chosen to contain all capacity values in the set of simulations at the particular discharge rate. For a particular voltage (A, B, or C), the summary plot shows the relationship between the linear combination of normalized inputs $\vw^T\vx$ and the capacity. The bottom row of subfigures is identical to the middle row except that the vertical scale is reduced (i.e., zoomed in) to elucidate the relationship between $\vw^T\vx$ and capacity. 

In all cases, the summary plots reveal a relationship between the linear combination of normalized inputs $\vw^T\vx$ and the capacity. The degree of spread around a univariate functional relationship varies. Figure \ref{fig:cap025Czoom}, which shows results for discharge rate 0.25C and voltage 3.1V, has the tightest univariate relationship; moreover, the relationship is linear. For this case, we are confident that capacity can be well approximated as $c_0 + c_1(\vw^T\vx)$, for some coefficients $c_0$ and $c_1$. Additionally, all weights are nearly zero except for the weight associated with anode porosity and the weight associated with anode Bruggeman coefficient. Figure \ref{fig:cap1Bzoom}, which shows discharge rate 1C at voltage 3.2V, has the largest spread around a univariate functional relationship. A global trend is apparent---and such a trend may be useful for a modeler seeking the range of capacities over the parameter values---but we are much less confident that a function of the form $g(\vw^T\vx)$, where $g$ is a univariate, scalar-valued function, is an appropriate approximation. Broadly, a univariate approximation appears more appropriate for (i) all voltages with discharge rate 4C and (ii) voltages near the ends of voltage range for discharge rates 0.25C and 1C. For intermediate voltages, the relationship between normalized inputs $\vx$ and capacity appears more complex than can be captured in the one-dimensional summary plot. 

Figures \ref{fig:capeig025}, \ref{fig:capeig1}, and \ref{fig:capeig4} show the first 5 of 19 eigenvalues from \eqref{eq:Heigs} in Algorithm \ref{alg:lsquad} for the three voltage regions of interest (A, B, and C) at each discharge rate (0.25C, 1C, and 4C), respectively. The gap between the first and second eigenvalues in each case suggests how dominant the one-dimensional subspace is compared to higher dimensional subspaces. (If gradients were available, the eigenvalues of a numerical estimate of $\mC$ from \eqref{eq:C} would suggest comparable information.) Compare the tightness of the trends in the summary plots to the eigenvalues gaps. For example, consider the 0.25C case. The first two eigenvalues in Figure \ref{fig:capeig025} associated with regions A, B, and C differ by two, one, and nearly four orders of magnitude, respectively. Compare this relationship between eigenvalue differences to the summary plots in Figures \ref{fig:cap025Azoom}, \ref{fig:cap025Bzoom}, and \ref{fig:cap025Czoom}. Larger differences between the first two eigenvalues correspond to tighter univariate trends in the summary plots. Smaller differences between the first two eigenvalues correspond to greater spread about a univariate mean function. (Note that ``mean'' here is not a statistical average or expectation over random quantities.) These observations are consistent. A large spread in the summary plot may suggest that the function (represented on the vertical axis) varies significantly along more than one direction in the 19-dimensional parameter space. A small eigenvalue gap strengthens evidence for this suggestion. For all tested capacities (i.e., at three different voltages and three different discharge rates) as functions of the 19 physical input parameters, the eigenvalue gaps from Figures \ref{fig:capeig025}, \ref{fig:capeig1}, and \ref{fig:capeig4} are consistent with the relative spread about a univariate trend in the summary plots in Figures \ref{fig:capacity025C}, \ref{fig:capacity1C}, and \ref{fig:capacity4C}. Together, these plots provide evidence of the degree to which each output of interest can be well approximated by a univariate function of the linear combination $\vw^T\vx$. 

When the summary plot reveals a nearly univariate functional relationship, the components of $\vw$ from \eqref{eq:w} can be treated as global sensitivity metrics for the input parameters with respect to the quantity of interest. In fact, their computation is similar to the regression coefficients proposed by \citet[Chapter 1.2.5]{saltelli2008global} except for the normalization; however, our interpretation differs significantly. \cite{Hadigol2015} estimated the Sobol' total sensitivity indices for the same model. Their estimation procedure first computed a degree 3 polynomial chaos expansion with $\ell_1$ regularization (i.e., lasso) and used the fitted surface to estimate the Sobol' indices. In contrast to the Sobol' total sensitivity indices used by \cite{Hadigol2015}, the components of $\vw$ are signed, and the signs can reveal useful insights into the input/output relationship. Consider univariate relationship revealed in Figure \ref{fig:cap025Czoom}, and note that the relationship is monotonic, i.e., increasing $\vw^T\vx$ increases capacity. The sign of $\vw$'s component associated with anode porosity $\epsilon_a$ is positive. Therefore, increasing $\epsilon_a$ increases capacity. By similar reasoning, the sign of $\vw$'s component associated with the anode Bruggeman coefficient $\brugg_a$ is negative, so increasing $\brugg_a$ will decrease capacity. The rate of increase or decrease is related to the components' magnitudes. 

Treating $\vw$'s components as sensitivity metrics, we observe several interesting qualities of the simulation outputs.
\begin{itemize}
\item The rankings on input parameters induced $\vw$'s components (i.e., with the magnitudes) are very similar to the rankings induced by the Sobol' total sensitivity indices from \cite{Hadigol2015}; the monotonic structure in the summary plots offers some insight into those similarities. 
\item Most of the 19 parameters are relatively unimportant in characterizing the relationship between inputs and capacity. This is consistent with the Sobol' index results from \cite{Hadigol2015}. 
\item For discharge rate 0.25C, a transition occurs in the parameter sensitivities around 3.7V. At lower voltages, the relationship between inputs and capacity involves only two parameters. At higher voltages, the relationship is still close to linear, but the number of parameters defining the linear relationship is larger.
\item The $\vw$ component associated with the anode Bruggeman coefficient changes sign as voltage decreases. Coupled with the summary plots, this implies that increasing $\brugg_a$ affects capacity in opposite directions, depending on voltage.
\item For discharge rate 4C, the sensitivities for porosities and Bruggeman coefficients are roughly opposite. This suggests a trade-off that a battery designer may exploit.  
\end{itemize}

\begin{figure}[!h]
\centering
\subfloat[Active subspace weights as function of $t^\ast$]{
\includegraphics[width=0.99\textwidth]{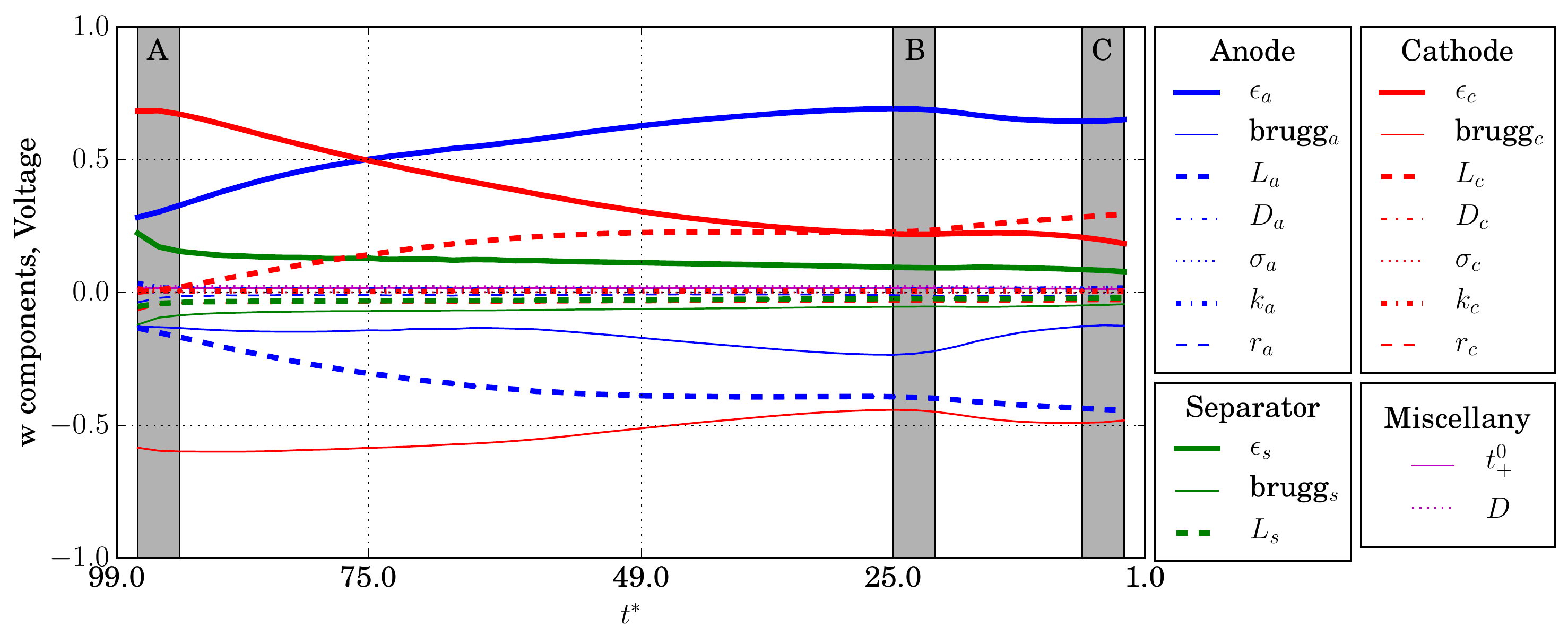}%
}
\\
\subfloat[A ($t^\ast=95$)]{
\includegraphics[width=0.33\textwidth]{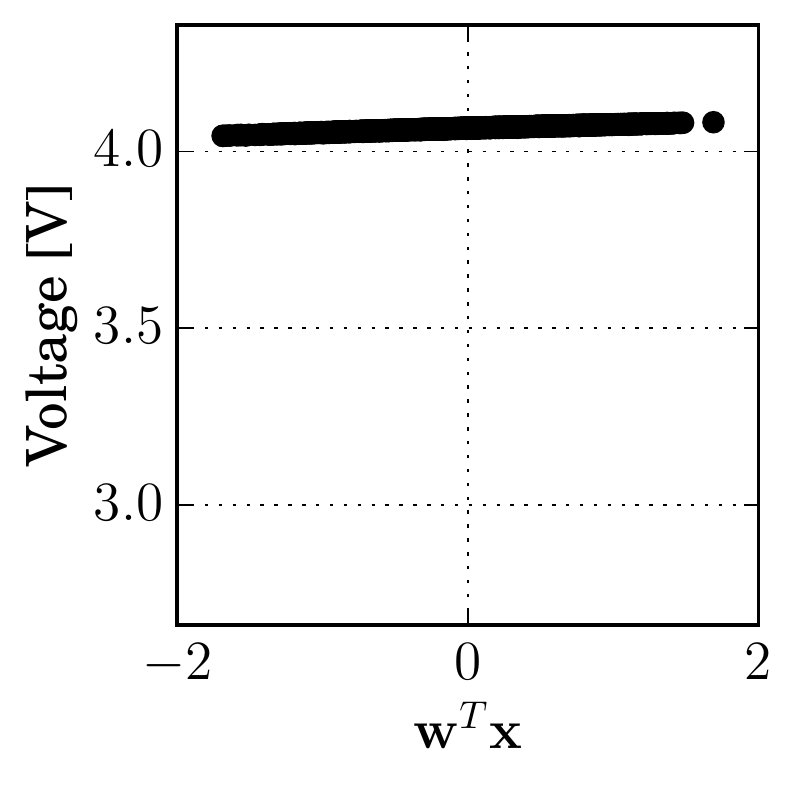}%
}
\subfloat[B ($t^\ast=23$)]{
\includegraphics[width=0.33\textwidth]{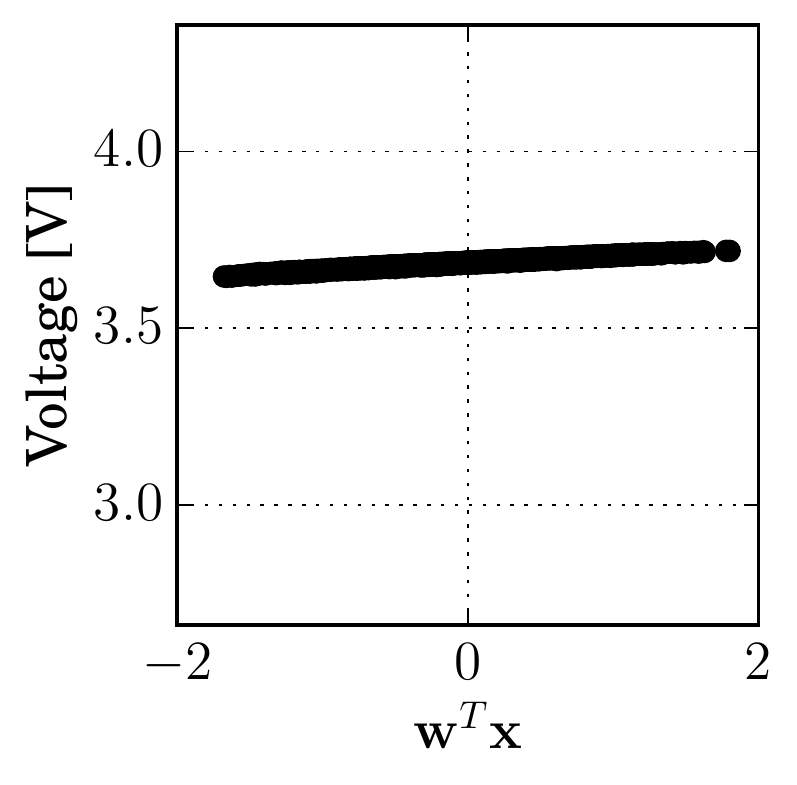}%
}
\subfloat[C ($t^\ast=5$)]{
\includegraphics[width=0.33\textwidth]{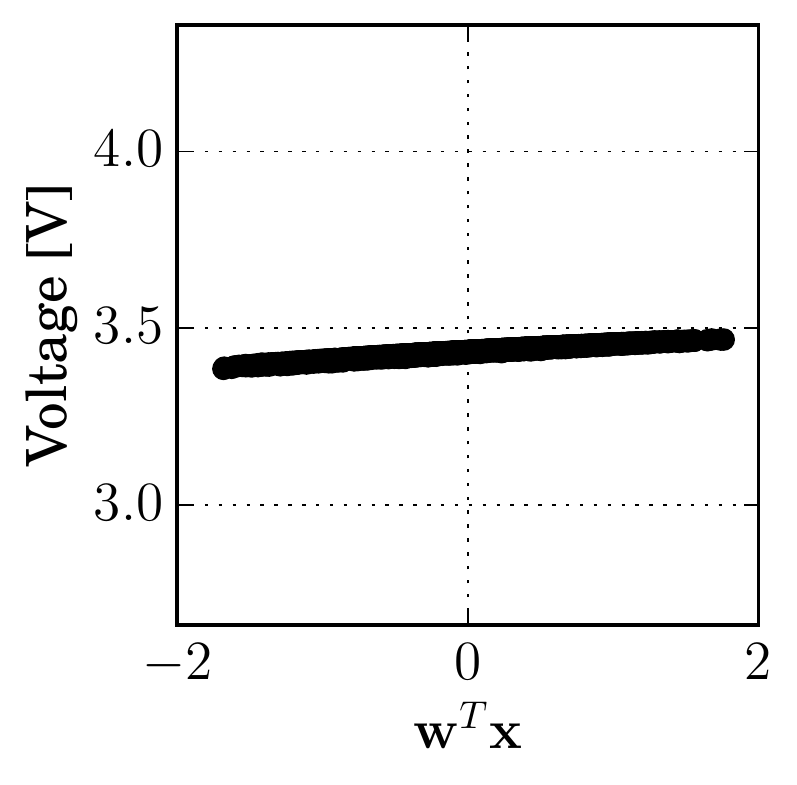}%
}
\\
\subfloat[A ($t^\ast=95$)]{
\includegraphics[width=0.33\textwidth]{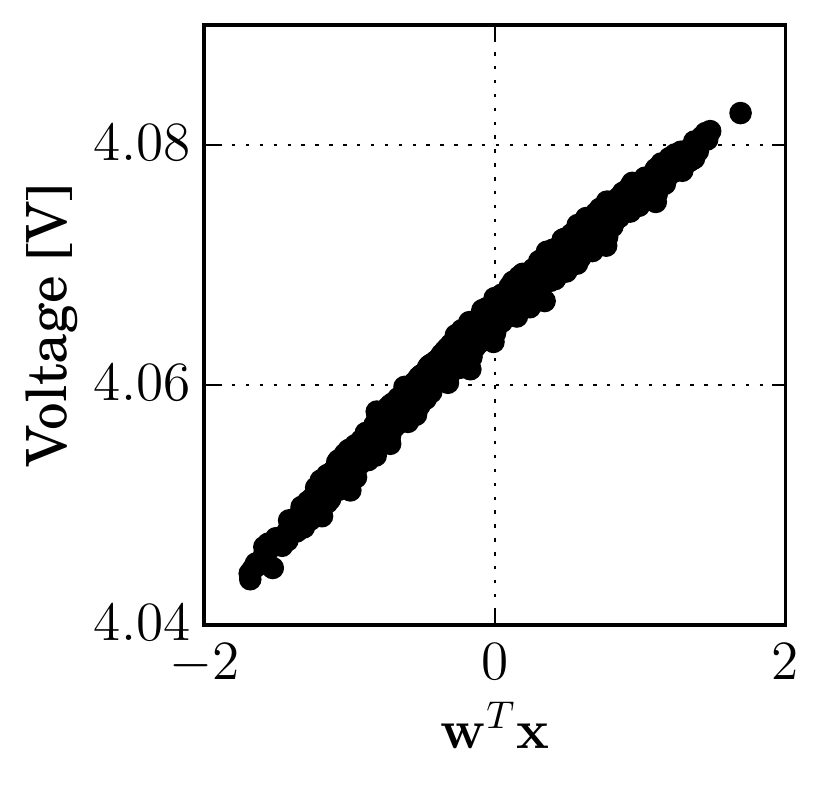}%
}
\subfloat[B ($t^\ast=23$)]{
\includegraphics[width=0.33\textwidth]{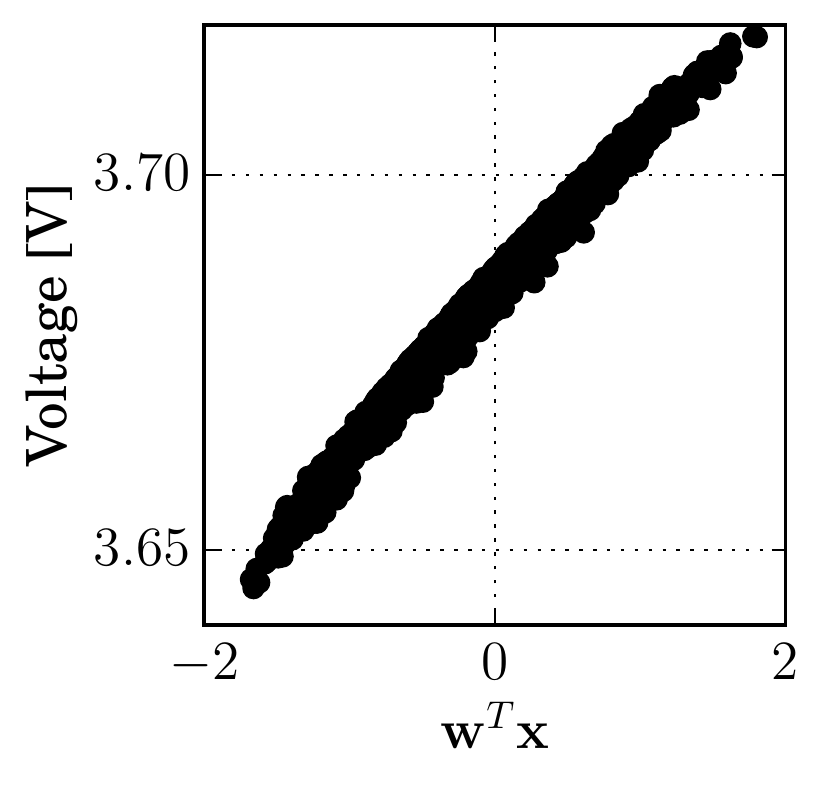}%
}
\subfloat[C ($t^\ast=5$)]{
\includegraphics[width=0.33\textwidth]{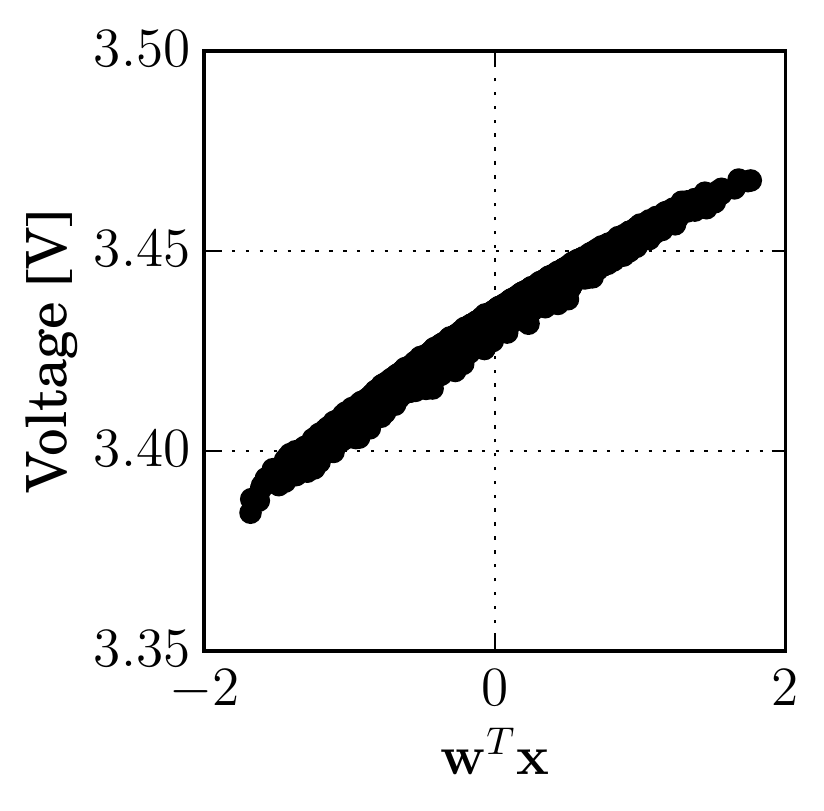}%
}
\caption{Results for voltage at discharge rate $0.25$C. The top figure shows the components of $\vw$ from Algorithm \ref{alg:lslin} as function of $t^*$. The middle row shows summary plots corresponding to the voltages labeled A, B, and C in the top figure. The bottom row is identical to the top row with the vertical axis zoomed to elucidate the relationship.}
\label{fig:voltage025C}
\end{figure}

\begin{figure}[!h]
\centering
\subfloat[Active subspace weights as function of $t^\ast$]{
\includegraphics[width=0.99\textwidth]{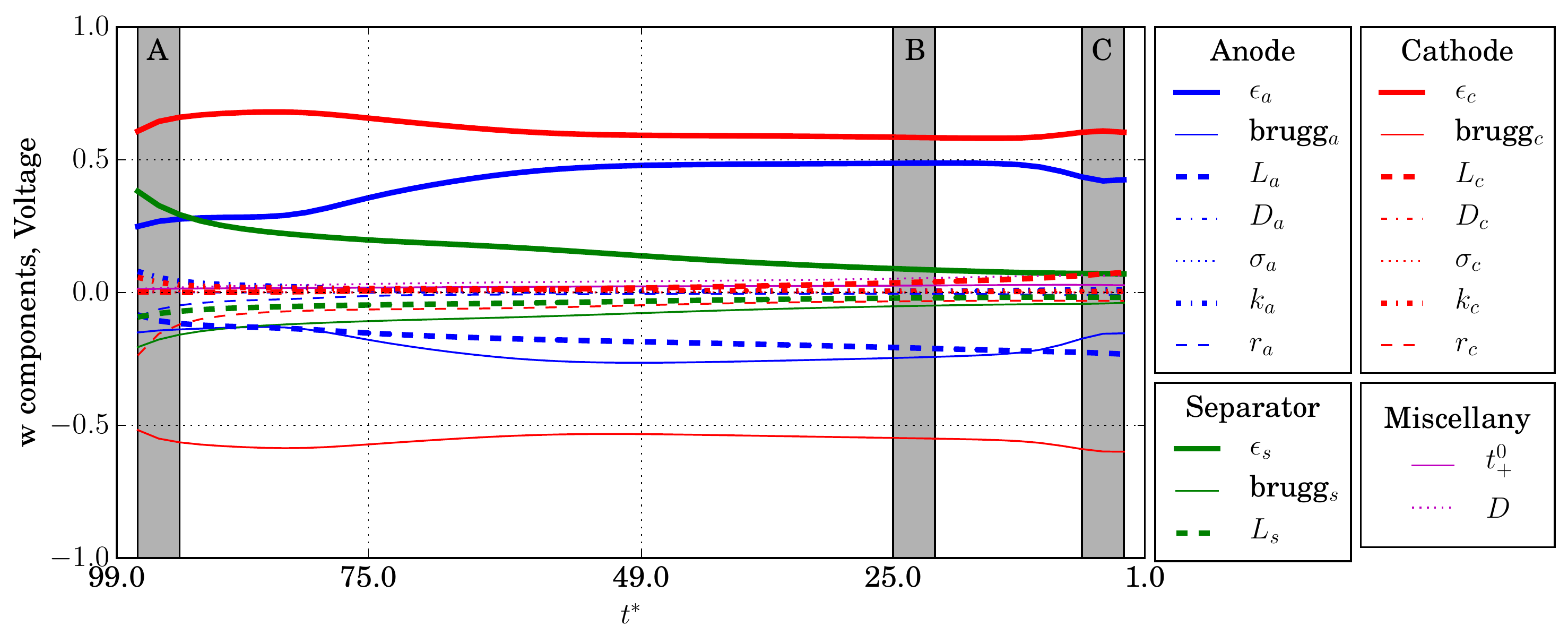}%
}
\\
\subfloat[A ($t^\ast=95$)]{
\includegraphics[width=0.33\textwidth]{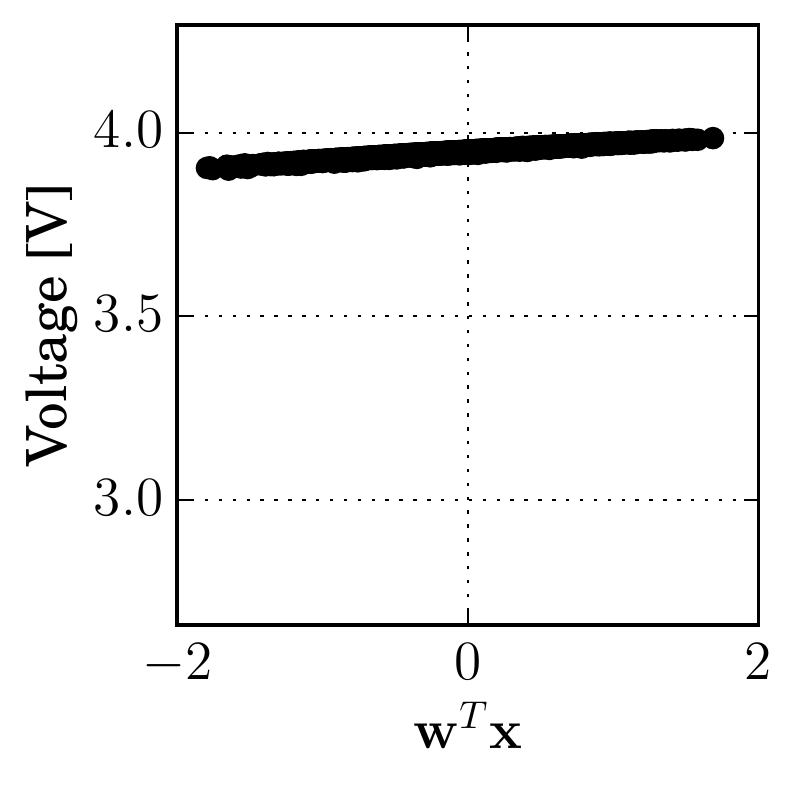}%
}
\subfloat[B ($t^\ast=23$)]{
\includegraphics[width=0.33\textwidth]{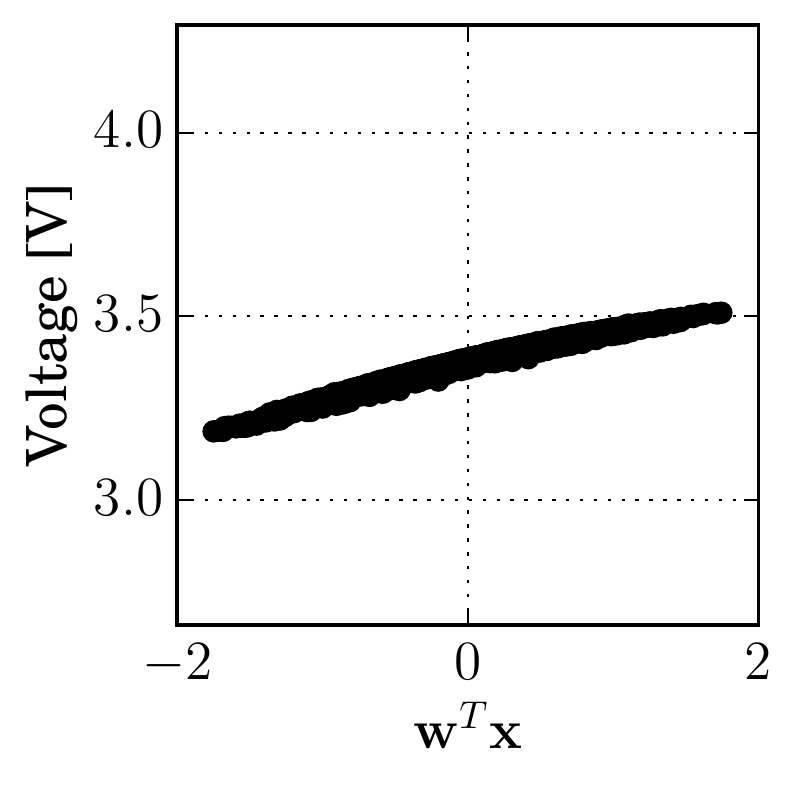}%
}
\subfloat[C ($t^\ast=5$)]{
\includegraphics[width=0.33\textwidth]{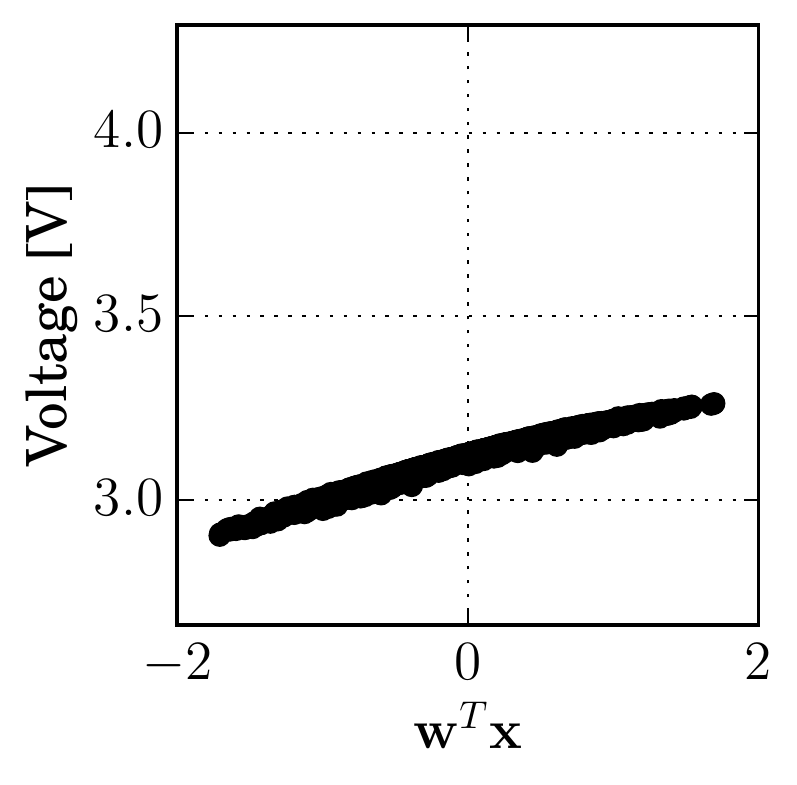}%
}
\\
\subfloat[A ($t^\ast=95$)]{
\includegraphics[width=0.33\textwidth]{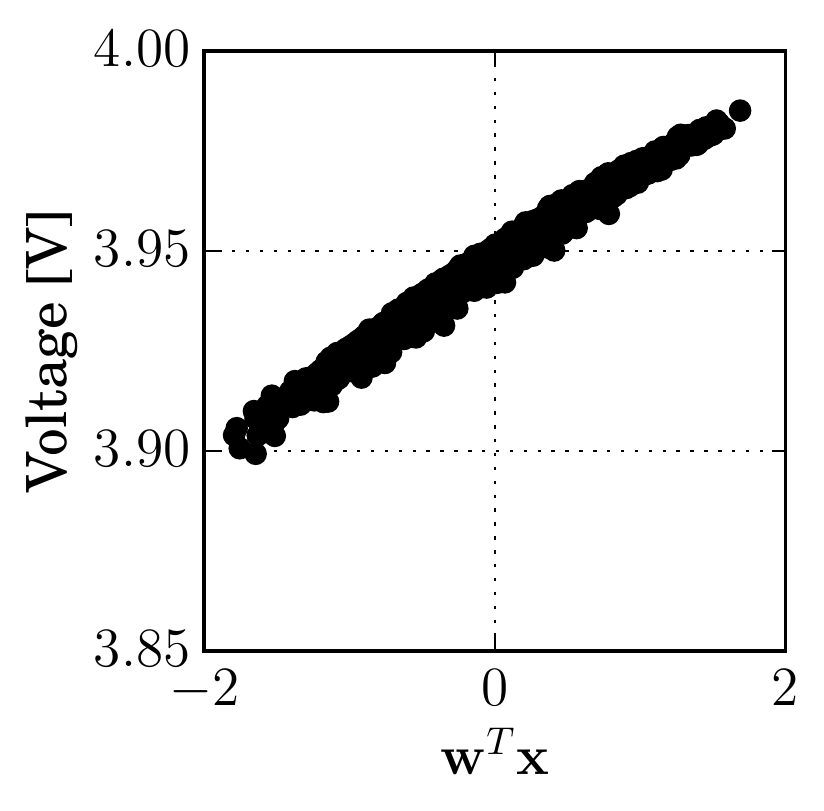}%
}
\subfloat[B ($t^\ast=23$)]{
\includegraphics[width=0.33\textwidth]{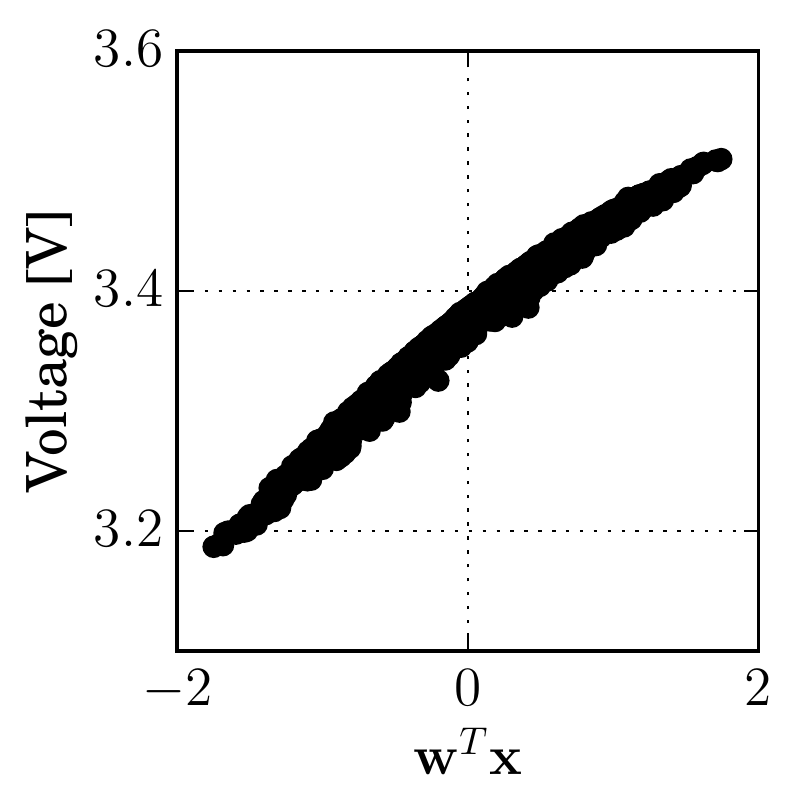}%
}
\subfloat[C ($t^\ast=5$)]{
\includegraphics[width=0.33\textwidth]{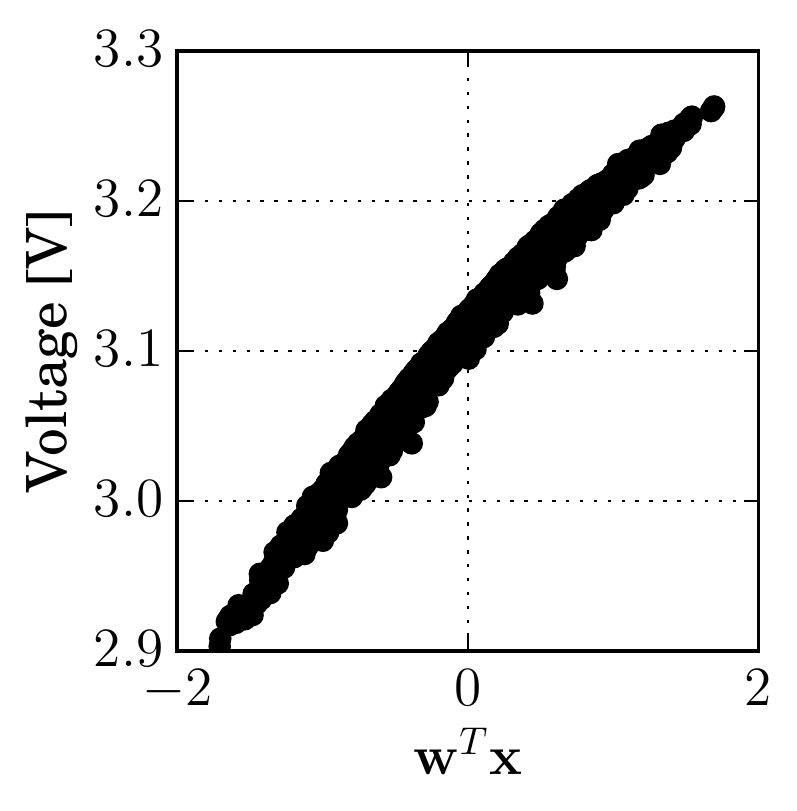}%
}
\caption{Results for voltage at discharge rate $1$C. The top figure shows the components of $\vw$ from Algorithm \ref{alg:lslin} as function of $t^*$. The middle row shows summary plots corresponding to the voltages labeled A, B, and C in the top figure. The bottom row is identical to the top row with the vertical axis zoomed to elucidate the relationship.}
\label{fig:voltage1C}
\end{figure}

\begin{figure}[!h]
\centering
\subfloat[Active subspace weights as function of $t^\ast$]{
\includegraphics[width=0.99\textwidth]{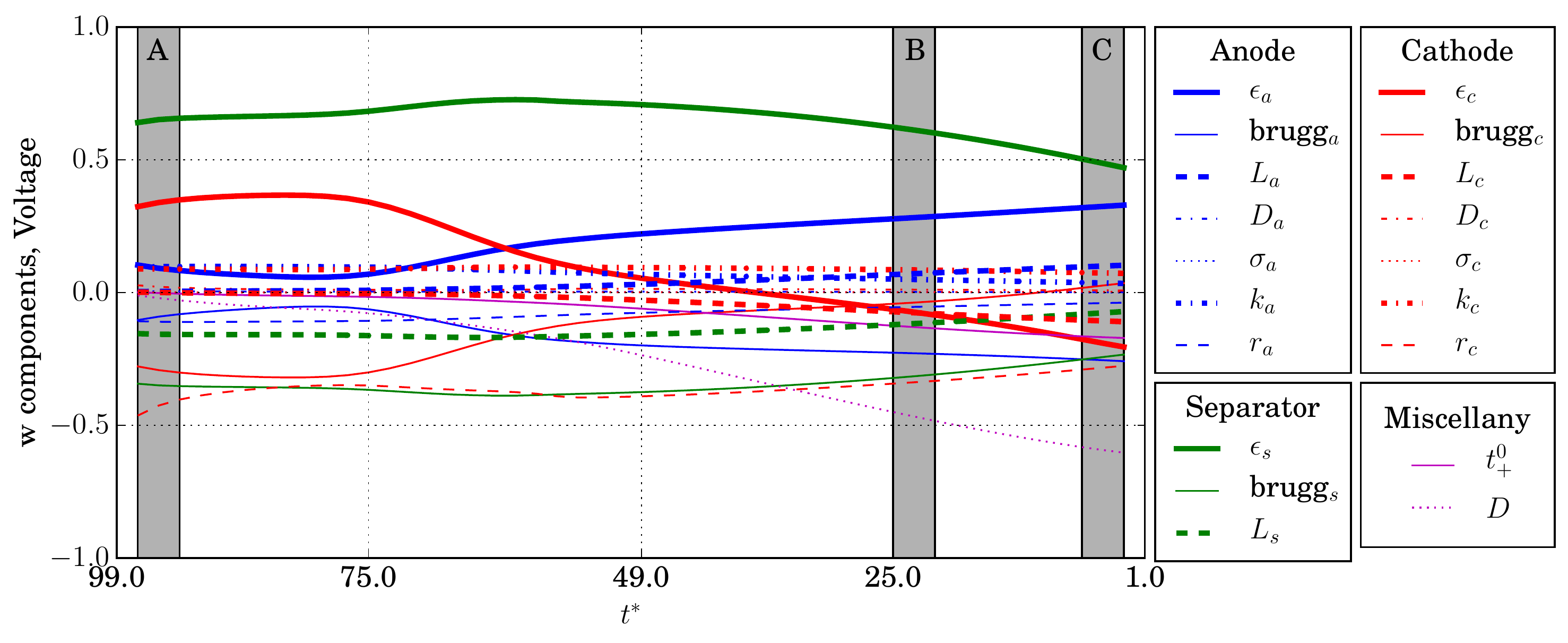}%
}
\\
\subfloat[A ($t^\ast=95$)]{
\includegraphics[width=0.33\textwidth]{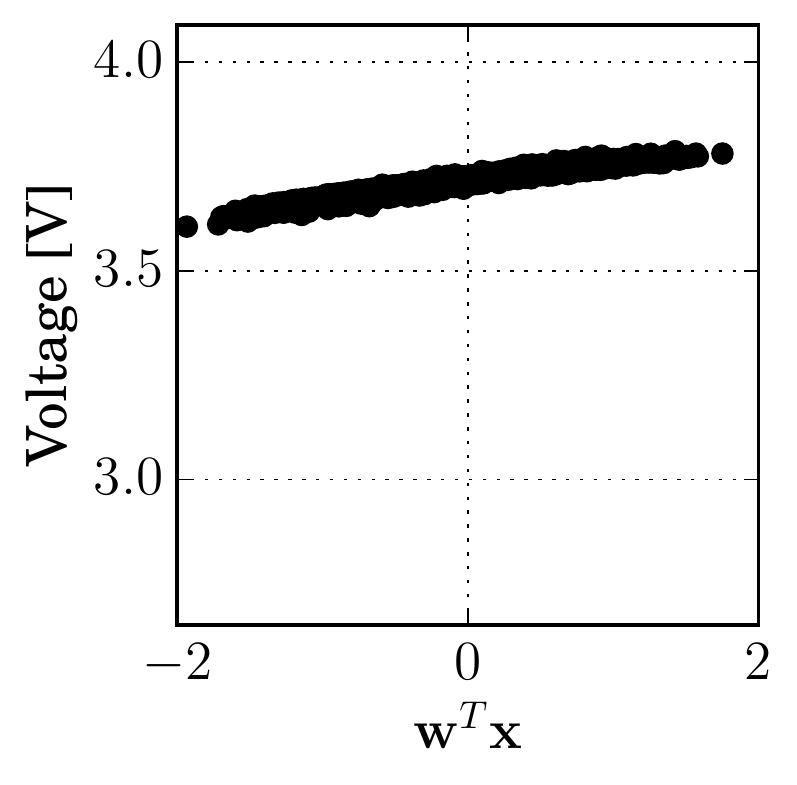}%
}
\subfloat[B ($t^\ast=23$)]{
\includegraphics[width=0.33\textwidth]{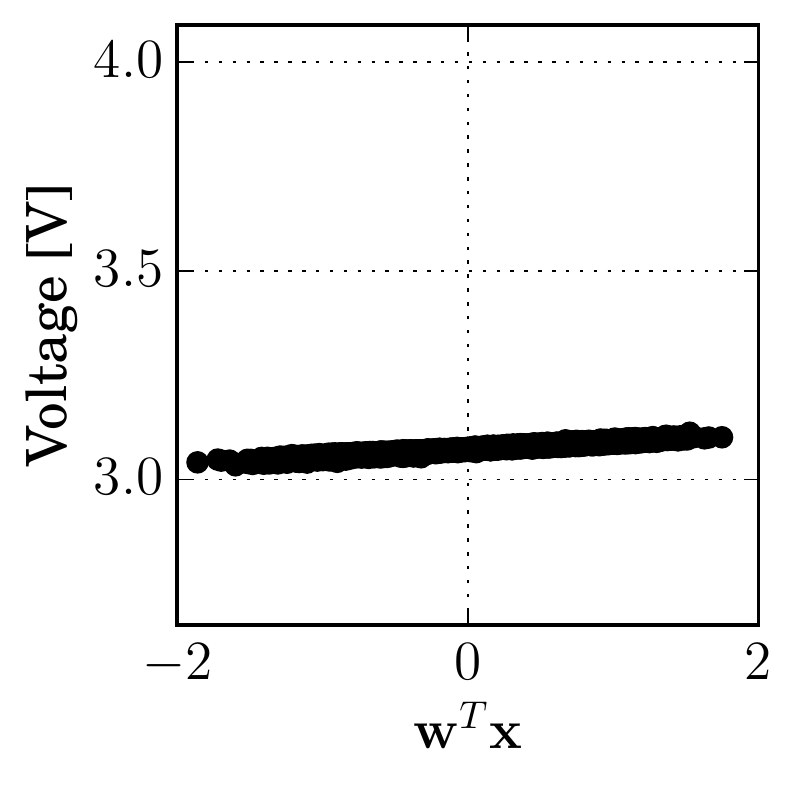}%
}
\subfloat[C ($t^\ast=5$)]{
\includegraphics[width=0.33\textwidth]{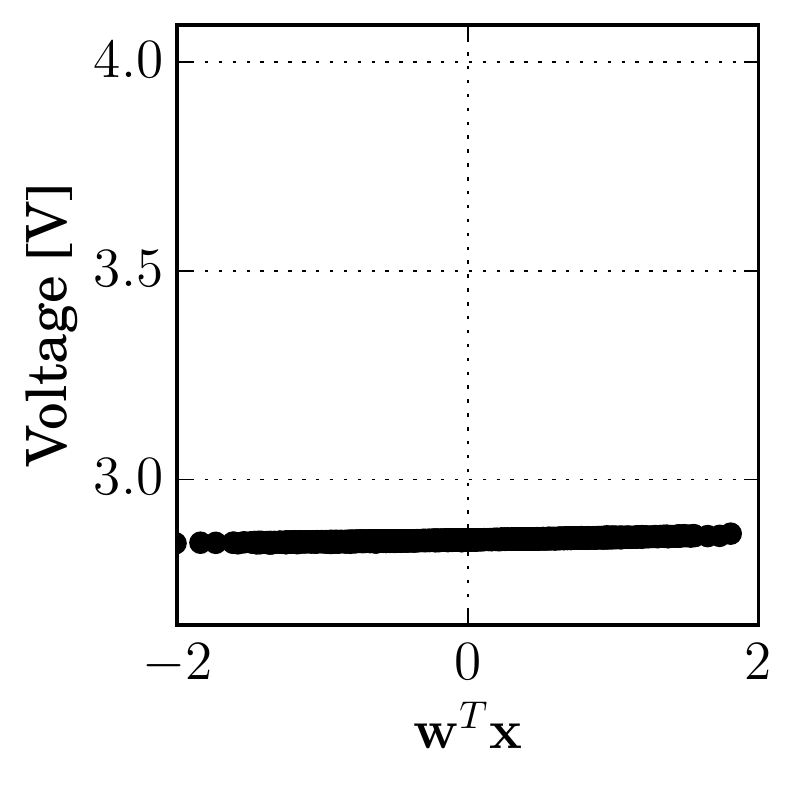}%
}
\\
\subfloat[A ($t^\ast=95$)]{
\includegraphics[width=0.33\textwidth]{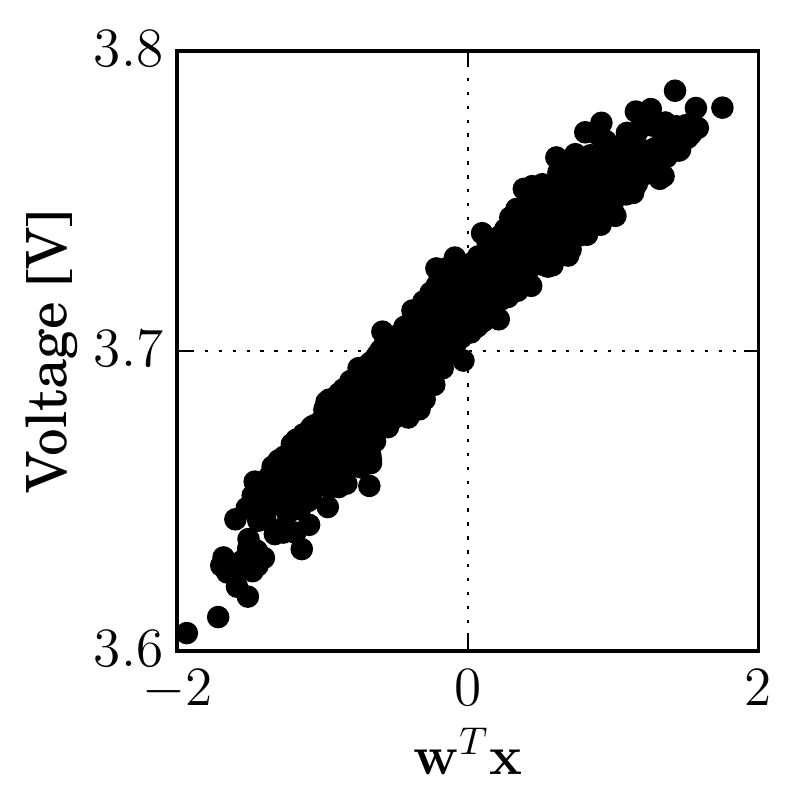}%
}
\subfloat[B ($t^\ast=23$)]{
\includegraphics[width=0.33\textwidth]{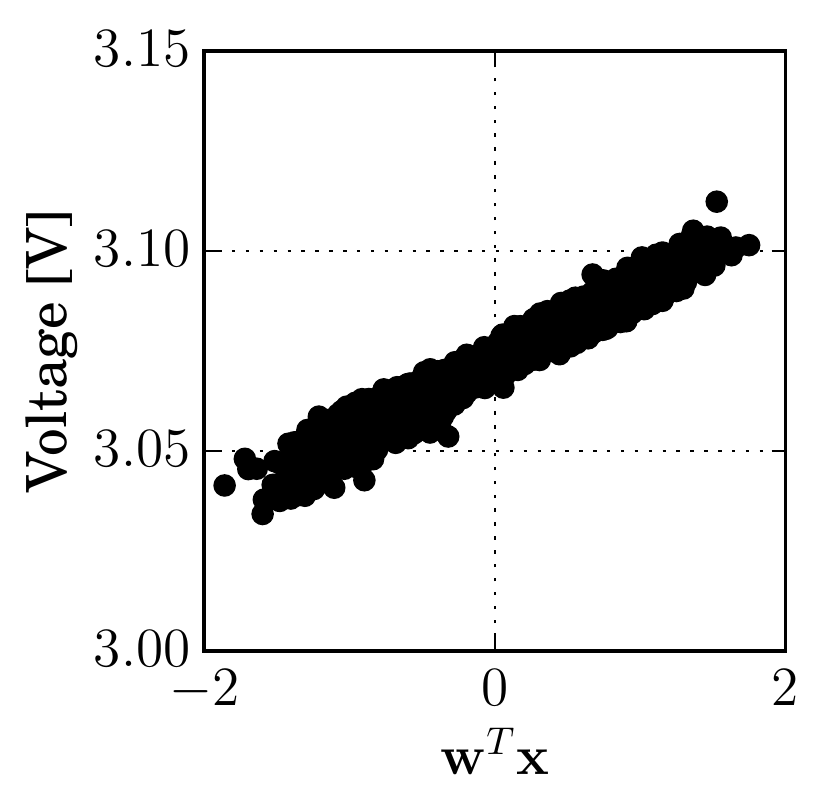}%
}
\subfloat[C ($t^\ast=5$)]{
\includegraphics[width=0.33\textwidth]{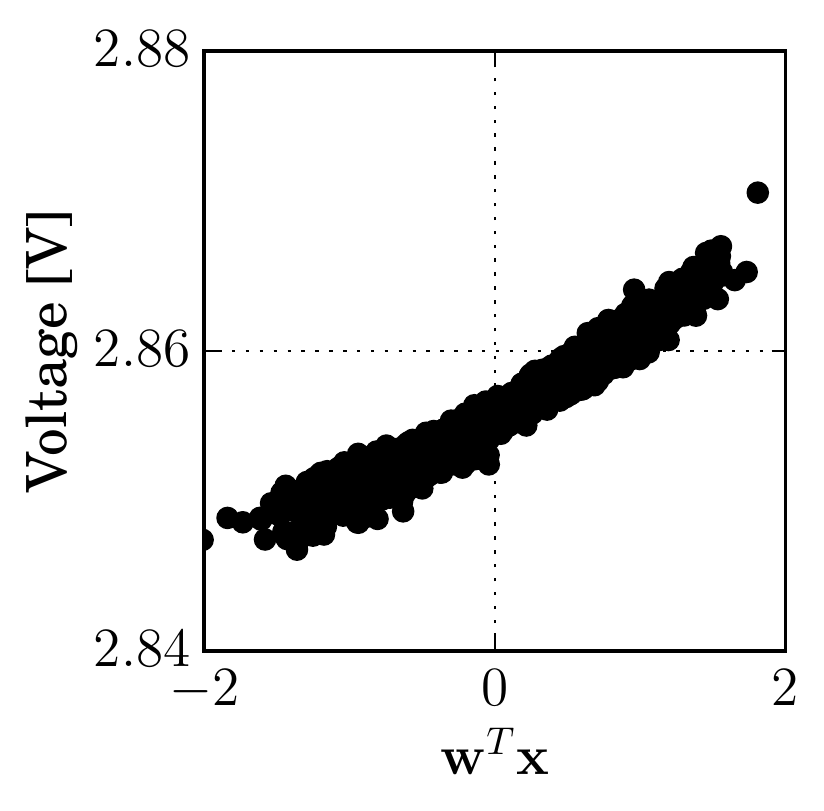}%
}
\caption{Results for voltage at discharge rate $4$C. The top figure shows the components of $\vw$ from Algorithm \ref{alg:lslin} as function of $t^*$. The middle row shows summary plots corresponding to the voltages labeled A, B, and C in the top figure. The bottom row is identical to the top row with the vertical axis zoomed to elucidate the relationship.}
\label{fig:voltage4C}
\end{figure}

Figures \ref{fig:voltage025C}, \ref{fig:voltage1C}, and \ref{fig:voltage4C} show the components of $\vw$ from \eqref{eq:w} and select summary plots for voltage over the discharge history, i.e., as a function of the scaled time $t^\ast$, for the three discharge rates. The format of the figures is identical to the figures for capacity. Compared to capacity, the relationship between the inputs and voltage is better modeled by a univariate function of $\vw^T\vx$ across the discharge rates and chosen $t^\ast$ values. Figures \ref{fig:voleig025}, \ref{fig:voleig1}, and \ref{fig:voleig4} show the first 5 of 19 eigenvalues from \eqref{eq:Heigs} at select $t^\ast$ values (A, B, and C) across the three discharge rates. The eigenvalue gaps between the first two eigenvalues are generally smaller for the 4C discharge rate than the other discharge rates; also, univariate relationships in the corresponding summary plots are much less pronounced. Combined, these plots provide consistent evidence for the degree to which voltage can be modeled as univariate function of $\vw^T\vx$. Again, the components of $\vw$ can be used as sensitivity metrics for the physical input parameters. The time histories (i.e., functions of $t^\ast$) of the sensitivity metrics are less variable than the capacity sensitivities as a function of voltage. 

These figures suggest the following insights into the model.
\begin{itemize}
\item Similar to voltage case, the rankings on input parameters induced $\vw$'s components are very similar to the rankings induced by the Sobol' total sensitivity indices from \cite{Hadigol2015}.
\item Most model parameters are not important when defining the important direction identified by Algorithm \ref{alg:lslin}. Generally, porosities and Bruggeman coefficients are among the most important parameters. This is consistent with the results from \cite{Hadigol2015}. 
\item For discharge rate 0.25C, the sizes of electrodes are important. This importance decreases as the discharge rate increases.
\item The picture for the highest discharge rate 4C is different from the others. The separator's porosity is by far the most important parameter. As the discharge process progresses, the salt diffusion coefficient suddenly becomes important; this is the only time a diffusion coefficient is important across all cases.
\end{itemize}

\begin{figure}[!h]
\centering
\subfloat[Capacity, $I=0.25C$]{
\label{fig:capeig025}
\includegraphics[width=0.37\textwidth]{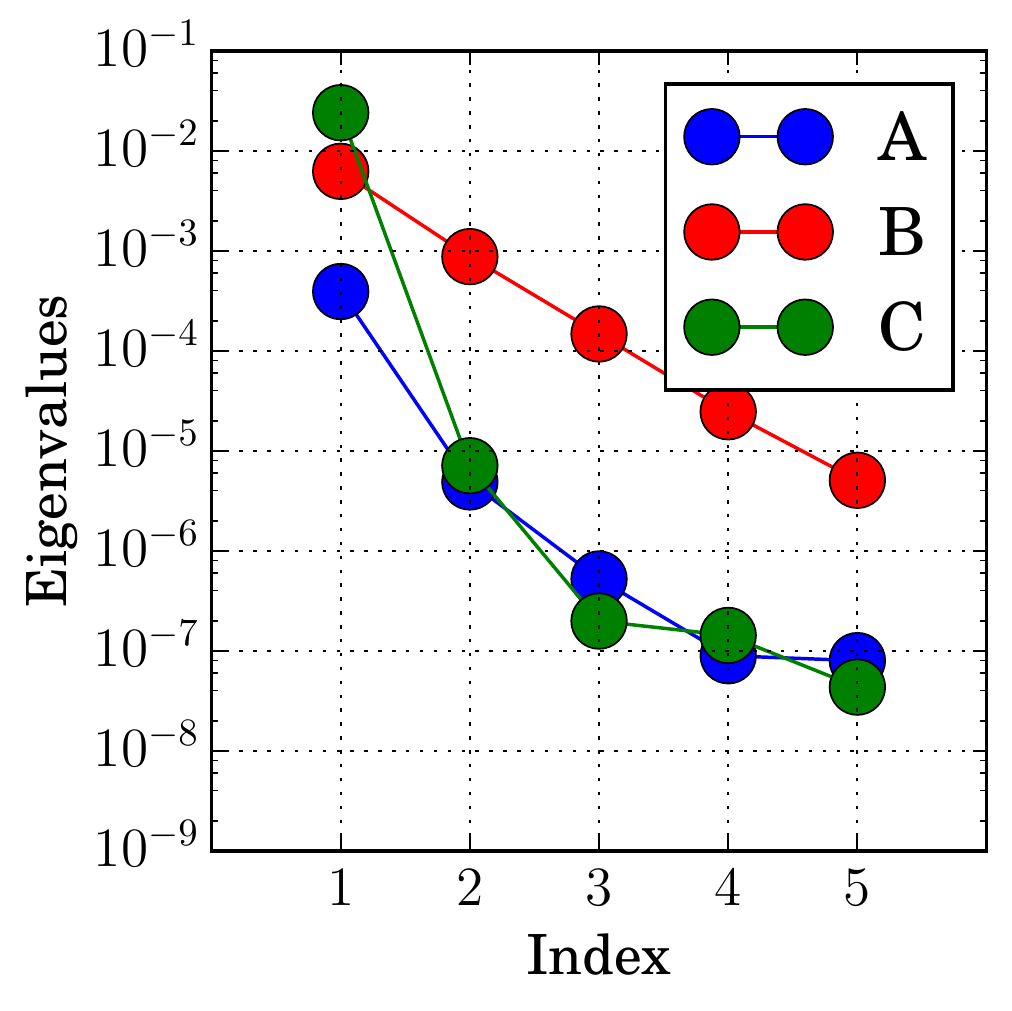}%
}
\subfloat[Voltage, $I=0.25C$]{
\label{fig:voleig025}
\includegraphics[width=0.37\textwidth]{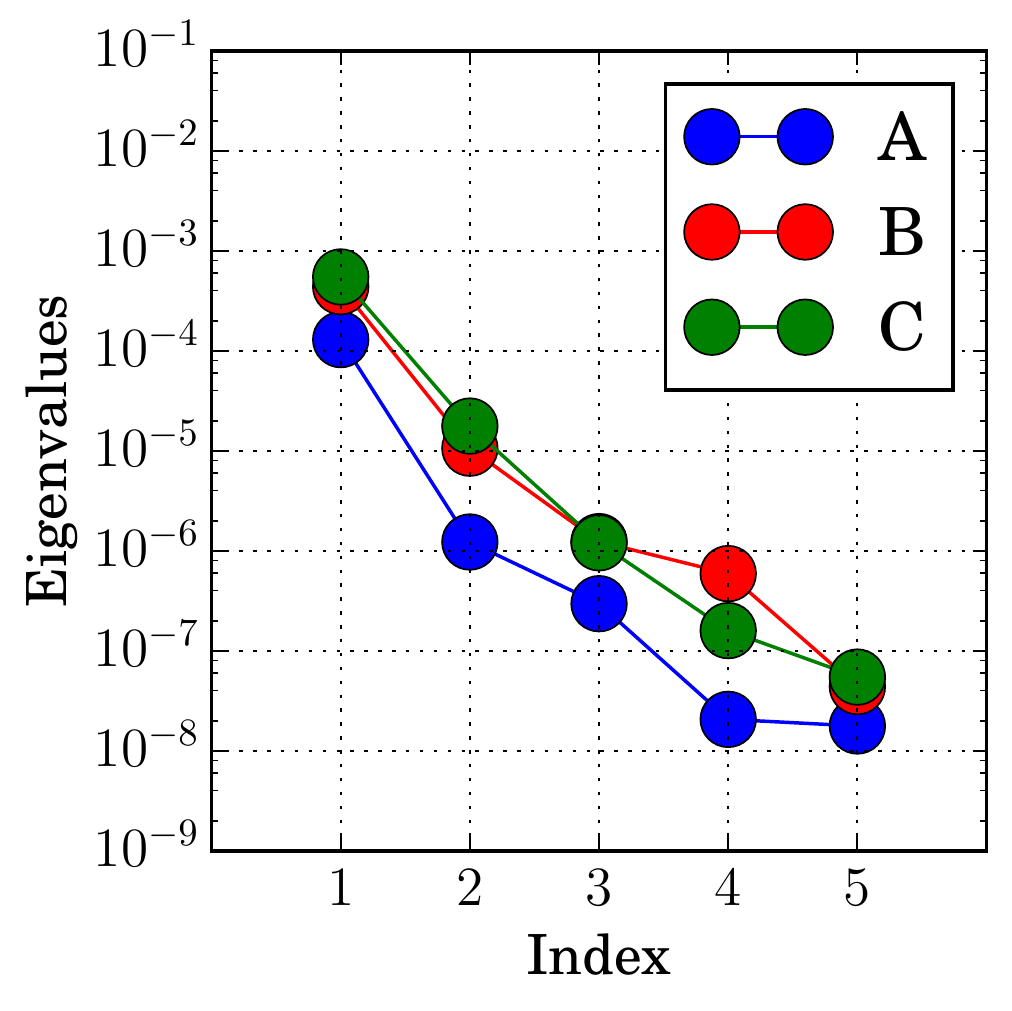}%
}
\\
\subfloat[Capacity, $I=1C$]{
\label{fig:capeig1}
\includegraphics[width=0.37\textwidth]{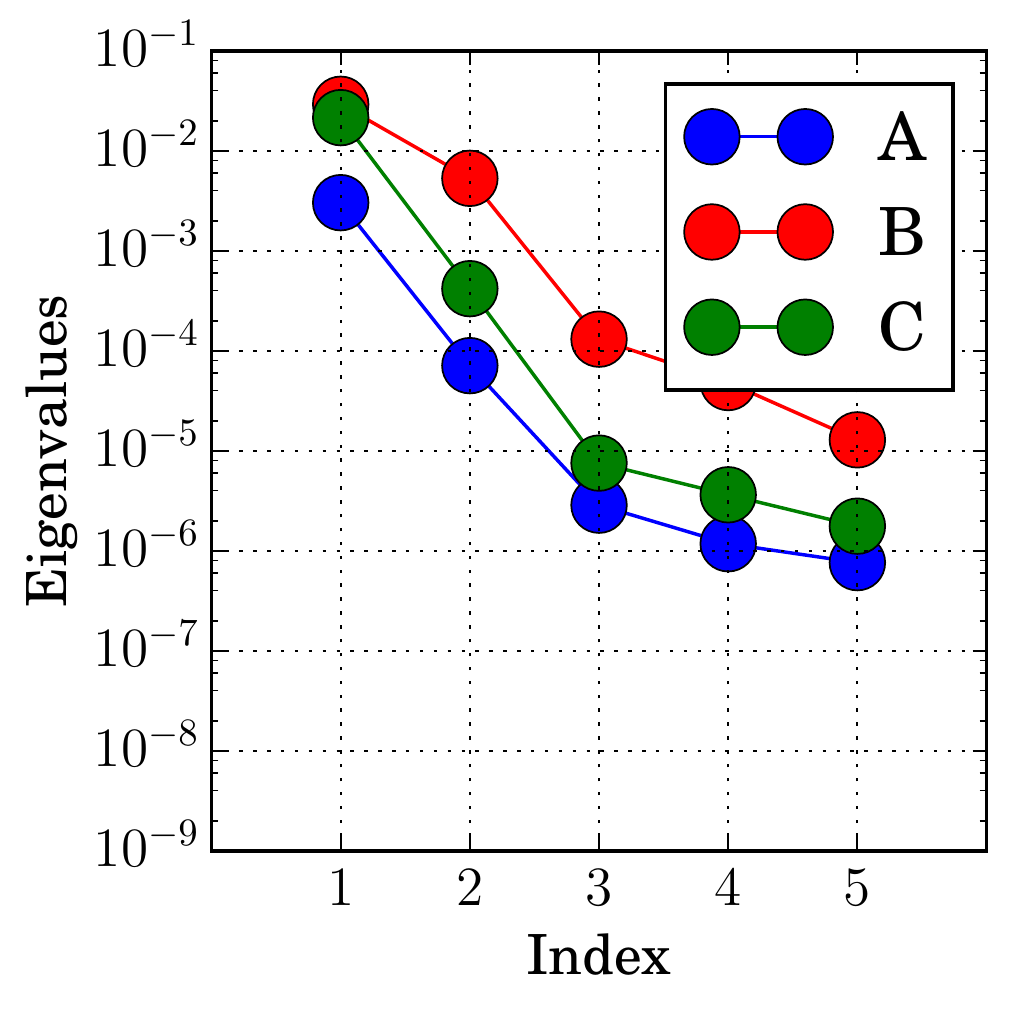}%
}
\subfloat[Voltage, $I=1C$]{
\label{fig:voleig1}
\includegraphics[width=0.37\textwidth]{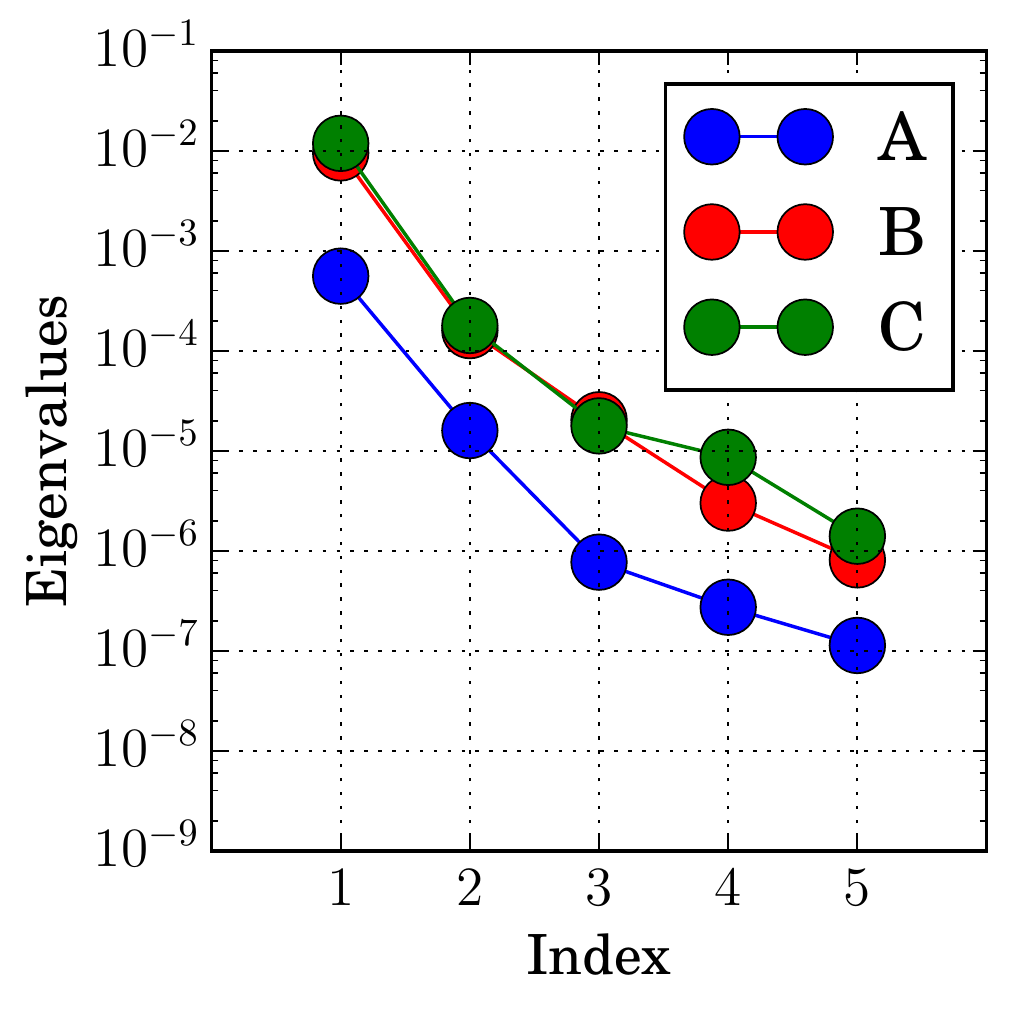}%
}
\\
\subfloat[Capacity, $I=4C$]{
\label{fig:capeig4}
\includegraphics[width=0.37\textwidth]{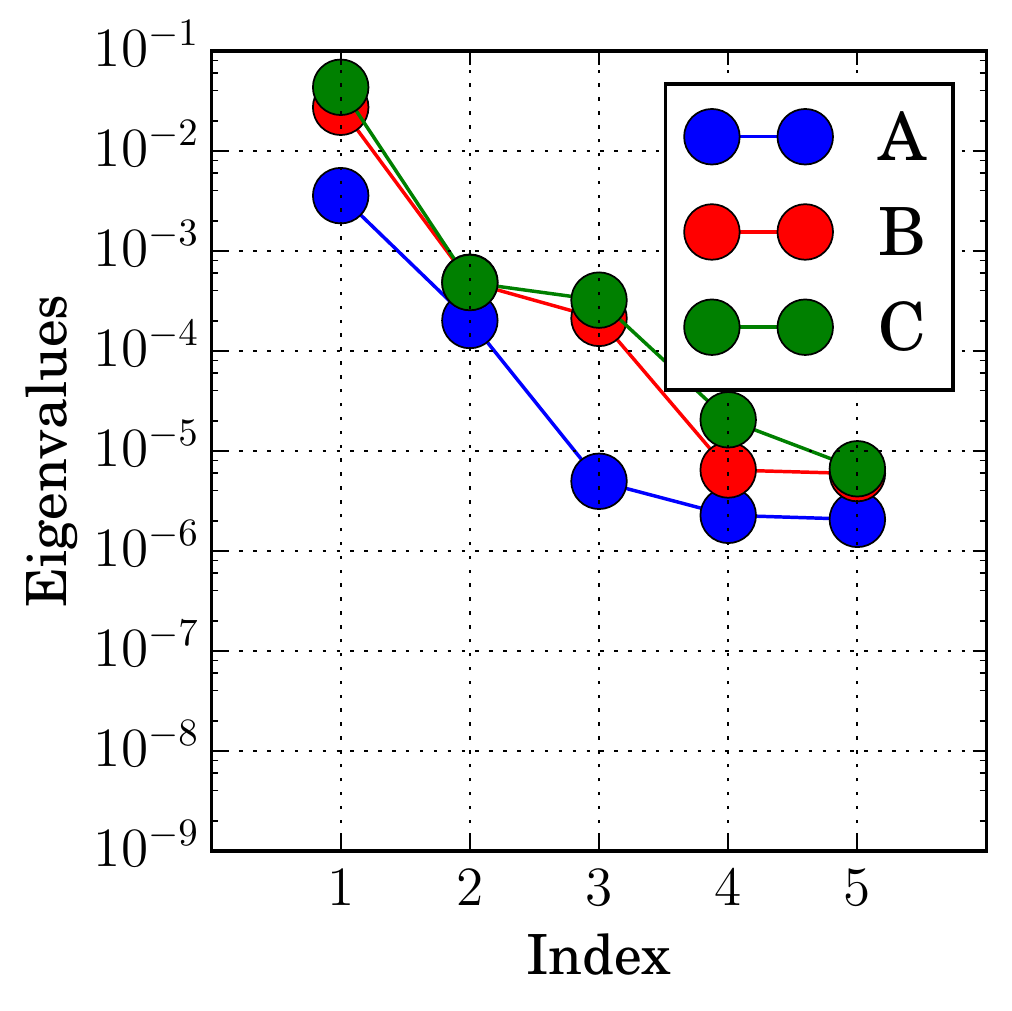}%
}
\subfloat[Voltage, $I=4C$]{
\label{fig:voleig4}
\includegraphics[width=0.37\textwidth]{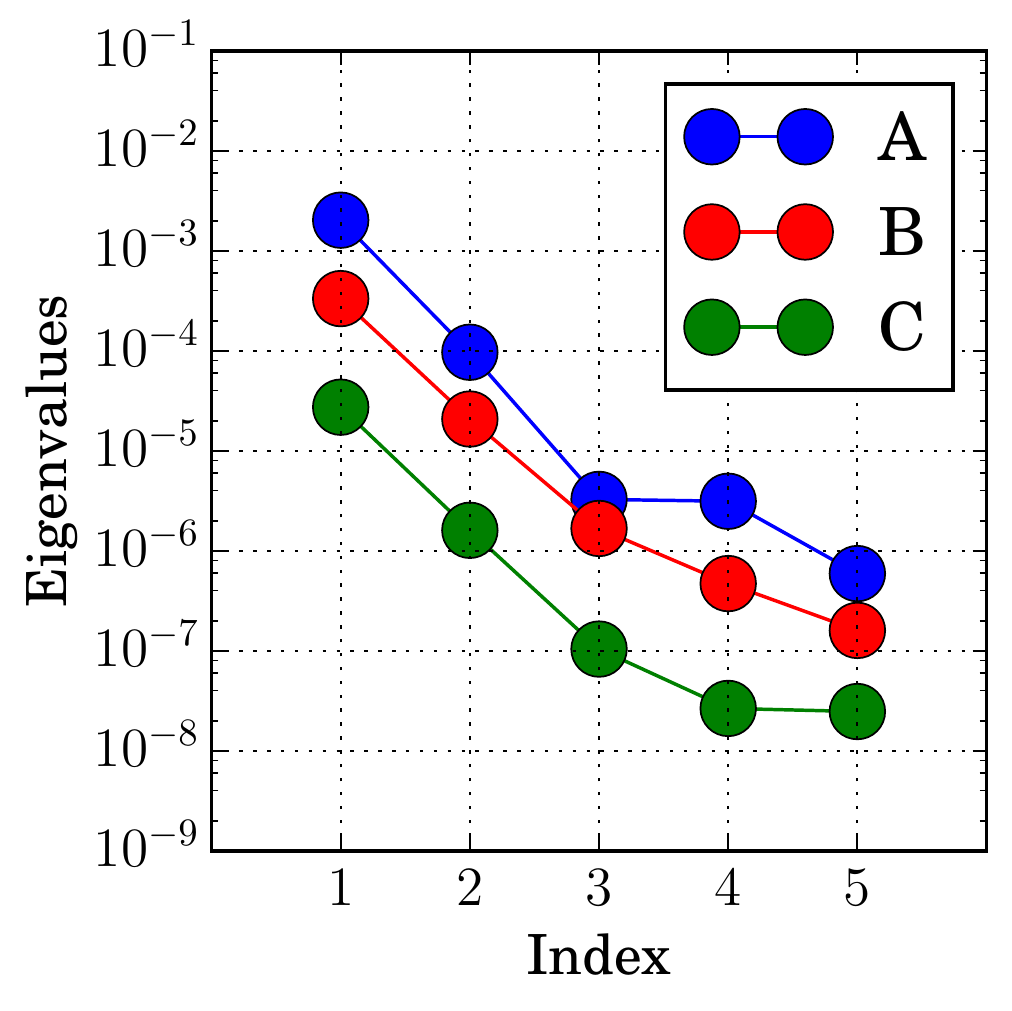}%
}
\caption{Eigenvalues from \eqref{eq:Heigs} in the quadratic model-based Algorithm \ref{alg:lsquad}. Figures \ref{fig:capeig025}, \ref{fig:capeig1}, and \ref{fig:capeig4} are for capacity as a function of voltage; see Figures \ref{fig:capacity025C}, \ref{fig:capacity1C}, and \ref{fig:capacity4C}, respectively. Figures \ref{fig:voleig025}, \ref{fig:voleig1}, and \ref{fig:voleig4} are for voltage over the discharge process; see Figures \ref{fig:voltage025C}, \ref{fig:voltage1C}, and \ref{fig:voltage4C}, respectively. 
}
\label{fig:quadeigs}
\end{figure}

\begin{figure}[!h]
\centering
\subfloat[Capacity, $I=0.25C$]{
\label{fig:capbr025}
\includegraphics[width=0.47\textwidth]{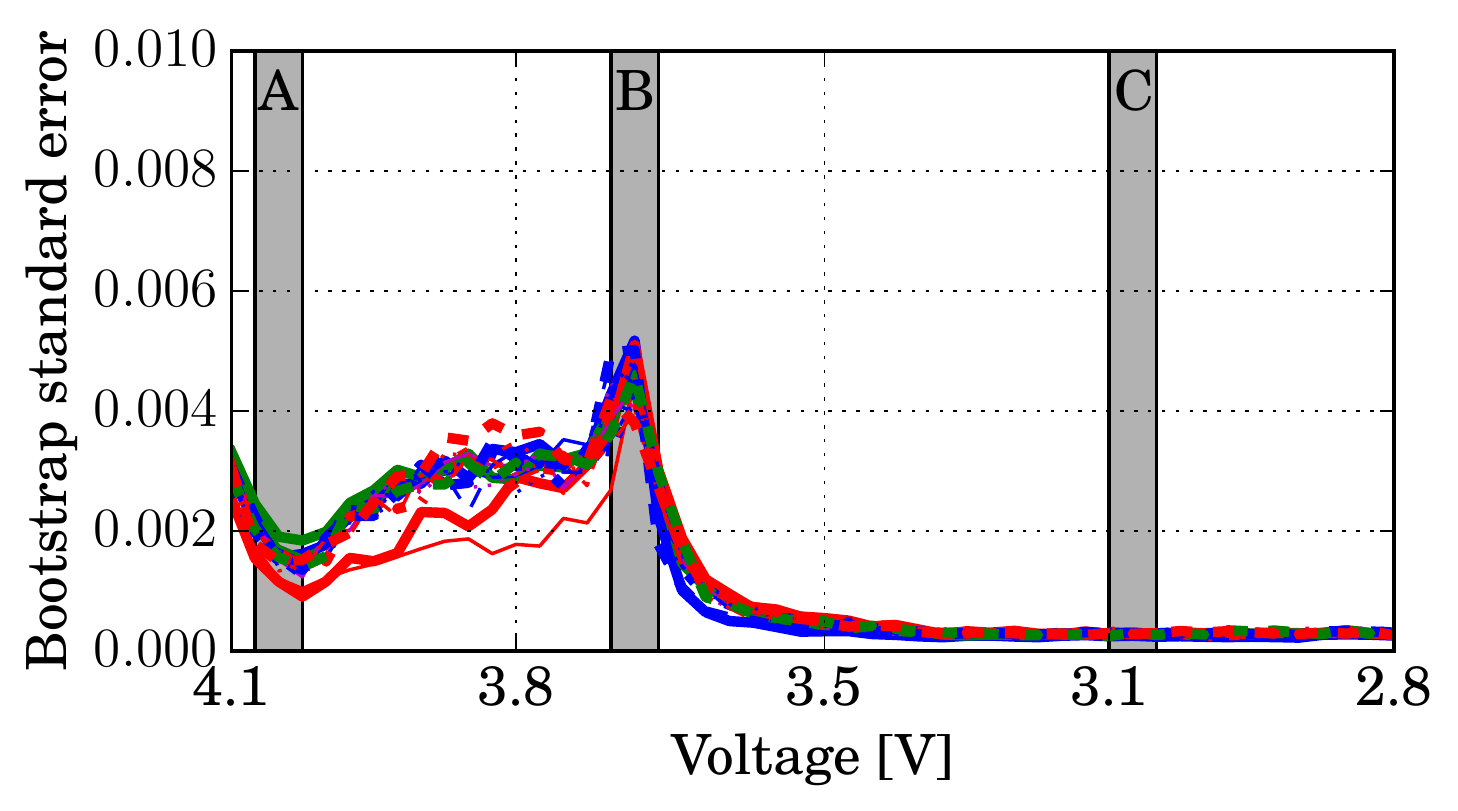}%
}
\subfloat[Voltage, $I=0.25C$]{
\label{fig:volbr025}
\includegraphics[width=0.47\textwidth]{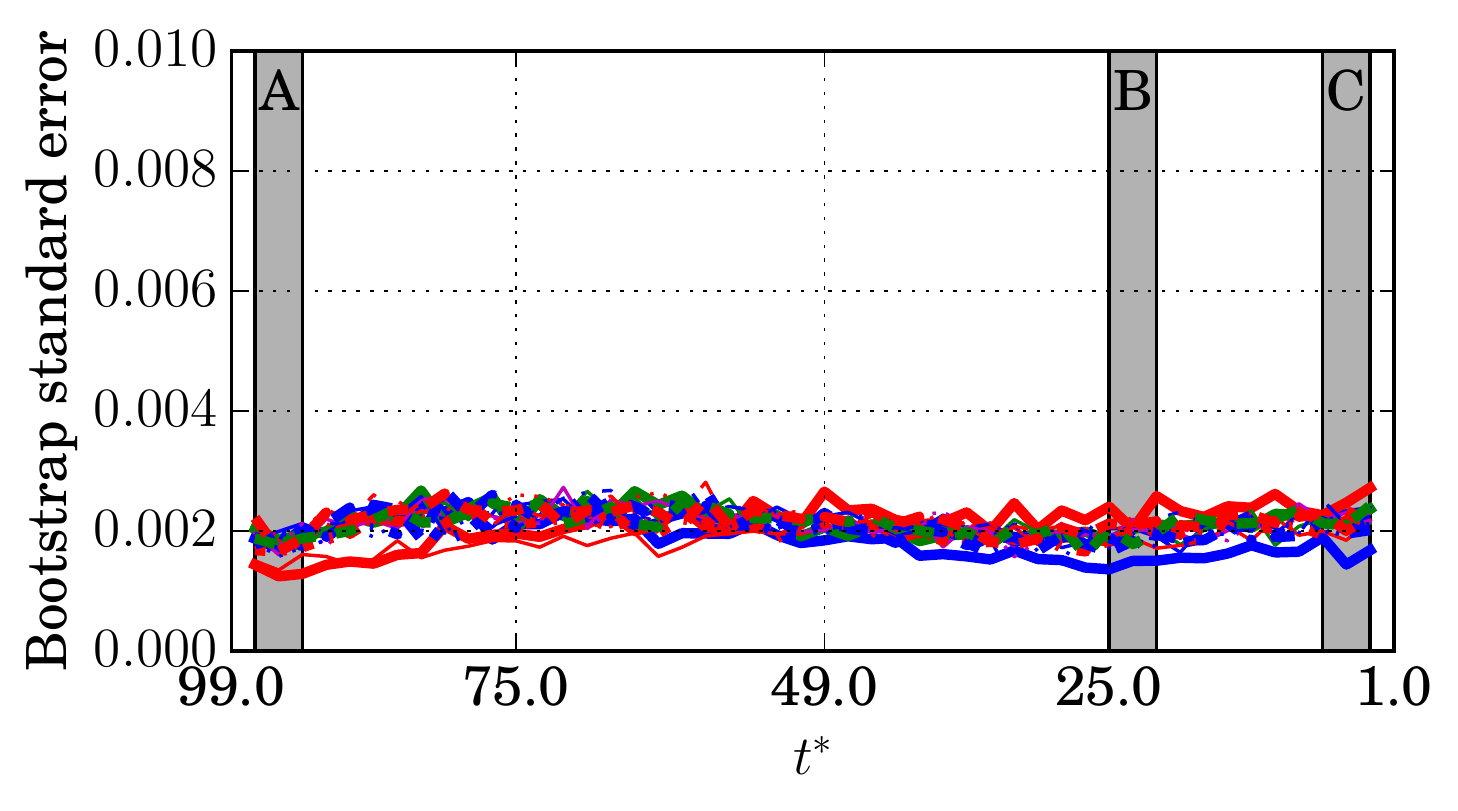}%
}
\\
\subfloat[Capacity, $I=1C$]{
\label{fig:capbr1}
\includegraphics[width=0.47\textwidth]{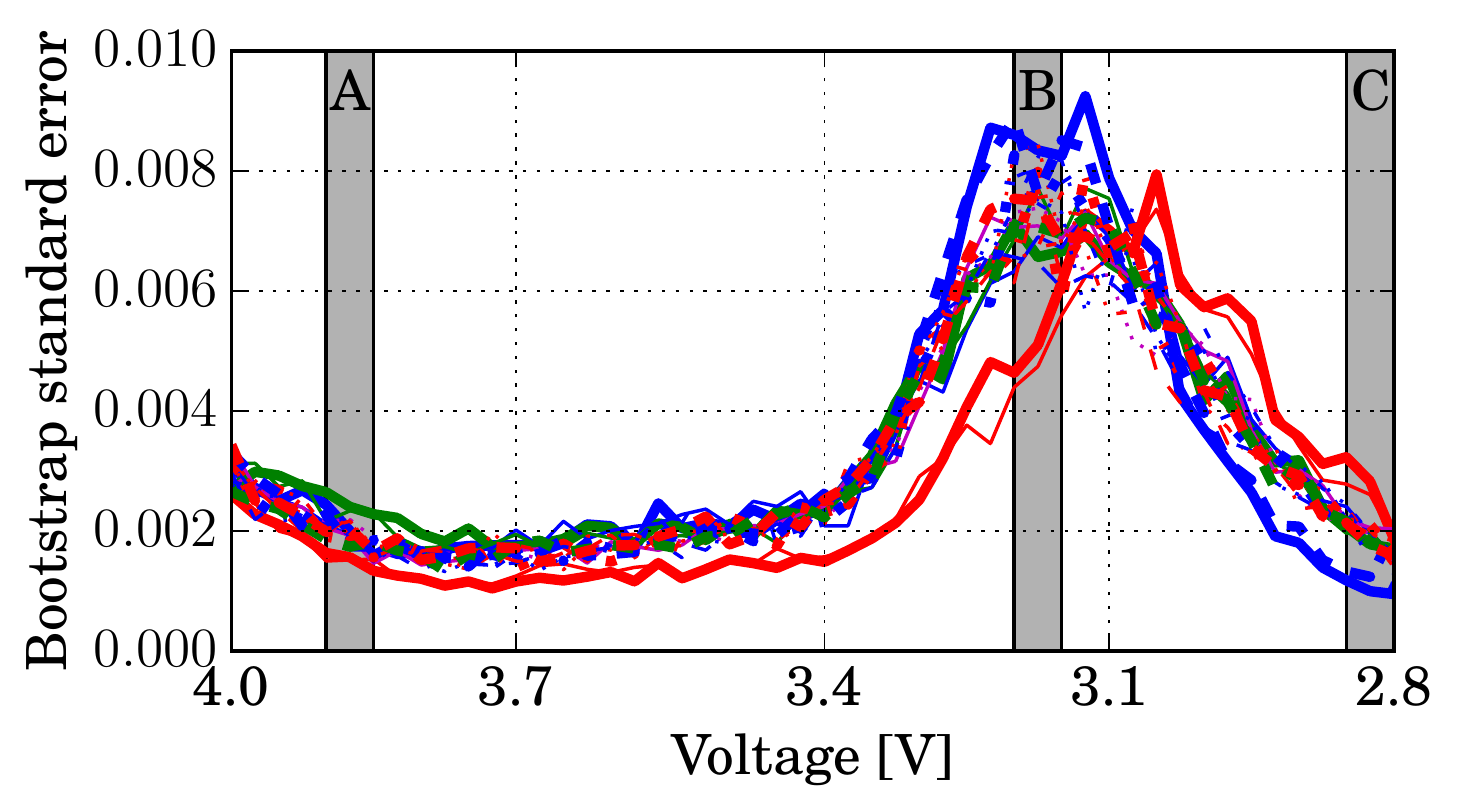}%
}
\subfloat[Voltage, $I=1C$]{
\label{fig:volbr1}
\includegraphics[width=0.47\textwidth]{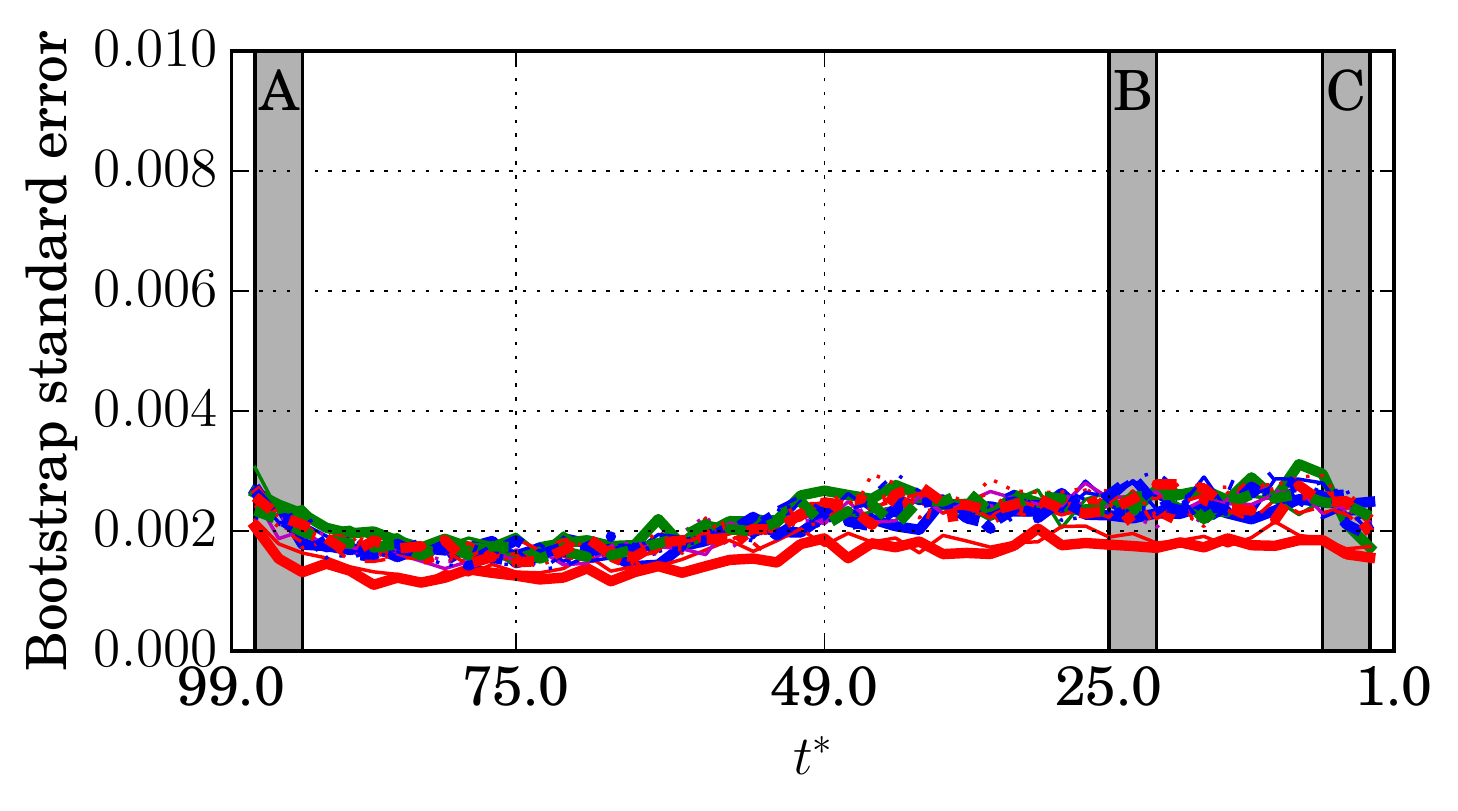}%
}
\\
\subfloat[Capacity, $I=4C$]{
\label{fig:capbr4}
\includegraphics[width=0.47\textwidth]{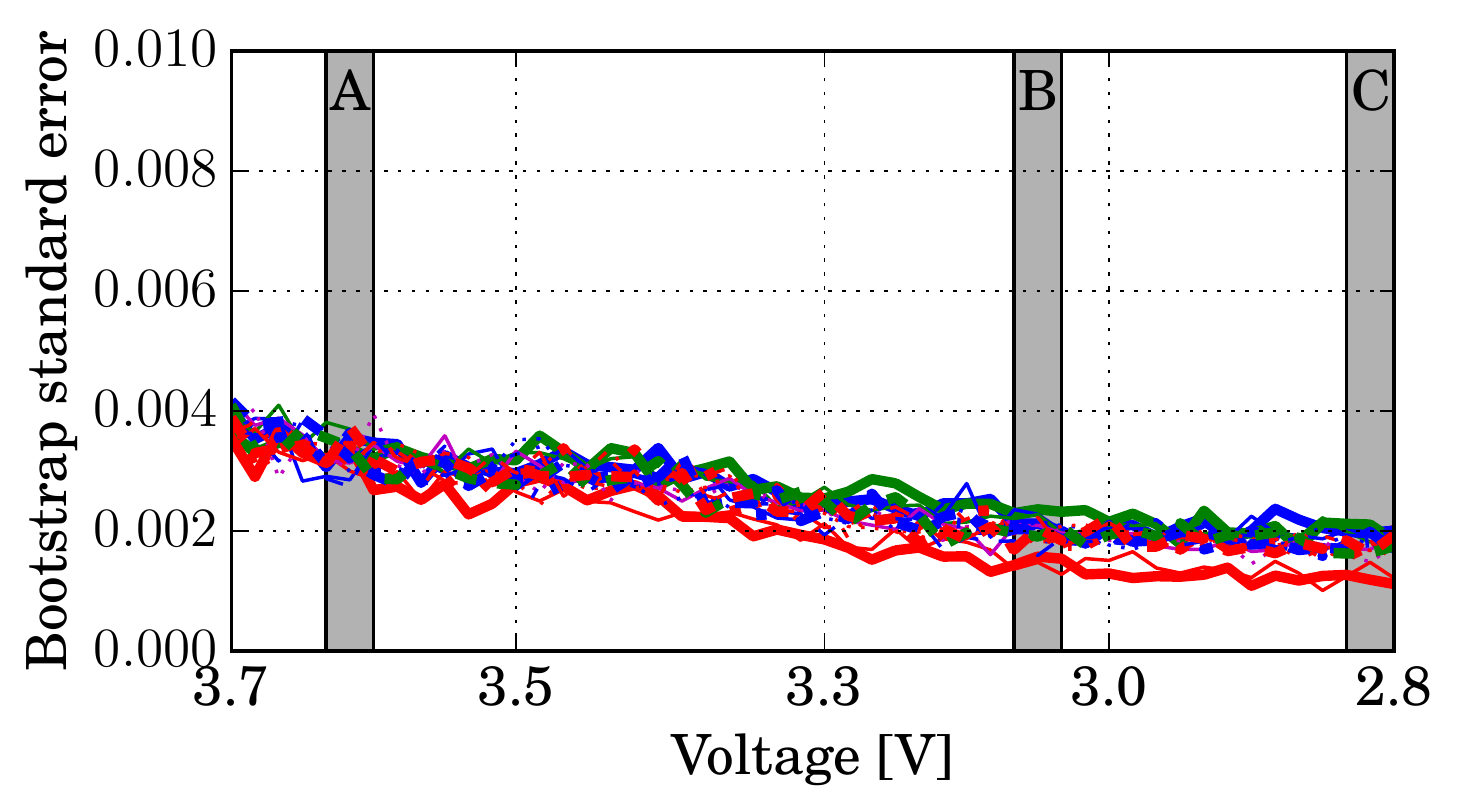}%
}
\subfloat[Voltage, $I=4C$]{
\label{fig:volbr4}
\includegraphics[width=0.47\textwidth]{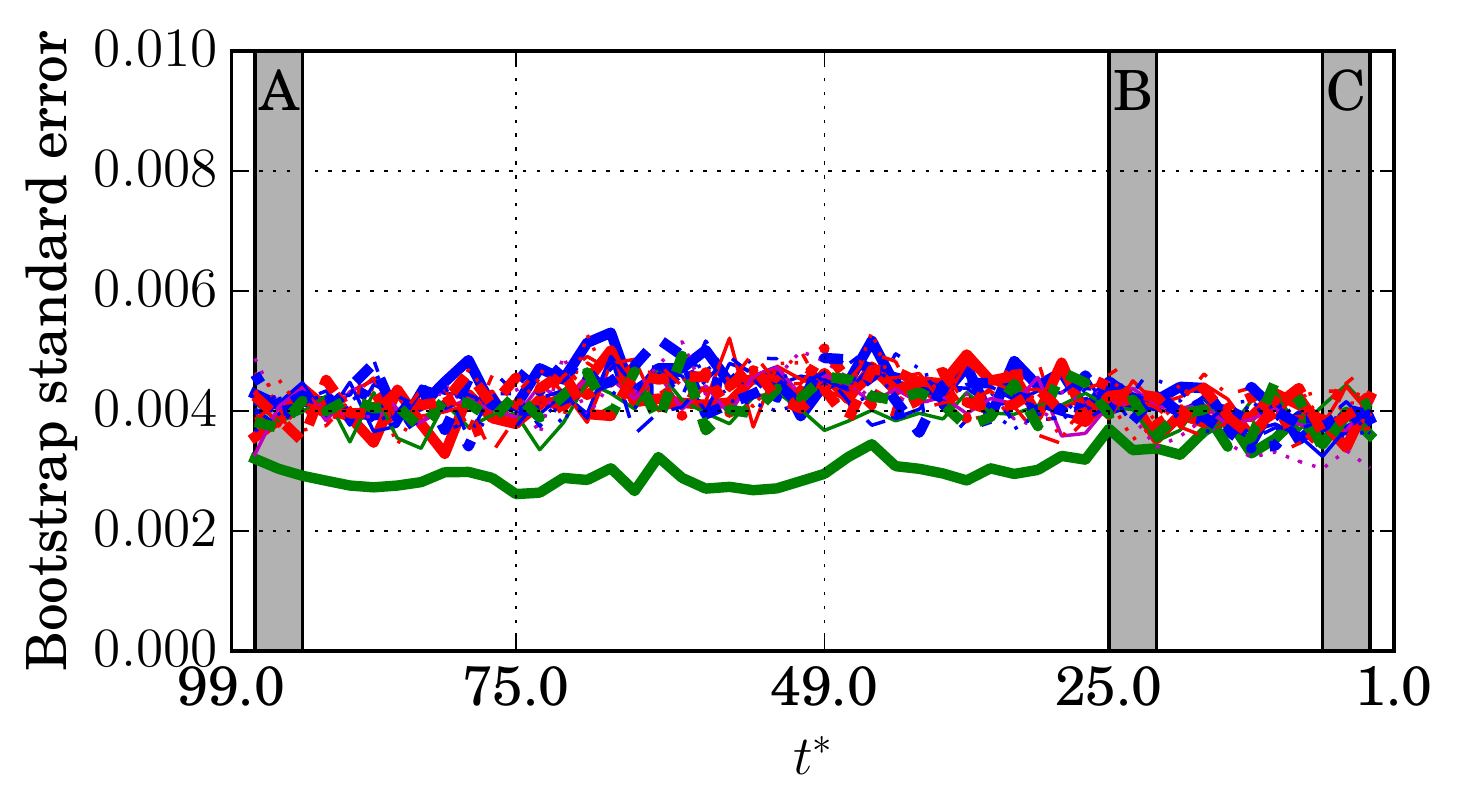}%
}
\caption{Bootstrap standard errors for the components of $\vw$. Figures \ref{fig:capbr025}, \ref{fig:capbr1}, and \ref{fig:capbr4} are for capacity as a function of voltage; see Figures \ref{fig:capacity025C}, \ref{fig:capacity1C}, and \ref{fig:capacity4C}, respectively. Figures \ref{fig:volbr025}, \ref{fig:volbr1}, and \ref{fig:volbr4} are for voltage over the discharge process; see Figures \ref{fig:voltage025C}, \ref{fig:voltage1C}, and \ref{fig:voltage4C}, respectively. 
}
\label{fig:boot}
\end{figure}

Figure \ref{fig:boot} shows the bootstrap standard errors on the components of $\vw$, where the standard errors are computed with 100 bootstrap replicates from the data set of 3600 runs. We stress that the bootstrap standard errors do not have the proper statistical interpretation, since there is no noise in the data. However, large bootstrap standard error does correspond to activity in the components of $\vw$ as a function of their independent coordinate (voltage or $t^\ast$). This indicates how the components of $\vw$ vary over the bootstrap replicates. Low errors suggest the coefficients are stable with respect to resampling (with replacement) from the data set. Notably, the regions of relatively large bootstrap error---region B in Figures \ref{fig:capbr025} and \ref{fig:capbr1}---correspond to summary plots that have the largest spread around a univariate functional relationship, i.e., where a univariate function of $\vw^T\vx$ is least appropriate. This connection is very intriguing, and we expect to explore this connection beyond the Li battery application in future work.

\section{Summary and conclusions}
\label{sec:conclusion}

Active subspaces are part of an emerging set of tools for discovering and exploiting a particular type of low-dimensional structure in functions of several variables. In particular, given (i) a scalar-valued function $f:\mathbb{R}^m\rightarrow\mathbb{R}$ and (ii) a probability density function associated with the function's inputs, the active subspace is defined by the $n$-dimensional eigenspace, with $n<m$, of a matrix-valued functional of $f$'s gradient. When gradients are not available, one may approximate the active subspace using models for $f(\vx)$; Algorithms \ref{alg:lslin} and \ref{alg:lsquad} use least-squares-fit linear and quadratic functions as the underlying model. In this paper, we extended these ideas to functions that also depend on another independent variable, such as time. 

We applied this approach to a set of 3600 input/output pairs from three varieties of a simulated lithium battery's discharge process, where each variety uses a different constant discharge rate. The model contained 19 input parameters, and we examined two output quantities: (i) capacity as a function of voltage and (ii) voltage as a function of time. In every case, there exists a one-dimensional active subspace in the 19-dimensional parameter space, though output variation orthogonal to the active direction changes across output quantities; evidence can be seen in both (i) the summary plots constructed with $\vw$ from the linear model-based Algorithm \ref{alg:lslin} and (ii) the eigenvalues from the quadratic model-based Algorithm \ref{alg:lsquad}. Therefore, the components of the vector that defines the active subspace can be used as sensitivity metrics for the 19 parameters. The components' behavior over the independent coordinate (voltage or time, respectively) reveals stages of activity in the output quantities. All insights derived from the sensitivity metrics is consistent with the Sobol' index study from \cite{Hadigol2015}. Some insights are entire novel, and we expect these analyses may aid Li battery designers that employ computational models.

\appendix 

\section{Lithium ion battery model equations}
\label{sec:goveq}


The lithium ion battery (LIB) simulations of this study are performed using the widely used Newman's electrochemical model~\citep{Newman75, DFN:93, Doyle96b}. Based on the porous electrode~\citep{West82} and concentrated solution~\citep{Newman75} theories, Newman's model describes the Li$^{+}$ ion transport and concentration through the electrolyte, the Li ion concentration in the solid phase, as well as the electric current carried by the electrolyte and the electrodes. At the anode (or cathode) particle surfaces, the charge transfer is described by the Butler-Volmer kinetic model~\citep{DFN:93}.

Table \ref{table:coupled_GE}, adopted from \cite{Hadigol2015}, presents the governing equations of Newman's model along with the associated boundary conditions for each equation. For computational efficiency, these non-linear, coupled equations are solved in a decoupled fashion described in \cite{Reimers13,Hadigol2015}. Following a similar notation as in \cite{Hadigol2015}, the field variables and the parameters of Table \ref{table:coupled_GE} are as follows:

\begin{itemize}
\item $c$: postive Li ion (Li$^+$) concentration in liquid phase [$\mathrm{mol} \cdot \mathrm{m^{-3}}$]
\item $c_s$: lithium concentration in solid phase [$\mathrm{mol} \cdot \mathrm{m^{-3}}$]
\item $c_s^{\mathrm{surf}}$: lithium concentration in solid phase at $r=r_s$ [$\mathrm{mol} \cdot \mathrm{m^{-3}}$]
\item $\phi_{e}$: Li$^+$ ion potential in liquid phase [$\mathrm{V}$]
\item $\phi_{s}$: electron potential in the solid phase [$\mathrm{V}$]
\item $\eta$: over-potential in electrodes [$\mathrm{V}$]
\item $L$: width of the cell [m]
\item $x$: distance from anode [m]
\item $r$: micro-scale distance from the center of solid particle [m]
\item $F$: Faraday's constant = 97484 [$\mathrm{C} \cdot \mathrm{mol}^{-1}$]
\item $t$: time [s]
\item $a$: active particle surface area per unit volume of electrode [$\mathrm{m}^{2} \cdot \mathrm{m^{-3}}$]
\item $\epsilon$: porosity of electrodes and stack
\item $j_{vol}$: volumetric reaction flux in the pore walls [$\mathrm{amp} \cdot \mathrm{m^{-3}}$]
\item $I$: total current density across the stack [$\mathrm{amp} \cdot \mathrm{m^{-2}}$]
\item $i_{ex}$: exchange current density of an electrode reaction [$\mathrm{amp} \cdot \mathrm{m^{-2}}$]
\item $T$: temperature [K]
\item $r_s$: solid particle size [m]
\item $t_+^0$: Li$^+$ transference number
\item $\tau$: tortuosity
\item $D_s$: diffusion coefficient of the solid phase [$\mathrm{m^{-2}} \cdot \mathrm{s^{-1}}$]
\item $D$: diffusion coefficient of the liquid phase [$\mathrm{m^{-2}} \cdot \mathrm{s^{-1}}$]
\item $\sigma$: electronic conductivity of the solid phase [$\mathrm{S} \cdot \mathrm{m^{-1}}$]
\item $\kappa$: electronic conductivity of the liquid phase [$\mathrm{S} \cdot \mathrm{m^{-1}}$]
\item $\kappa_D$: liquid phase diffusional conductivity [$\mathrm{S} \cdot \mathrm{m^{-1}}$]
\item $k$: reaction rate constant [$\mathrm{m}^4 \cdot \mathrm{mol} \cdot \mathrm{s}$]
\item Subscript $a$: anode
\item Subscript $s$: separator
\item Subscript $c$: cathode
\item Subscript $e$: electrolyte
\item $\text{eff}$: effective value
\end{itemize}
Several of the preceeding parameters appear in Table \ref{tab:inputs} with modeled distributions.

\begin{table}[h]
\caption{Governing equations of LIB used in this study, adapted from \cite{Hadigol2015}.} 
\centering
\resizebox{\textwidth}{!}{%
\begin{tabular}[t]{ >{\normalsize} p{2cm}  >{\normalsize} c  c }   
\hline \hline
  & \textbf{Governing equation} & \textbf{Boundary conditions} \\  
\hline \hline
 Electrolyte phase diffusion & 
\begin{minipage}{9cm} \begin{equation} \label{eqn:c} \begin{aligned} \frac{\partial ( \epsilon c )}{\partial t} = \nabla (\epsilon D^{\mathrm{eff}} \nabla c) + \frac{1-t_+^0}{F} j_{vol} \end{aligned} \end{equation} \end{minipage}
&  \begin{minipage}{6cm} \begin{equation} \begin{aligned} \nabla c|_{x=0} = \nabla c|_{x=L} = 0 \end{aligned}  \nonumber \end{equation} \end{minipage}  \\ 
\hline
Solid phase diffusion & 
\begin{minipage}{9cm} \begin{equation} \label{eqn:c_s} \begin{aligned} \frac{\partial c_s }{\partial t} = \frac{1}{r^2} \frac{\partial}{ \partial r} \Big( D_s r^2 \frac{\partial}{ \partial r} c_s \Big) \end{aligned} \end{equation} \end{minipage}
&  \begin{minipage}{6cm} \begin{eqnarray} && \nabla c_s|_{r=0} = 0 \nonumber \\ &&  \nabla c_s|_{r=r_s} = -\frac{j_{vol}}{a F D_s}  \nonumber \end{eqnarray} \end{minipage} \\ 
\hline
Liquid phase potential & 
\begin{minipage}{9cm} \begin{equation} \label{eqn:phi_e} \begin{aligned} \nabla ( \kappa^{\mathrm{eff}} \nabla \phi_{e}) - \nabla (\kappa_D^{\mathrm{eff}} \nabla \ln c) + j_{vol} = 0 \end{aligned} \end{equation} \end{minipage}
& \begin{minipage}{6cm} \begin{eqnarray} && \nabla \phi_{e}|_{x=0} = \nabla \phi_{e}|_{x=L} = 0 \nonumber \\ && \phi_{e}|_{x=L} = 0  \nonumber \end{eqnarray} \end{minipage} \\ 
\hline
Solid phase potential & 
\begin{minipage}{9cm} \begin{equation} \label{eqn:phi_s} \begin{aligned} \nabla ( \sigma^{\mathrm{eff}} \nabla \phi_{s}) - j_{vol} = 0 \end{aligned} \end{equation} \end{minipage} 
&  \begin{minipage}{6cm} \begin{eqnarray}  && \nabla \phi_{s}|_{x=0} = \nabla \phi_{s}|_{x=L} = \frac{-I}{\sigma^{\mathrm{eff}}} \nonumber \\
&&  \nabla \phi_{s}|_{x=L_a} = \nabla \phi_{s}|_{x=L_a+L_s} = 0 \nonumber \end{eqnarray} \end{minipage} \\ 
\hline
Reaction kinetics & \multicolumn{2}{l }{ \begin{minipage}{10cm} \begin{eqnarray} \label{eqn:j_vol} \begin{aligned} && j_{vol} = a i_{ex} \Big[ \exp \Big( \frac{0.5 F \eta}{RT} \Big) - \exp \Big( - \frac{0.5 F \eta}{RT} \Big)\Big] \\ && i_{ex} = Fk(c_s^{\mathrm{surf}})^{0.5}(c_{s,max} - c_s^{\mathrm{surf}})^{0.5}(c)^{0.5} \end{aligned} \end{eqnarray} \end{minipage} }   \\
\hline
\end{tabular}}
\label{table:coupled_GE} 
\end{table}




\newpage

\bibliographystyle{natbib}
\bibliography{batteries}

\end{document}